\documentclass[a4paper,12pt]{article}
\usepackage[utf8]{inputenc}
\usepackage{cancel}
\usepackage{ulem}
\usepackage{amsfonts}
\usepackage{amssymb}
\usepackage{graphicx}
\usepackage{amsmath,bm}
\usepackage{enumerate}
\usepackage{mathtools}
\usepackage{tikz} \usetikzlibrary{calc}
\usepackage[countmax]{subfloat}
\usepackage{movie15}
\usepackage{xcolor}
\usepackage{float}
\usepackage{color}
\usepackage{here}
\usepackage{cite}
\usepackage{subcaption}
\usepackage{mwe}
\usepackage{lscape}
 
\newcommand{\floor}[1]{\lfloor {#1} \rfloor}
\usepackage{mathrsfs}
\usepackage{float,epsfig}
\usepackage{dcolumn}% Align table columns on decimal point
\usepackage{pgfplots}
\usepackage{bm}% bold math
\usepackage{amsmath,amssymb,amsthm}
\usepackage[colorlinks=true,linkcolor=blue,citecolor=red]{hyperref}
\newcommand{\iu}{{i\mkern1mu}}

\usepackage{animate}

\definecolor{lime}{HTML}{A6CE39}
\newcommand{\orcidicon}{%
	\begin{tikzpicture}
	\draw[lime, fill=lime] (0,0)
	circle [radius=0.16]
	node[white] {{\fontfamily{qag}\selectfont \tiny ID}};
	\draw[white, fill=white] (-0.0625,0.095)
	circle [radius=0.007];
	\end{tikzpicture}   \hspace{-2mm}
}
\newcommand\orcidHasan{{\href{https://orcid.org/0000-0001-7408-0910}{\orcidicon}}}
\newcommand\orcidKarima{{\href{https://orcid.org/0000-0001-5419-8516}{\orcidicon}}}
\newcommand\orcidFaical{{\href{https://orcid.org/0000-0002-2977-0821}{\orcidicon}}}
\title{\bf Riemann Surfaces and Winding Numbers of R\'enyi Phase Structure of Charged-Flat Black Holes 
}

\author{
F. Barzi\orcidFaical\!\!$^{1,3}$\thanks{faical.barzi@edu.uiz.ac.ma (Corresponding author)},  
 H.  El Moumni\orcidHasan\!\!$^1$\thanks{h.elmoumni@uiz.ac.ma}, K. Masmar\orcidKarima\!\!$^{1,2}$\thanks{karima.masmar@gmail.com}
\\
{\small $^{1}$ LPTHE, Physics Department, Faculty of Sciences, Ibnou Zohr University, Agadir, Morocco. }\\
{\small $^{2}$Laboratory of  High Energy Physics and Condensed Matter
HASSAN II University,}\\{\small Faculty of Sciences Ain Chock, Casablanca, Morocco.}\\
{\small $^{3}$CRMEF, Regional Center for Education and Training Professions Marrakesh, Morocco.
}
}

\date{\today}
\begin{document} 
	\maketitle
\begin{abstract}
It's widely recognized that the free energy landscape captures the essentials of thermodynamic phase transitions. In this work, we extend the findings of \cite{Xu:2023vyj} by incorporating the nonextensive nature of black hole entropy. Specifically, the connection between black hole phase transitions and the winding number of Riemann surfaces derived through complex analysis is extended to the Rényi entropy framework. This new geometrical and non-extensive formalism is employed to predict the phase portraits of charged-flat black holes within both the canonical and grand canonical ensembles.

Furthermore, we elucidate novel relations between the number of sheets comprising the Riemann surface of the Hawking-Page and Van der Waals transitions and the dimensionality of black hole spacetimes. Notably, these new numbers are consistent with those found for charged-AdS black holes in Gibbs-Boltzmann statistics, 
providing another significant example of the potential connection between the cosmological constant and the nonextensive Rényi parameter.

		{\noindent }
\end{abstract}
	%\addcontentsline{toc}{section}{\nameref{appendix}}
	\tableofcontents
	
	%\newpage
	%\newpage

	%	 \tableofcontents

\section{Introduction}
Several recent studies \cite{Biro:2011ncf, Czinner:2015eyk} have raised significant concerns regarding the suitability of standard Gibbs-Boltzmann (GB) statistics, particularly in systems characterized by long-range interactions such as black holes. Furthermore, authors \cite{Promsiri:2020jga,Barzi:2022ygr,Nakarachinda:2021jxd,Li:2020khm,Cimidiker:2023kle,Chunaksorn:2022whl,Cimdiker:2022ics,Hirunsirisawat:2022fsb,Sriling:2021lpr,Nojiri:2021czz,Promsiri:2021hhv,Samart:2020klx} have unveiled the limitations of traditional stability analyses in the context of black hole systems, attributing these challenges to the non-additivity of entropy, as observed in the Bekenstein-Hawking entropy. These issues indicate that employing GB statistics within self-gravitating systems may yield incomplete results.

Unlike typical systems where entropy scales with volume, black hole entropy is proportional to its surface area. This fundamental difference emphasizes the importance of investigating black hole thermodynamics through the principles of nonextensive statistics and entropy. 
A novel approach arises by adopting a weaker composition rule, particularly Abe's rule \cite{PhysRevE.63.061105}
\begin{equation}\label{Abe_rule}
H_\lambda(S_{AB})=H_\lambda(S_A)+H_\lambda(S_B)+\lambda H_\lambda(S_A)H_\lambda(S_B).
\end{equation}
Here, $H$ represents a differentiable function of entropy $S$, $\lambda$ is a real parameter, and $A$ and $B$ are two independent systems. One example of nonextensive entropy, Tsallis entropy \cite{Tsallis:1987eu,Tsallis:2012js}, is expressed as follows:
\begin{equation}\label {abe}
S_q=\frac{1-\sum_{i=1}^\Omega p_i^q}{q-1}\equiv S_{T},\quad q\in \mathbb{R}.
\end{equation}
With $\Omega \in \mathbb{N}$ as the total number of configurations and $p_i$ as their probabilities, the standard GB entropy $S_{GB} = -\sum_{i=1}^\Omega p_i \ln p_i$ is easily recovered by setting $q \to 1$. Here, defining $H_\lambda(S) = S_T$ and $\lambda = 1-q$, the pseudo-additive rule emerges as
\begin{equation}\label{biro_van}
S_T(A\cup B)=S_T(A)+S_T(B)+(1-q)S_T(A)S_T(B).
\end{equation}
 In this pseudo-additive framework, the definition of an empirical temperature remains elusive. Biró and Van \cite{Biro:2011ncf} propose addressing this by introducing the formal logarithm of entropy, corresponding to the Rényi entropy
\begin{equation}\label{Renyi_def}
L(S_T)=\frac{1}{1-q}
\ln\left[1+(1-q)S_T\right]\equiv S_R. 
\end{equation}
This parameter was initially introduced to accommodate the non-additivity of Bekenstein-Hawking entropy as expressed by \eqref{biro_van}. However, it also imparts to the Rényi entropy a notable additive composition law, such as

%By substituting the non-additive composition rule \eqref{nonAd} for $S_\text{T}$, we have
\begin{eqnarray}
S_\text{R}(A\cup B) &=& \frac{1}{\lambda}\ln \left[ 1+ \lambda \left(S_\text{T}(A)+S_\text{T}(B)+\lambda S_\text{T}(A)S_\text{T}(B) \right) \right] \nonumber \\
&=& \frac{1}{\lambda} \ln \left[ \left( 1+\lambda S_\text{T}(A) \right)\left( 1+\lambda S_\text{T}(B) \right) \right]  \\
&=& \frac{1}{\lambda} \ln \left( 1+\lambda S_\text{T}(A) \right)+ \frac{1}{\lambda} \ln \left( 1+\lambda S_\text{T}(B) \right) \nonumber \\ \nonumber
&=& S_\text{R}(A)+S_\text{R}(B).
\end{eqnarray}
%\begin{equation}
 %   S_{R}(A\cup B)=S_{R}(A)+S_{R}(B),\label{composition_Renyi}
%\end{equation}
This is elaborated through Eqs.\eqref{biro_van} and \eqref{Renyi_def}. Thus, Rényi entropy effectively resolves the non-extensive characteristics of black holes while maintaining the additivity of their entropy.

Furthermore, the Rényi entropy $S_R=\frac{1}{\lambda}\ln(1+\lambda S_T)$ is consistent with the zeroth law of thermodynamics and facilitates a well-defined formulation of empirical temperature
\begin{equation}
\frac{1}{T_R}=\frac{\partial S_R (E)}{\partial E},
\end{equation}
Here, $E$ represents the energy of the system. This alternative entropic framework shows potential in tackling the complexities presented by nonextensive systems, offering a means to model black hole thermodynamics more effectively within the context of nonextensivity. In addition, the potential connection between the nonextensive parameter $\lambda$ and the cosmological constant, initially suggested in \cite{Promsiri:2020jga}, gains support from studies involving charged black holes \cite{Hirunsirisawat:2022fsb,Barzi:2022ygr,Barzi:2023mit}, de Rham-Gabadadze-Tolley black holes \cite{Chunaksorn:2022whl}, Bardeen black holes \cite{Wang:2023lmr}, among others. Additional evidence is also found from a topological perspective \cite{Barzi:2023msl} and chaotic behavior \cite{Barzi:2024bbj}.

A recent development in the study of critical phenomena in black hole thermodynamics is the incorporation of topology \cite{Wei:2021vdx,Wei:2022dzw,Yerra:2022coh,Wu:2022whe,Wu:2023xpq,Gogoi:2023qku,Ali:2023jox,Bai:2022klw,Du:2023wwg,Zhu:2024jhw,Hazarika:2024cpg}. The method introduced in \cite{Wei:2022dzw}, known as the off-shell free energy approach, allows for the examination of black hole thermodynamics by viewing black hole solutions as topological defects in their thermodynamic space. Such topological approches have been extended to Rényi entropy formalism in \cite{Barzi:2023msl}. Moreover, in \cite{Xu:2023vyj}, the authors introduced a novel approach for assessing phase transitions in thermodynamic systems. By employing complex analysis, they revealed the intricate structures of phase transitions in black hole thermodynamics, providing a clear and intuitive representation of these transitions. Through the examination of the winding number related to these complex structures, they were able to associate the order of phase transition behavior of black hole systems with the number of foliations of the corresponding Riemann surfaces. This work has been extended to Lovelock gravity in \cite{Wang:2023qxw}.

We aim to leverage the properties of analytical functions to examine the universal traits of Rényi thermodynamical phase transitions in charged-flat black holes spacetimes. Such an approach defines the counterpart of these transitions in the complex domain from a nonextensive perspective.

The outline of our paper is as follows. In Sec. II, we briefly introduce black hole thermodynamics from the Rényi perspective. Next, we explore the relationship between the winding number and black hole thermodynamics through a complex analysis approach. In Sec. IV, we present the connection between the Maxwell Area law and the Braggs-Williams formalism. The Rényi-Hawking-Page phase transition of four-dimensional charged-flat black holes is examined from a complex analysis perspective in Sec. V. Subsequently, we investigate the Van der Waals phase transition in the canonical ensemble. In Sec. VII, we extend the previous discussions to arbitrary-dimensional spacetimes and establish new relationships between the number of sheets constituting the Riemann surface of the Hawking-Page and Van der Waals transitions and the dimensionality of black hole spacetimes. Finally, the last section is devoted to a summary and discussion.

\section{Review of Rényi thermodynamics of a 4d-charged-flat black hole}

	First, we consider the spherically symmetric Reissner-Nordström black hole with mass $M$ and charge $Q$. The line element is given by
	\begin{eqnarray}
	ds^2 = -f(r)dt^2 + \frac{dr^2}{f(r)} + r^2d\Omega^2,
	\end{eqnarray}
	where $d\Omega^2 = d\Theta^2 + \sin^2\Theta d\Phi^2$ is the square of a line element on 2-sphere and the blackening function $f(r)$ stands for
	\begin{eqnarray}
	f(r)=1-\frac{2M}{r} + \frac{Q^2}{r^2}.
	\end{eqnarray}
	It has been realized  that the outer and inner horizons are 
	roots of $f(r)$, and are given by  
	\begin{eqnarray}
	r_{\pm}=M \pm \sqrt{M^2-Q^2}.
	\end{eqnarray}
	The black hole event horizon is associated with the position $r_h=r_+$. This permits to express the mass $M$ in terms of the horizon radius $r_h$ and the charge $Q$ as
	\begin{eqnarray}
	M = \frac{r_h}{2}\left( 1 + \frac{Q^2}{r_h^2} \right). \label{mass}
	\end{eqnarray}
	While the charge $Q$  generates the bulk gauge field
	\begin{eqnarray}
	A = A_tdt = -\left(\frac{Q}{r}-\Phi \right)dt,  \label {bh4}
	\end{eqnarray}
	and we can obtain the electric potential $\Phi$ 
	by setting $A_t=0$ at the horizon $r=r_h$. Thus, we have
	\begin{equation}
	\Phi = \frac{Q}{r_{h}} = \frac{Q}{M + \sqrt{M^2 - Q^2}}. \label {bh6}
	\end{equation}

According to \cite{Promsiri:2020jga,dilaton}, the  R\'enyi entropy $S_R$ is the formal logarithm of Bekenstein-Hawking one $S_{BH}$ taken as the Tsallis entropy such as, 
\begin{equation}
S_R=\frac{1}{\lambda}\ln(1+\lambda S_{BH}). \label{bh17}
\end{equation}
It's important to note that in the limit where the nonextensivity parameter $\lambda$ approaches zero, the standard Gibbs-Boltzmann statistics are recovered: $S_R \overset{\lambda \rightarrow 0}{\longrightarrow} S_{BH}$. In this paper, we assume $\lambda$ is small and positive, $0 < \lambda \ll 1$, under the premise that non-extensive effects are first-order corrections to classical extensive statistical mechanics.
\paragraph{} The Rényi entropy generalizes the Gibbs-Boltzmann entropy by incorporating non-extensive effects, leading to a first-order correction in $\lambda$ to the RN-flat black hole temperature. The Rényi temperature $T_R$ can be expressed as \cite{Promsiri:2020jga},
\begin{eqnarray}
T_R = \frac{1}{\partial{S_R/\partial{M}}} &=& T_H(1+\lambda S_{BH})\label{Tr}\\ 
&=& \frac{(r_h^{2}-Q^2)(1+ \lambda \pi r_h^{2})}{4\pi r_h^{3}}. \label{bh25}
\end{eqnarray}
Where $T_H = \frac{r_h^2 - Q^2}{4\pi r_h^3}$ and $S_{BH} = \pi r_h^2$ are the Hawking temperature and the Bekenstein-Hawking entropy of the RN-flat black hole, respectively. In the grand canonical ensemble, the mass $M$ and Rényi temperature are expressed as
\begin{equation}\label{key}
T_R=\displaystyle  \frac{ \left(\pi \lambda r_{h}^{2} + 1\right)\left( 1-\phi^{2}\right)}{4 \pi r_{h}},
\end{equation}
\begin{equation}\label{key}
M=\displaystyle \frac{r_{h} \left(\phi^{2} + 1\right)}{2}.
\end{equation}
Moreover, in the Rényi extended phase space \cite{Promsiri:2020jga}, the nonextensivity parameter $\lambda$ is related to the Rényi thermodynamic pressure, the electric potential $\Phi$, and the charge $Q$ as follows
\begin{eqnarray}\label{p-lam}
P_R = \frac{3\lambda (1-\Phi^2)}{32}%, \ \ \ V=\frac{4}{3}\pi r_h^3. 
= \displaystyle \frac{3 \lambda \left(r_{h}^{2}-Q^{2}\right)}{32r_{h}^{2}}.
\end{eqnarray}
While, its conjugate quantity  is the thermodynamic volume, given by $V=\frac{4}{3}\pi r_h^3$\cite{Barzi:2023mit}.  By combining all these quantities, we can express the first law of Rényi thermodynamics and its associated Smarr formula as
\begin{eqnarray}
dM = T_RdS_R + VdP_R + \Phi dQ 
\qquad \text{ and }\qquad
M = 2T_RS_R - 2P_RV + \Phi Q. \label{Smarr_rényi_mod}
\end{eqnarray}

%\begin{equation}\label{key}
%P_R=\displaystyle \frac{3 \lambda \left(r_{h}^{2}-Q^{2}\right)}{32r_{h}^{2}}
%\end{equation}
%\begin{equation}\label{key}
% V=\frac{4\pi}{3}r_h^3
%\end{equation}
In this sense, we define the specific volume
$v=\frac{8l_p^2}{3}r_h=\frac{8}{3}r_h$, in which $l_p$ is the Planck's length.
Therefore, the equation of state for the black hole thermodynamic fluid can be expressed as follows
\begin{equation}\label{bhr1}
P_R=  \frac{T_R}{v}  - \frac{2}{3 \pi v^{2}} + \frac{128 Q^{2}}{27 \pi v^{4}}.
\end{equation}
This equation of state resembles the Van der Waals (VdW) type, suggesting that the charged Rényi-flat black hole behaves analogously to a VdW fluid and exhibits first and second-order (liquid/gas) phase transitions \cite{Barzi:2023mit}. Specifically, the black hole fluid undergoes a second phase transition at a critical temperature $T_{c} = \frac{\sqrt{6}}{18 \pi Q}$, derived by solving the following system:
\begin{equation}
	\left(\frac{\partial P_R}{\partial v}\right)_{T_c,Q}=
	\left(\frac{\partial^2 P_R}{\partial v^2}\right)_{T_c,Q}=0.
	\end{equation}
After outlining various thermodynamic aspects of the RN-flat black hole, including quantities related to Rényi statistics, we now shift our focus to exploring their topological characteristics through the foliation of the Riemann surface using complex functions. This investigation starts in the grand canonical ensemble and proceeds into the canonical ensemble using Rényi's non-Boltzmannian formalism.

\section{Multi-valued complex functions: Short review}	

\begin{figure}[!ht]
	\centering
	\begin{tabbing}
		\hspace{0cm}
		\centering
	\includegraphics[scale=0.46]{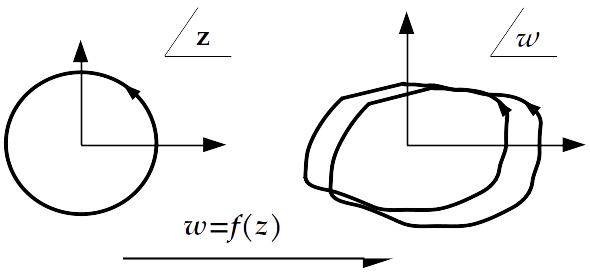}
	\hspace{1.2cm}	\includegraphics[scale=0.46]{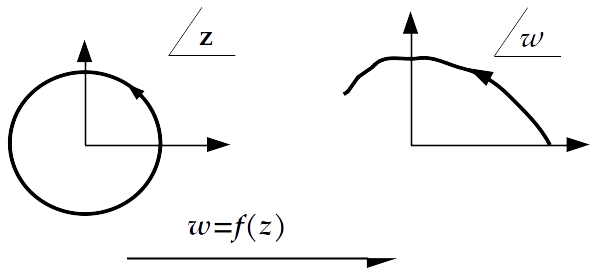}
	\end{tabbing}
	\caption{\footnotesize \it Schematic of complex functions: \textbf{Single-valued function (Left:)} as the antecedent $z$ goes around the origin once in the domain space of $f(z)$, the image $w$ may go around the origin many times in the image space of $f(z)$. \textbf{Multi-valued function (Right:)} as the antecedent $z$ goes around the origin once in the domain space of $f(z)$, the image $w$ does not and revisits the same antecedents for different images.}
	\label{fig:complex-function}
\end{figure}

\begin{figure}[!ht]
	\centering
	\begin{tabbing}
		\hspace{+2cm}
		\centering
		\includegraphics[scale=.38]{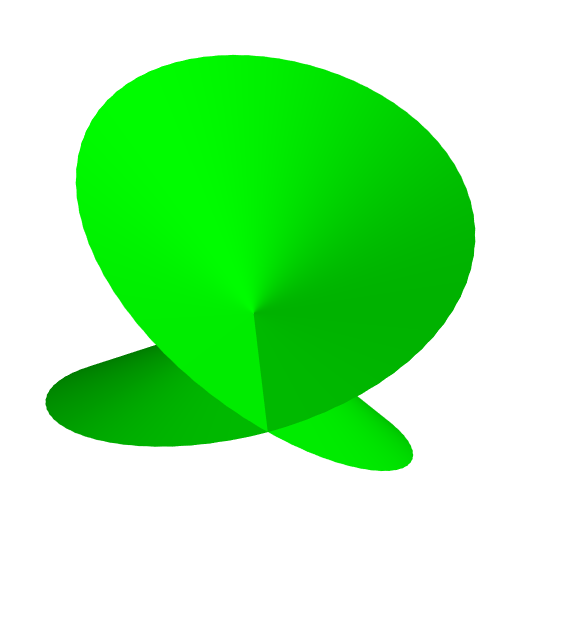}
		\includegraphics[scale=.38]{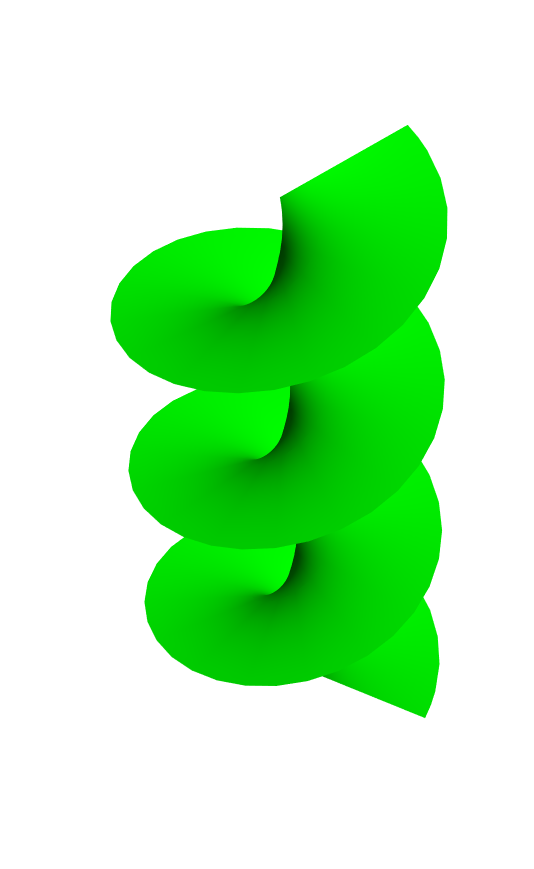}
	\end{tabbing}
	\vspace{-1cm}
	\caption{\footnotesize\it  Some Riemann surfaces of multi-valued complex functions.\textbf{(Left:) }square root function $f(z)=\sqrt{z}$ with one branch cut and two Riemann sheets.\textbf{(Right:)} logarithm function, $f(z)=\log(z)$, which possesses an infinite number of Riemann sheets. }
	\label{fig:fig_cmplx_examples}
\end{figure}

In \cite{Xu:2023vyj}, the authors misinterpreted single-valued complex functions as multi-valued ones, the analytic functions "$\psi(z)$" they used are well defined single-valued function and therefore they have trivial Riemann surfaces. Riemann surfaces are geometric constructions designed for multi-valued functions, based on the existence of branch cuts, which are absent in single-valued complex functions. The concept of the winding number likely misled the authors into confusing the two types of functions. Concretely, a function $f(z)$ is multi-valued if an antecedent $z$ in the domain of $f(z)$ has more than one image $w$. However, if a single image in the range of $f(z)$ has more than one antecedent, it is still a single-valued function. Here are two examples to illustrate the difference and clear up any confusion. For $f(z)=w=z^2$, the image $w$ has two antecedents at least, $z$ and $-z$, but it is single-valued since no $z$ has multiple images. In contrast, $f(z)=w=\sqrt{z}$, is multi-valued, because $z=r e^{i\theta}=r e^{i\theta+2i\pi}$, has two distinct images $w=\sqrt{r}e^{i\frac{\theta}{2}}$ and $w=\sqrt{r}e^{i\frac{\theta}{2}+i\pi}=-\sqrt{r}e^{i\frac{\theta}{2}}$. An other example of a multi-valued complex function is $f(z)=\log(z)$, it is straightforward that for a given $z=r e^{i\theta}=r e^{i\theta+2 i p\pi}$, we get $w=\log{r}+i\theta+2 i p\pi$, where $p\in \mathbb{Z}$.  Thus, $z$ has an infinite number of images, which translates to an infinite number of Riemann sheets and branch cuts along the positive real axis. In Fig.\ref{fig:fig_cmplx_examples}, we depicted the Riemann surfaces of the square root and logarithm complex functions. This shows that the winding numbers for these multi-valued functions are $2$ and $\infty$, respectively.

As a general rule, all polynomial and rational complex functions are single-valued. The analytic functions in \cite{Xu:2023vyj} fall into these two categories, being either polynomials or rational functions. Therefore, Riemann surfaces cannot be directly used to visualize the winding numbers. In the present work, we propose a possible association with Riemann surfaces that allows for such visualization.

Furthermore, in complex analysis, the winding number of a single-valued complex function is determined by
\begin{equation}\label{def_num_sheet}
\mathcal{W}=\mathcal{N}-\mathcal{P},
\end{equation}
where $\mathcal{N}$ and $\mathcal{P}$ are the number of zeros and poles of the function, respectively. The winding number represents the number of turns around the origin executed in the image space as one turn around the origin is completed in the domain space. This implies that if a complex function $f$ has a winding number, $\mathcal{W}>1$, then necessary many points in the domain space have the same image and therefore, $f^{-1}$ must be multi-valued. It results\textit{ that the winding number $\mathcal{W}$ of a singled-valued complex function $f$ is equal to the number of sheets $\mathcal{S}$ of its multi-valued inverse function $f^{-1}$}.
\begin{equation}
    \mathcal{S}(f^{-1})=\mathcal{W}(f)
\end{equation}

This observation will be used later to associate Riemann surfaces with Hawking-Page and Van der Waals phase (VdW) transitions. However, before delving into the complex analysis of these phase transitions, we will briefly introduce the Bragg-Williams formalism.
\section{Braggs-Williams approach: concise introduction}

\subsection{Motivations}
Herein, we list some general motivations why researchers might explore formalisms from condensed matter physics, such as Bragg-Williams, in the context of black hole thermodynamics. First, from the analogy perception:  the mathematical structures used in condensed matter physics, like the Bragg-Williams formalism, may be adapted to describe or model certain features of black holes since the black holes are nothing more than thermodynamic systems \cite{Kubiznak:2016qmn} and exhibiting diverse behaviors like those of different substances we encounter in everyday life,  such as Van der Waals fluids, reentrant phase transitions, and triple points, etc.

Recalling the entropy and phase transitions concepts, the Bragg-Williams formalism is often used to describe entropy and phase transitions in condensed matter systems so it's legitimate to extend it to the black hole framework for a deep understanding of the black hole systems from the topological point of view. Moreover, within the
	holographic dualities: The AdS/CFT correspondence and other holographic dualities suggest connections between certain gravitational theories (like those describing black holes) and certain quantum field theories. In fact, the exploration of holography's implications for systems with finite density has given rise to the birth of the "Anti-de-Sitter/Condensed Matter Theory" (AdS/CMT) conjecture. This conjecture suggests that the gravitational dynamics within the bulk space encode information about a condensed matter universe, exhibiting essential similarities to conventional paradigms. Notably, this condensed matter system displays features such as a metallic state characterized by Drude-type transport \cite{Zaanen:2018edk}, a superconducting transition, and high-temperature superconductivity—collective properties that diverge from those of individual fermions. Importantly, these phenomena are governed by "Planckian dissipation" \cite{Zaanen:2004etq}.
	Besides, since certain features of black hole thermodynamics are thought to emerge from underlying microscopic degrees of freedom, researchers may explore analogies with condensed matter systems to understand emergent phenomena and describe them mathematically.

%Researchers may leverage tools from condensed matter physics in the holographic context, aiming to gain insights into the behavior of black hole thermodynamics.

It's important to note that the use of formalisms lent from condensed matter physics in the study of black hole thermodynamics often involves developing new theoretical frameworks or adapting existing ones. This is an active area of research, and the motivations and approaches can vary among researchers exploring these connections.  In this study and in a previous one to our humble knowledge we are constructing the first steps in the association of Rényi statistics, Brag-Williams topology, and black hole thermodynamics formalisms.
\subsection{Setup}\label{sec3.2}
The Bragg-Williams formalism can be effectively illustrated through the widely recognized 3-dimensional Ising Model involving $N$ interacting spins \cite{Podgornik}, featuring a paramagnetic-ferromagnetic phase transition. In this model, each lattice site is linked to a spin variable $s_k$ with potential values of $\pm 1$. The interaction Hamiltonian governing such a system is expressed as follows
	\begin{equation}
	H=-\beta\sum_{kk'} s_k s_{k'},\label{ising_hami}
	\end{equation}
	In the given expression, where $\beta > 0$ serves as the coupling constant, and the summation encompasses the nearest neighbors, a pivotal concept emerges: the system's \textit{order parameter}. In this context, the average spin or \textit{magnetization} denoted by $\sigma = \langle s\rangle$ serves as the relevant order parameter, exerting control over the system's phase transitions. Expressed in terms of this order parameter, the internal energy $U= \langle H\rangle$ and the entropy $S$ of the system read as follows
	\begin{eqnarray}
	U&=&-\frac{1}{2}\beta \gamma N \sigma^2,\\
	S&=&N\ln(2)-\frac{N}{2}(1+\sigma)\ln(1+\sigma)-\frac{N}{2}(1-\sigma)\ln(1-\sigma).
	\end{eqnarray}
	Here $\gamma$ indicates the number of nearest neighbors to each site. One proceeds by introducing the \textit{Bragg-Williams free energy} such as,
	\begin{eqnarray}
	\nonumber
	f(\sigma,T)&=&U-TS\\
	&=&-\frac{N}{2}\beta \gamma \sigma^2-NT\ln(2)+\frac{NT}{2}(1+\sigma)\ln(1+\sigma)+\frac{NT}{2}(1-\sigma)\ln(1-\sigma)
	\end{eqnarray}
	\begin{figure}[H]
		\centering
		\centering
		\includegraphics[scale=.45]{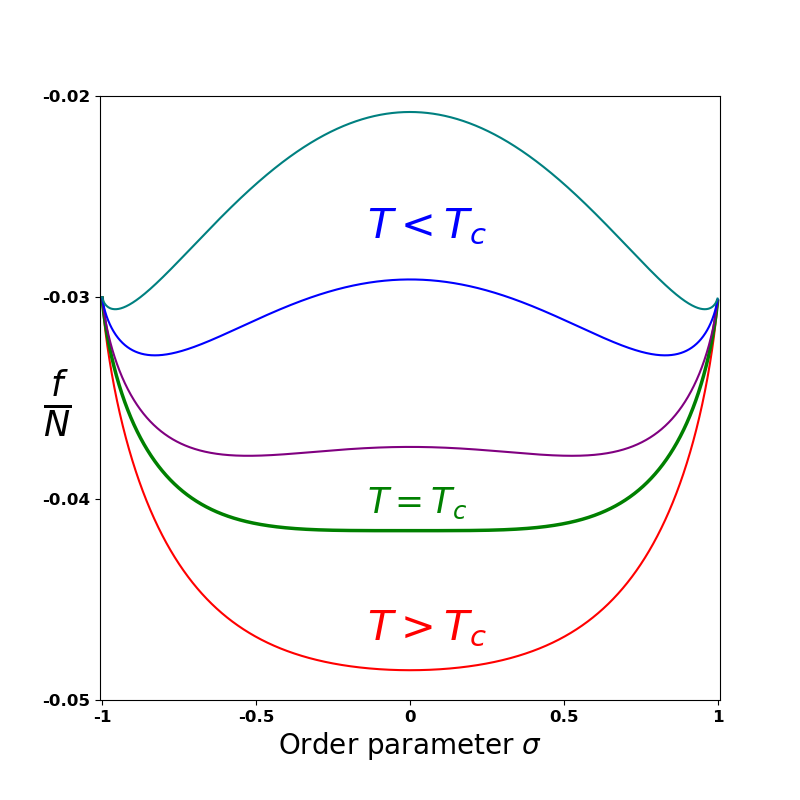}%\>
		\vspace{-0.7cm}
		\caption{\footnotesize\it Depiction of the Bragg-Williams free energy per spin as a function of the order parameter $\sigma$. Above the critical temperature $T_c$ (red line at $T/T_c=1.15$), there is a trivial minimum at $\sigma=0$ representing the stable paramagnetic state. Below $T_c$ (purple line at $T/T_c=0.9$, blue line at $T/T_c=0.7$, and teal line at $T/T_c=0.5$), non-zero minima arise, revealing the stable ferromagnetic state while the paramagnetic one becomes a maximum and thus an unstable state. The critical behavior is represented by the green line. The Ising system's parameters are $\beta=0,01$ and $\gamma=6$.}\label{fig_ising}
	\end{figure}
	We note a key assumption of the Bragg-Williams approach, as for all mean field theories, that is the substitution in the energy $H$, Eq.\eqref{ising_hami}, of the variables $s_k$ by their average $\sigma$, the order parameter, \textit{neglecting spatial fluctuations throughout the spin system}. Even with such a drastic averaging the formalism retains its predicting power. It is easily shown by plotting the Bragg-Williams  free energy against the order parameter $\sigma$ for different temperatures, as depicted in Fig.\ref{fig_ising}, 
	the emergence of a critical behavior at the temperature $T_c=\beta\gamma$ known as the \textit{Curie temperature}, with the function $f$ presenting a single minimum at $\sigma=0$ above $T_c$, and two minima at non-zero order parameter below it. This Mexican hat shape is signaling \textit{paramagnetic-ferromagnetic phase transitions} which is an archetypal example of a second-order phase transition according to Ehrenfest's classification and characterized by a continuous change of the order parameter across $T_c$. We note also \textit{the inflection points} within \textit{the spinodal regions} which separate \textit{the equilibrium points}, the unstable paramagnetic state is given by the maximum at $\sigma=0$ from the stable ferromagnetic states displayed by the minima at $\sigma\neq 0$. The other points on the $f-\sigma$ plane can be interpreted as \textit{fluctuating intermediate states} where the spin system is driven by thermal and quantum fluctuations towards the nearest equilibrium state either stable or unstable. Furthermore, the critical temperature is computed by requiring simultaneously a vanishing first and second derivatives of the Bragg-Williams free energy,
	\begin{equation}
	\frac{\partial f}{\partial \sigma}= 0, \quad \frac{\partial^2 f}{\partial \sigma^2}= 0.
	\end{equation}
	It is worth mentioning that Landau's mean field theory which was applied previously to the study of black hole phase transitions \cite{Guo2020hpp,Barzi:2023mit} can be derived from the Bragg-Williams approach. This is achieved by focusing on the most important region of the phase diagram which is the one near $T=T_c$, there, the order parameter is approaching zero. An expansion of the Braggs-Williams free energy around $\sigma=0$ gives, for the present Ising system,
	\begin{equation}\label{landau}
	f(\sigma,T)=-NT\ln(2)+\frac{\mathbf{(T-T_c)}}{\mathbf{2}}\sigma^2+\frac{T}{12}\sigma^4+\frac{T}{30}\sigma^6+O(\sigma^8).
	\end{equation}
	The main feature of Eq.\eqref{landau} is the sign change of the coefficient of the $\sigma^2$-term, above and below $T_c$. This sign reversal shifts the position of the minimum of $f$ from $\sigma=0$ for $T>T_c$ to $\sigma\neq 0$ for $T<T_c$ thus unveiling the second-order phase transition. In contrast, the BW formalism permits an investigation of the phase profile of the system not only at small values of the order parameter but for all possible values. Moreover, \textit{the applicability of Landau's theory fails to capture first-order phase transitions}, while the BW theory applies equally well to first, second, and higher-order ones because it does not impose any continuity requirements on the free energy at phase transition points.
	\subsection{Equivalence of the Maxwell Area law and the Braggs-Williams formalism}
\begin{figure}[!htb]
		\centering
		\begin{tabbing}
		\hspace{-2.4cm}
		\includegraphics[scale=0.42]{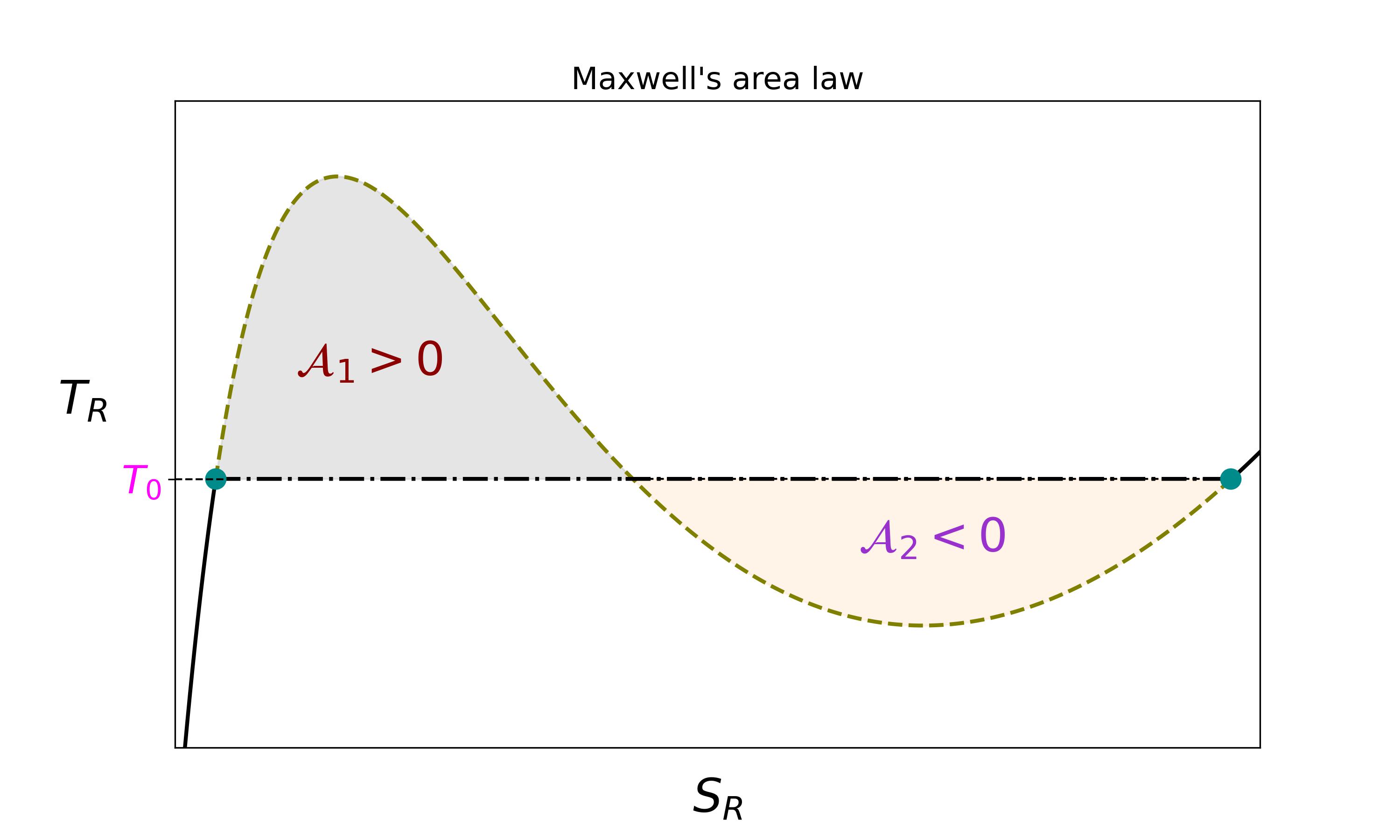}
		\hspace{-1cm}
		\includegraphics[scale=0.42]{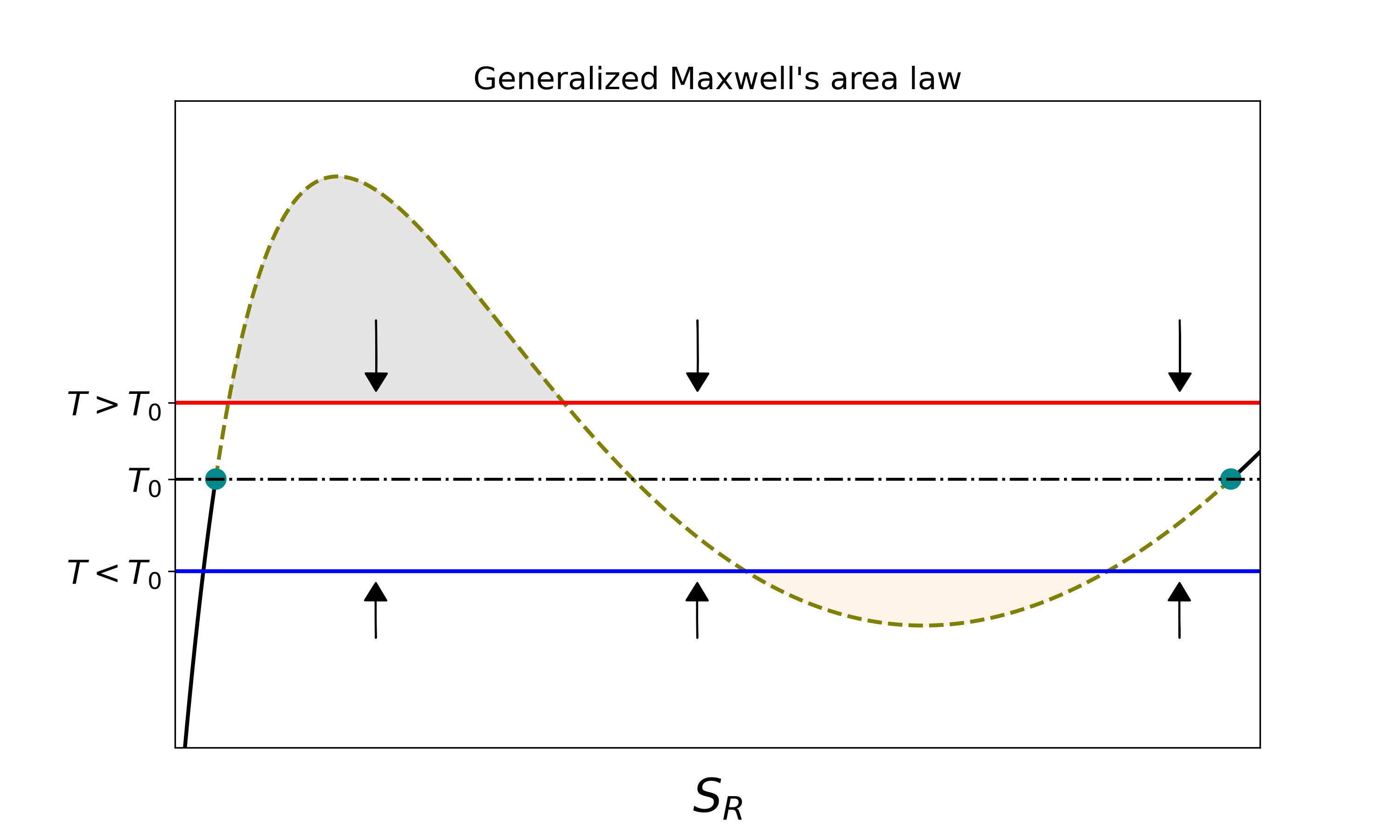}
		\end{tabbing}
		\vspace{-0.6cm}
		\caption{\footnotesize Maxwell's equal area law}
		\label{fig:maxwell-law}
	\end{figure}

	One of the most powerful constructions in inspecting the phase transitions of a thermodynamic system is the well-known Maxwellian area law which provides an algorithm to compute physical phase transitions by geometrical means. Specifically, as illustrated in Fig.\ref{fig:maxwell-law}, The oscillating behavior in the $T_R-S_R$ diagram is replaced by a plateau indicating the physical temperature $T_0$ at which the first order liquid/gas phase transition effectively takes place. This temperature is given by demanding the equality,
	\begin{equation}\label{Maxwell_law}
	\int(T_R-T_0)\;dS_R=0
	\end{equation}
	
	Geometrically, this equation traduces the requirement of the equality of the area above and below the plateau at $T_0$. If we define a generalized function $f$ such as,
	\begin{equation}
	f(r_h,T)=\int(T_R-T)\;dS_R
	\end{equation}
	where the temperature $T$ is treated as a free parameter that would coincide with physical temperatures only at phase transition points. Interestingly, $f(r_h,T)$ is nothing but the Braggs-Willaims free energy and the condition \eqref{Maxwell_law} is equivalent to the statement,
 \begin{equation}
     f(r_h,T)=0.
 \end{equation}
 \paragraph{}In other words, \textit{the zeros of the Braggs-Williams free energy are the phase transition points of a thermodynamic system}.
	\paragraph{}In the subsequent sections we extend the applicability of the Bragg-Williams approach to the charged R\'enyi-flat black hole, through analytic continuation, to the powerful domain of complex analysis in the hope of revealing hidden features in the dynamics of the Hawking-Page and Van-der-Waals phase transitions.
	
\section{R\'enyi-Hawking-Page phase transition of the 4-dimensional charged-flat black holes from complex analysis perception}

\subsection{ Complexification of the R\'enyi-Bragg-Williams free energy in the grand canonical ensemble }

The Gibbs-Williams free energy of the charged R\'enyi-flat black hole in the grand canonical ensemble is computed as
\begin{align}
f_\phi(r)&=M-t S_R-\phi Q\\
&=\displaystyle \frac{r(1-\phi^2)}{2}- t \frac{\log{\left(\lambda \pi  r^{2} + 1 \right)} }{\lambda}
\end{align}
where $r$, $t$ and $\phi$ are free parameters\footnote{We abbreviated the full dependence of $f_\phi$ on $r$, $t$ and $\phi$ to just $r$, for convenience.}. The complexification of the free energy $f_\phi$ amounts to an analytic continuation to the set of complex horizon radii through the substitution, $r\longrightarrow z$, thus we get,
\begin{equation}\label{eq1}
f_\phi(z)=\displaystyle \frac{z(1-\phi^2)}{2}- t \frac{\log{\left(\lambda \pi  z^{2} + 1 \right)} }{\lambda}
\end{equation}
We proceed by taking the limit of small nonextensivity parameter $\lambda$  together with a scaling the parameters $ z $, $ t $ and $ \phi $ as well as the free energy $ f_\phi $ such as $z\longrightarrow \lambda^{-\frac{1}{2}}z$, $t\longrightarrow \lambda^{\frac{1}{2}}t$, $\phi\longrightarrow \phi$ and $f_\phi\longrightarrow \lambda^{-\frac{1}{2}}f_\phi$, we obtain,
\begin{equation}\label{eq2}
f_\phi(z)=\displaystyle \frac{z \left(1- \phi^{2} + \pi^{2} t z^{3} - 2 \pi t z \right)}{2}
\end{equation}
It is worth noting that due to the smallness of the nonextensivity parameter $\lambda\ll1$, the scaling of the complex horizon radius adopted in Eq.\eqref{eq2}, brings its values to the vicinity of $z=0$. Thus the neighborhood of $z=0$ is the most relevant region of parameter space to investigate the complex structure of black hole phase transitions.

\paragraph{}If we decompose $z$ as, $z=x+\iu y$, then the free energy $f_\phi$ can be expressed as, $f_\phi=\chi(x,y)+\iu \psi(x,y)$, with
\begin{align}\label{key}
\begin{cases}
\chi(x,y)=&\displaystyle \frac{x}{2} - \frac{\phi^{2} x}{2} - \pi t \left(x^{2} - y^{2}\right) + \frac{\pi^{2} t \left(x^{4} - 6 x^{2} y^{2} + y^{4}\right)}{2} \\\\
\psi(x,y)=&\displaystyle \frac{y \left(- \phi^{2} + 4 \pi^{2} t x \left(x^{2} - y^{2}\right) - 4 \pi t x + 1\right)}{2}
\end{cases}
\end{align}
It is both interesting and mathematically reassuring to note that the functions $\chi$ and $\psi$ satisfy the two-dimensional Laplace equation, such that
\begin{equation}\label{key}
\nabla^2\chi=0 \text{ and, } \nabla^2\psi=0.
\end{equation}

On one hand, this observation ensures that $f_\phi(z)$ satisfies the Cauchy-Riemann conditions for a {\it holomorphic} function, meaning it is a well-defined differentiable complex function. On the other hand, it allows the interpretation of $f_\phi(z)$ as a complex energy flow potential of the charged-flat black hole, with its real and imaginary parts representing the velocity potential and stream function of such a flow, respectively. In fluid mechanics, such a flow is called a {\it potential flow}; these flows are inviscid, irrotational, and incompressible. The velocity vector field, $\mathbf{u} = (u_x, u_y)$, of the flow is given through the \textit{velocity potential} or the \textit{stream function} as,
\begin{equation}\label{key}
\begin{cases}
{\bm u}={\bm \nabla }\chi\\
{\bm u}={\bm \nabla}\times\psi\\
\end{cases}\implies
\begin{cases}
u_x=\partial_x\chi=\partial_y\psi\\
u_y=\partial_y\chi=-\partial_x\psi\\
\end{cases}
\end{equation}
also outside singularity points, the vorticity is defined as 
\begin{equation}\label{key}
{\bm \omega}={\bm \nabla}\times{\bm u},
\end{equation} 
is null, ${\bm \omega} = {\bm 0}$, throughout the flow field. Within the present fluid mechanics picture, the expression of $f_\phi(z)$, Eq.\eqref{eq2}, can be decomposed into a linear superposition of complex flow potentials of well-known and basic flows, namely, a superposition of different \textit{flows in a sector}. A flow in a sector of angle $\displaystyle \frac{\pi}{n}$, where $n \in \mathbb{Z}$, is represented by a complex flow potential of the form $f(z) = C\:z^n$, where $C$ is a constant. A notable example of such a flow is the uniform flow with $n=1$. A close inspection reveals that $f_\phi(z)$ is a superposition of three flows in a sector, $n=1, 2, \text{ and }  4$
\begin{equation}\label{eq_flow_superp}
\begin{cases}
f_1(z)=C_1z : \quad \text{\;uniform flow},\\\\
f_2(z)=C_2z^2 : \quad\text{flow in a sector of angle}\; \displaystyle\frac{\pi}{2}\; \text{ and}\\\\
f_4(z)=C_4z^4 : \quad\text{flow in a sector of angle}\; \displaystyle \frac{\pi}{4}.
\end{cases}
\end{equation}
Here, the constants $C_1= \frac{1-\phi^2}{2}$, $C_2=-\pi t$ and $C_4=\frac{\pi^2 t}{2}$, measure the speed of the flow far from their respective sectors.

\begin{figure}[H]
	\begin{tabbing}
		\hspace{-2.3cm}
		\centering
		%		\hspace{-2.3cm}%\=\kill
		\includegraphics[scale=.34]{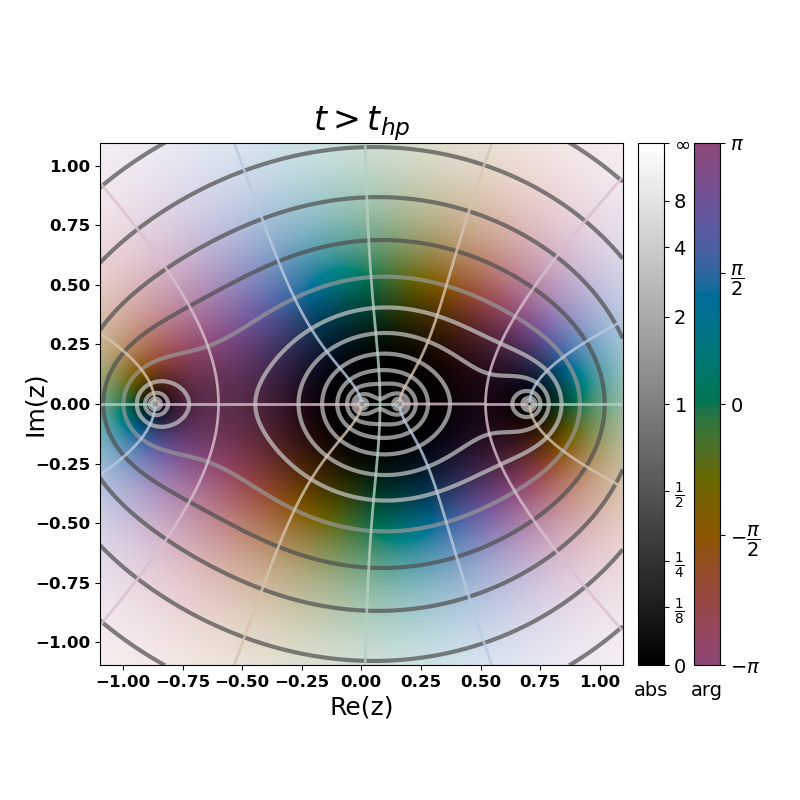}%\>
		\hspace{-0.4cm}
		\includegraphics[scale=.34]{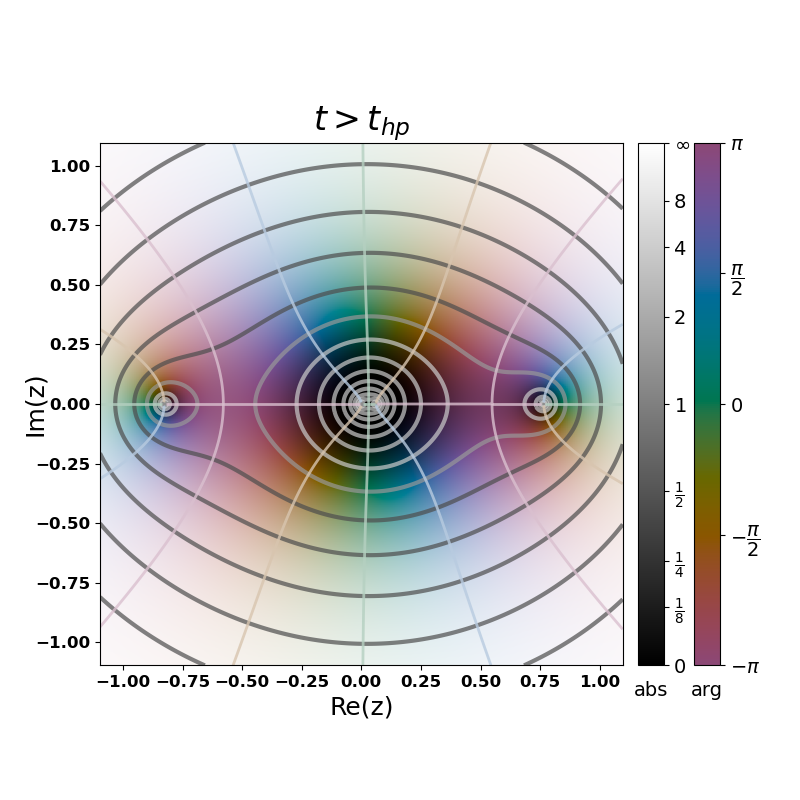}%\\
		\hspace{-0.4cm}
		\includegraphics[scale=.34]{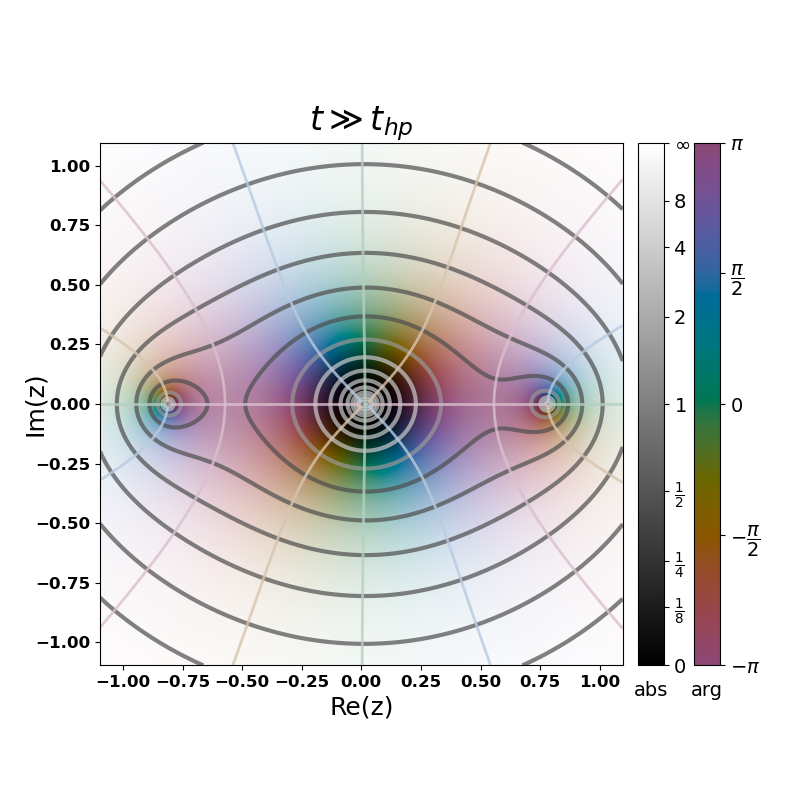}%\\
	\end{tabbing}
	\vspace{-0.7cm}
	\caption{\footnotesize \it  R\'enyi-Hawking-Page phase transition: The absolute value and phase of the complex Bragg-Williams free energy for the 4d-charged-flat black hole as the temperature gradually increases. The electric potential has been set to
  $\phi=0.5$}	\label{fig:fig_d4_1}.
\end{figure}

\paragraph{} Using the representation of the complex function, Eq.\eqref{eq1}, as a complex number, Fig.\ref{fig:fig_d4_1}
 illustrates the absolute value and phase of the complexified Braggs-Williams free energy of the R\'enyi charged-flat black hole as the temperature gradually increases. The study reveals the complex dynamics involved in the formation of the zeros of the free energy, indicating the Hawking-Page phase transition at $t = t_{hp}$. Unlike the usual depiction of the Hawking-Page transition as an isolated point on the real line or within the phase profile, this process is shown to be continuous. Specifically, two complex vortex points, located above and below the real line, emerge well below $t \ll t_{min}$. These points move towards the real line and eventually coalesce to form the real phase transition point when the temperature reaches $t_{hp}$. As the temperature rises above $t_{hp}$, the Hawking-Page phase point splits into two distinct phase points, corresponding to the transitions between thermal radiation (ThR) and large black hole (LBH), and between small black hole (SBH) and LBH, respectively \cite{Barzi:2023msl}. High above $t_{hp}$, the phase transition points move further apart from each other as the SBH$\longleftrightarrow$LBH phase transition point proceeds slowly toward $z=0$ with the LBH state wining in stability while the ThR$\longleftrightarrow$LBH transition point ultimately dominates the phase profile of the R\'enyi charged-flat black hole at $t \gg t_{hp}$. This smooth progression offers a new perspective on the Hawking-Page phase transition. As the temperature increases, complex conjugate phase transition points are generated within the thermal radiation phase. Their eventual coalescence leads to the Hawking-Page phase transition.

\begin{figure}[H]
	\begin{tabbing}
		\hspace{-2.3cm}
		\centering
		%		\hspace{-2.3cm}%\=\kill
		\includegraphics[scale=.34]{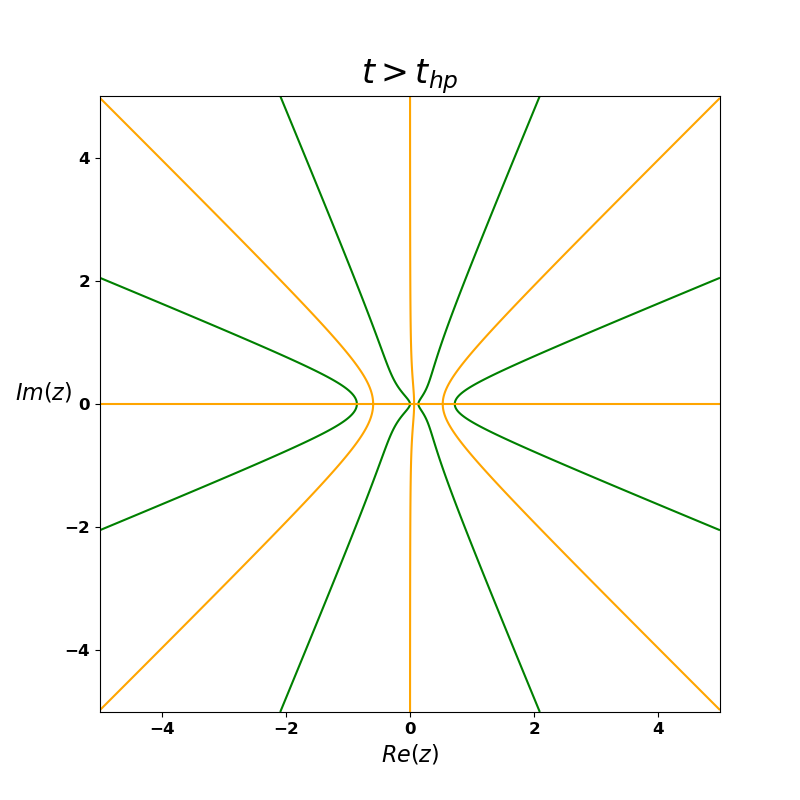}%\>
		\hspace{-0.4cm}
		\includegraphics[scale=.34]{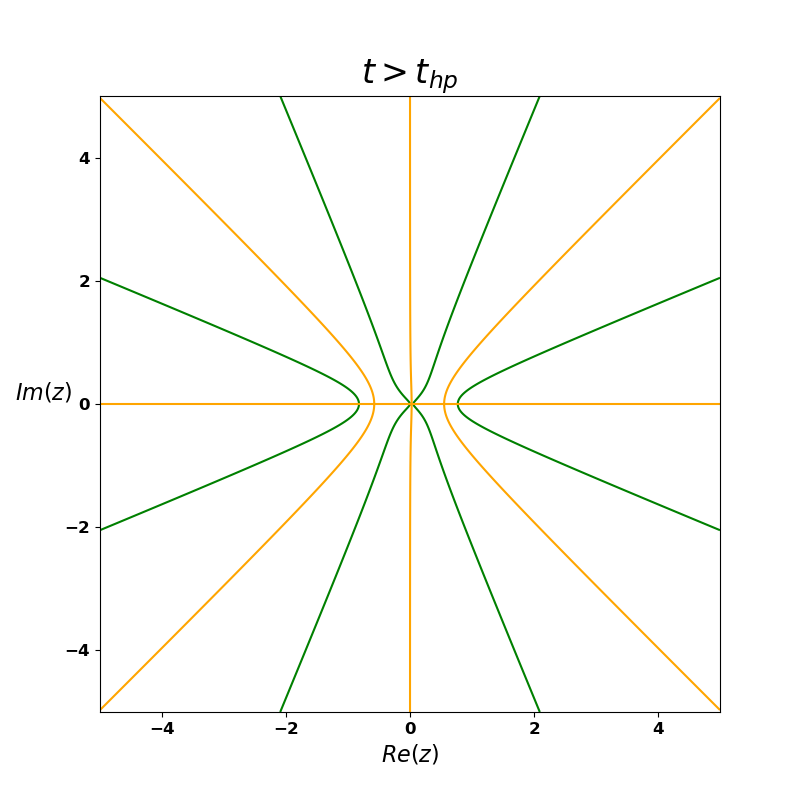}%\\
		\hspace{-0.4cm}
		\includegraphics[scale=.34]{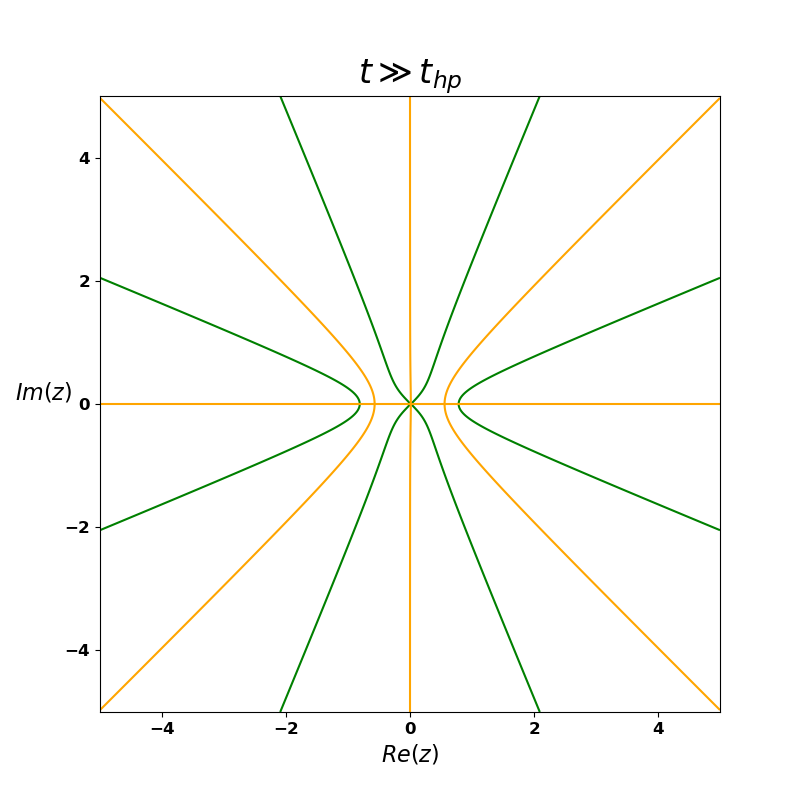}%\\
	\end{tabbing}
	\vspace{-0.7cm}
	\caption{\footnotesize \it  R\'enyi-Hawking-Page phase transition: The evolution of the zeros of the real part $\chi=0$ (\textbf{green lines}) 
 and the zeros of the imaginary part $\psi=0$ (\textbf{orange lines}) of the complex free energy for the 4d-charged-flat black hole as the temperature gradually increases. The electric potential has been set to
  $\phi=0.5$}	\label{fig:fig_d4_2}.
\end{figure}

Figure \ref{fig:fig_d4_2} illustrates the dynamics of the zeros of the real and imaginary components of the complex free energy in the grand canonical ensemble as the temperature increases. The green lines represent the zeros of the real part, $\chi(x,y)=0$, while the orange lines denote the zeros of the imaginary part, $\psi(x,y)=0$. The complex free energy vanishes at the intersections of these green and orange lines. The Hawking-Page phase transition is indicated by their intersection on the positive real axis at $t=t_{hp}$.

\begin{figure}[H]
	\centering
	\begin{tabbing}
		\centering
		\hspace{-2.5cm}%\=\kill
		\includegraphics[scale=.37]{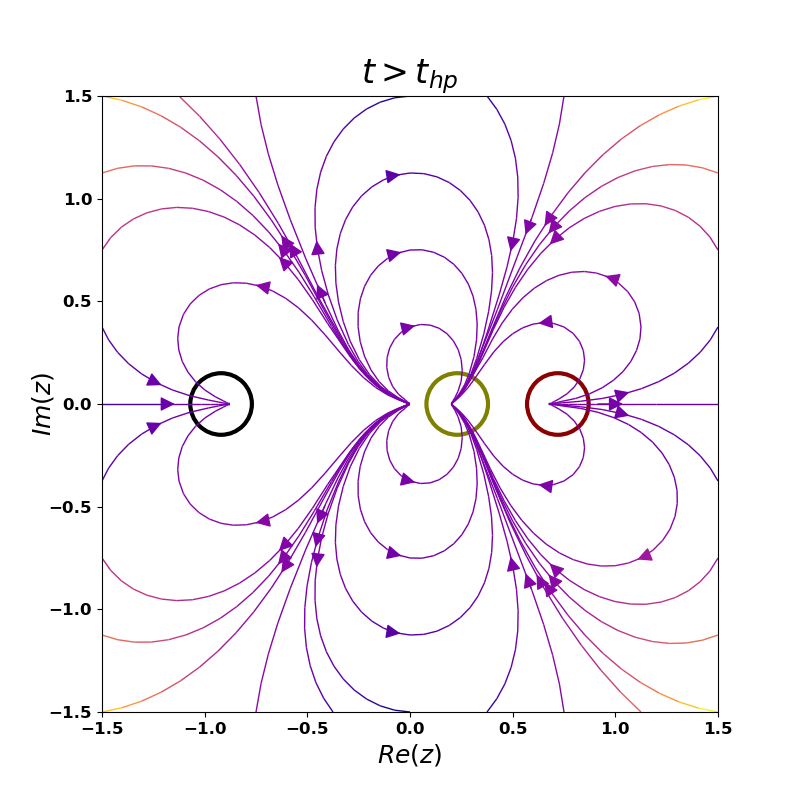}\label{}%\>
		\hspace{-0.85cm}
		\includegraphics[scale=.37]{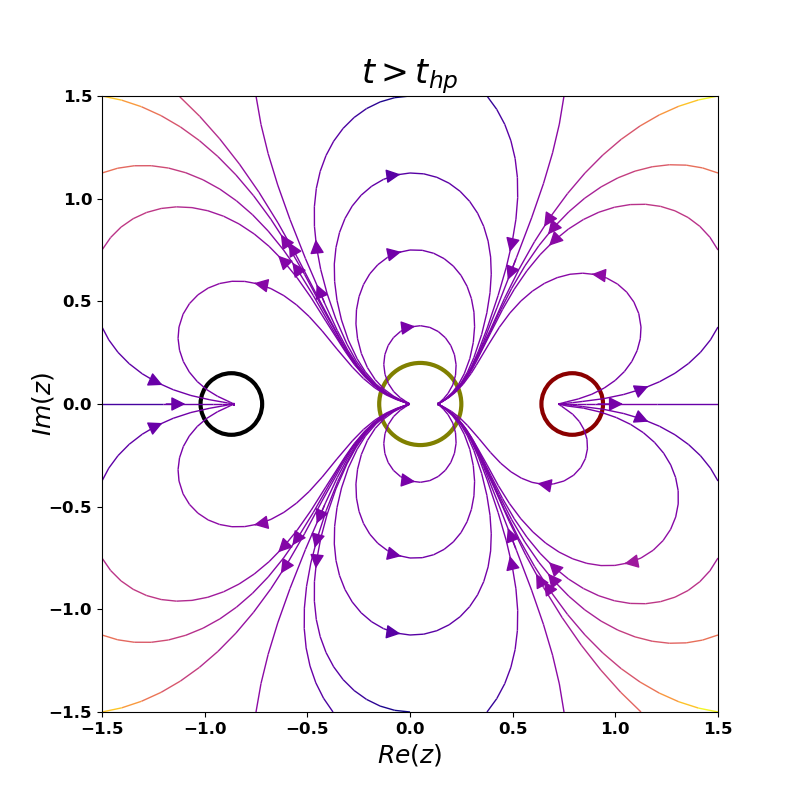},
		\hspace{-1.1cm}
		\includegraphics[scale=.37]{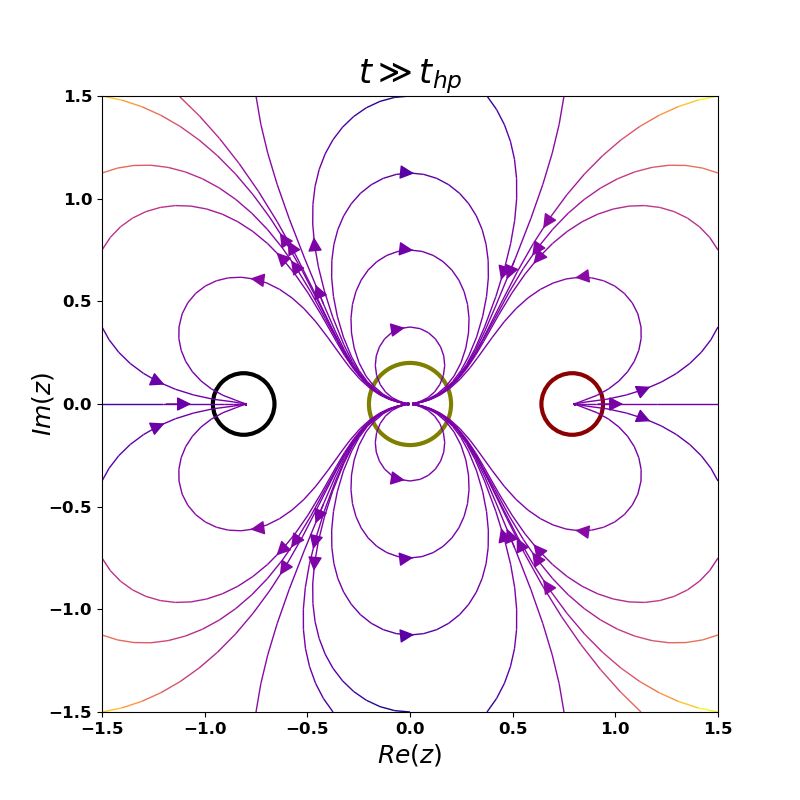}
	\end{tabbing}
	\vspace{-0.7cm}	
	\caption{\it \footnotesize R\'enyi-Hawking-Page phase transition: The energy flow vector illustrations for the 4d-charged-flat black hole as the temperature gradually increases. Colored circles indicate relevant sources and sinks in the energy flow. The electric potential has been set to
  $\phi=0.5$}\label{fig:fig_d4_3}
\end{figure}

\paragraph{} Additionally, the depiction of the complex free energy as a vector field in Fig.\ref{fig:fig_d4_3} illustrates the energy flow of the black hole as the temperature steadily increases. This visualization reveals the distortion and reconnection of field lines, leading to the formation of a knot on the real line at $t=t_{hp}$, which signals a Hawking-Page phase transition. Furthermore, the flow exhibits three sink points, with two of them having a positive real part $Re(z)>0$, along with a source point at $z=0$. The source point represents the thermal radiation phase with no black hole state, from which the energy flow is directed towards the Hawking-Page complex conjugate phase points with $Re(z)>0$.

\newpage

\begin{figure}[H]
	
	\vspace{-0.8cm}
	\centering
	\begin{tabbing}
		\centering
		\hspace{-2.3cm}%\=\kill
		\includegraphics[scale=.34]{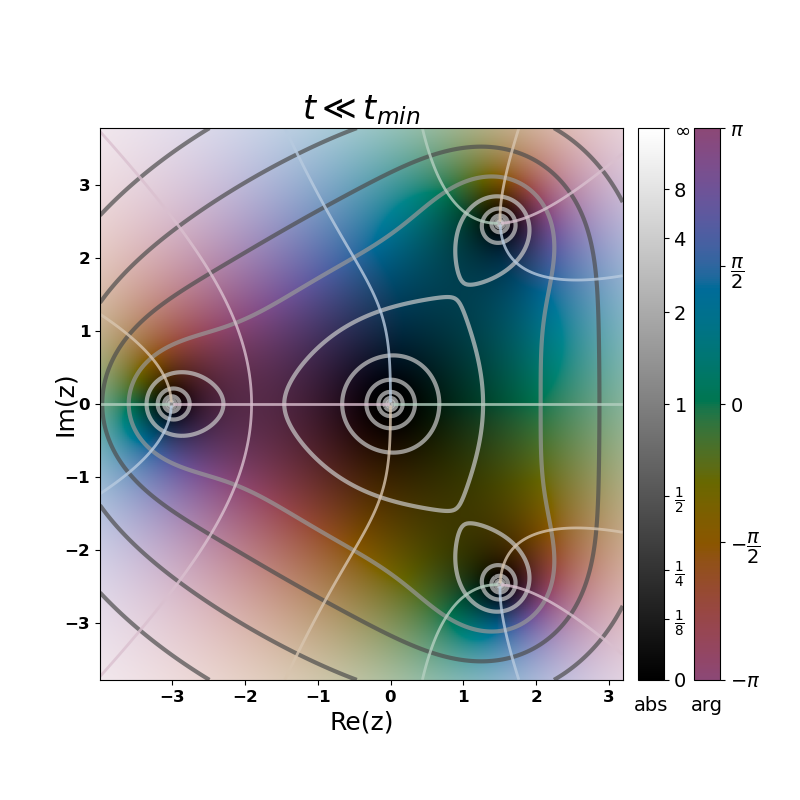}%\>
		\hspace{-0.4cm}
		\includegraphics[scale=.348]{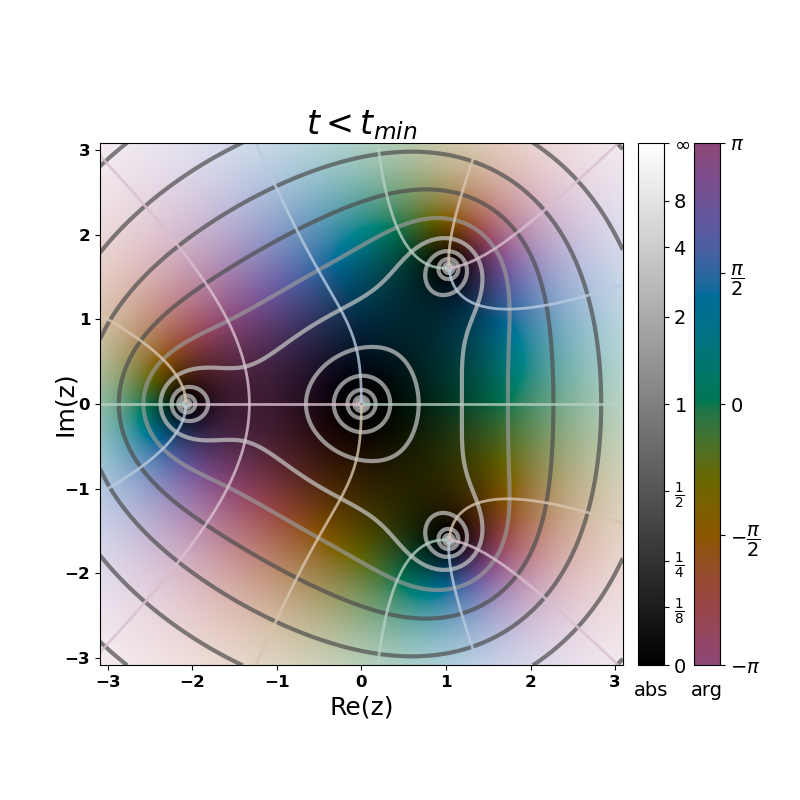}%\\
		\hspace{-0.4cm}
		\includegraphics[scale=.345]{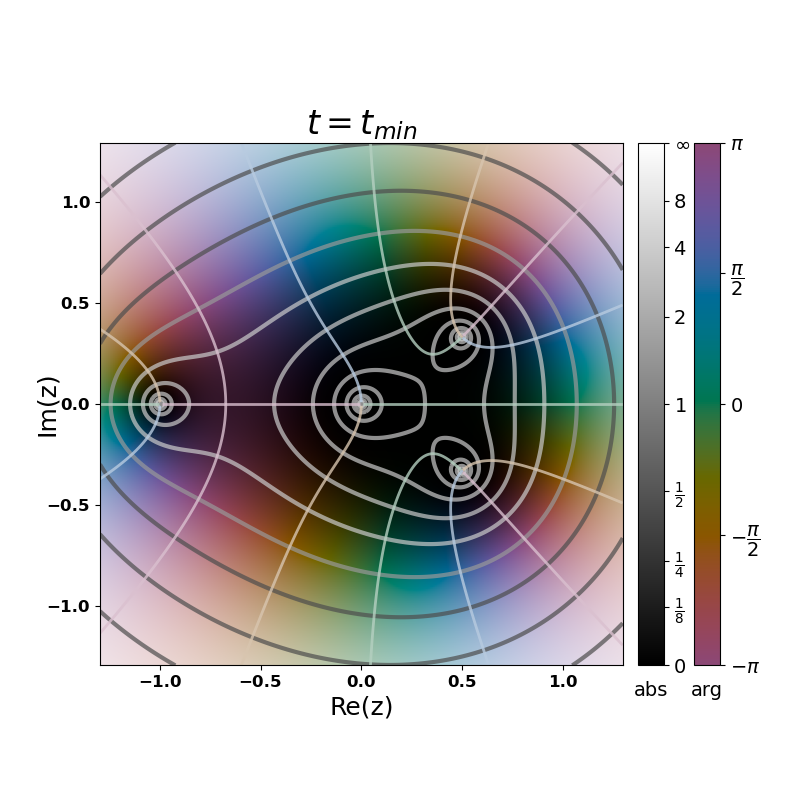}
	\end{tabbing}
	\vspace{-0.7cm}

\end{figure}
\vspace{-1.5cm}

\begin{figure}[H]
	\begin{tabbing}
		\hspace{-2.3cm}
		\centering
		%		\hspace{-2.3cm}%\=\kill
		\includegraphics[scale=.335]{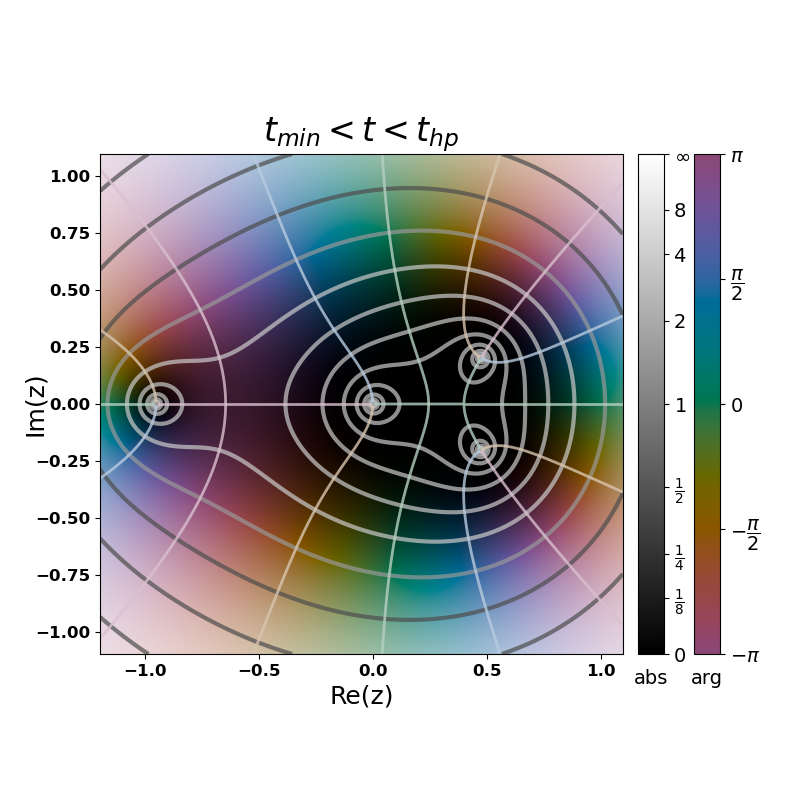}%\>
		\hspace{-0.4cm}
		\includegraphics[scale=.34]{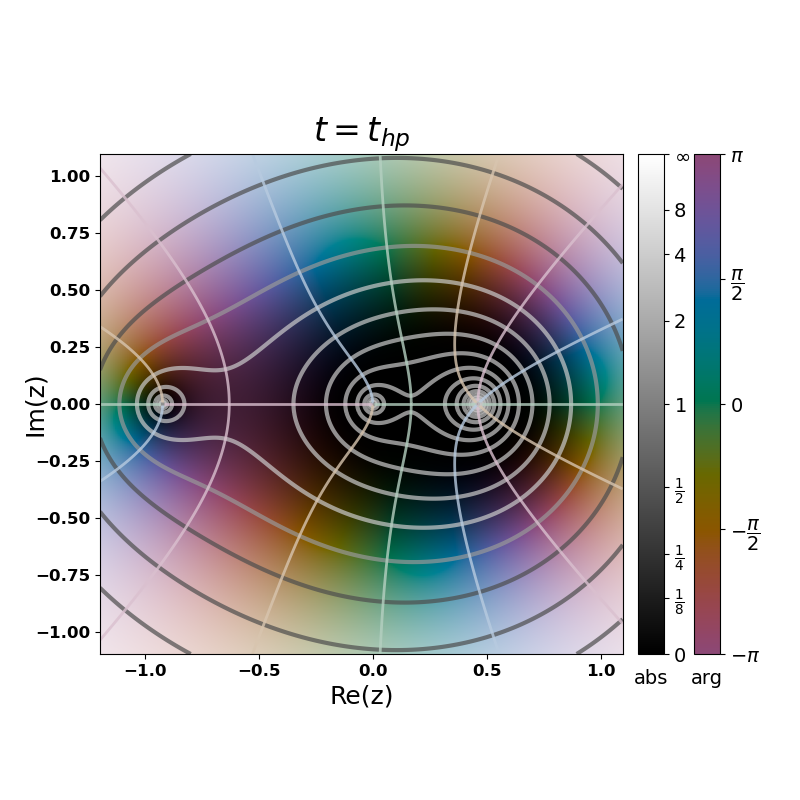}%\\
		\hspace{-0.4cm}
		\includegraphics[scale=.34]{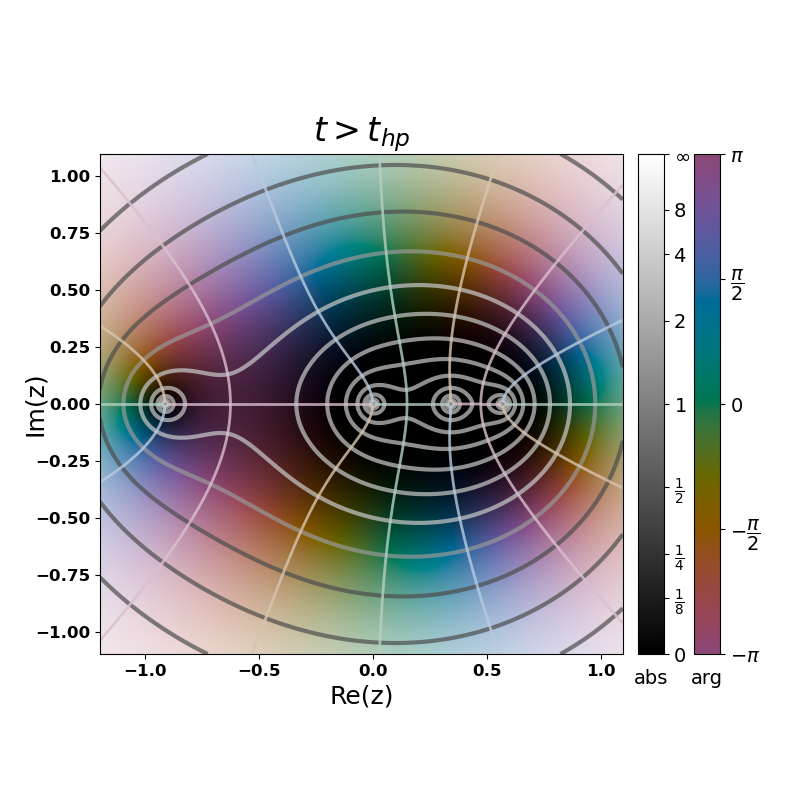}%\\
	\end{tabbing}
\end{figure}

\vspace{-1.8cm}

%%%%%%%%%%%%%%%%%%%%%%%%%%%%%%%%%%%%5
\newpage

\begin{figure}[H]
	
	\vspace{-0.8cm}
	\centering
	\begin{tabbing}
		\centering
		\hspace{-2.3cm}%\=\kill
		\includegraphics[scale=.34]{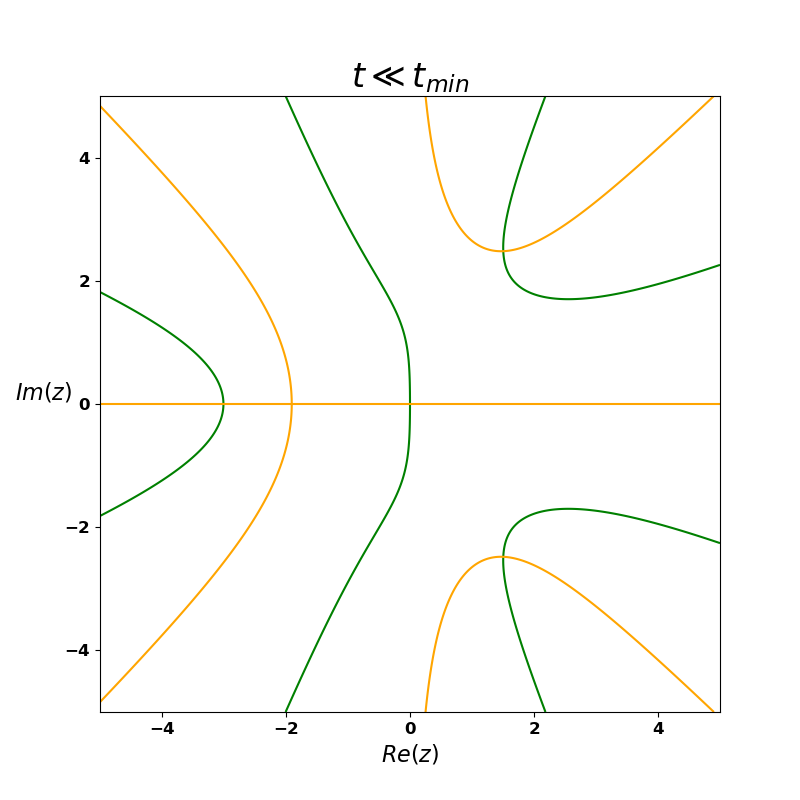}%\>
		\hspace{-0.4cm}
		\includegraphics[scale=.348]{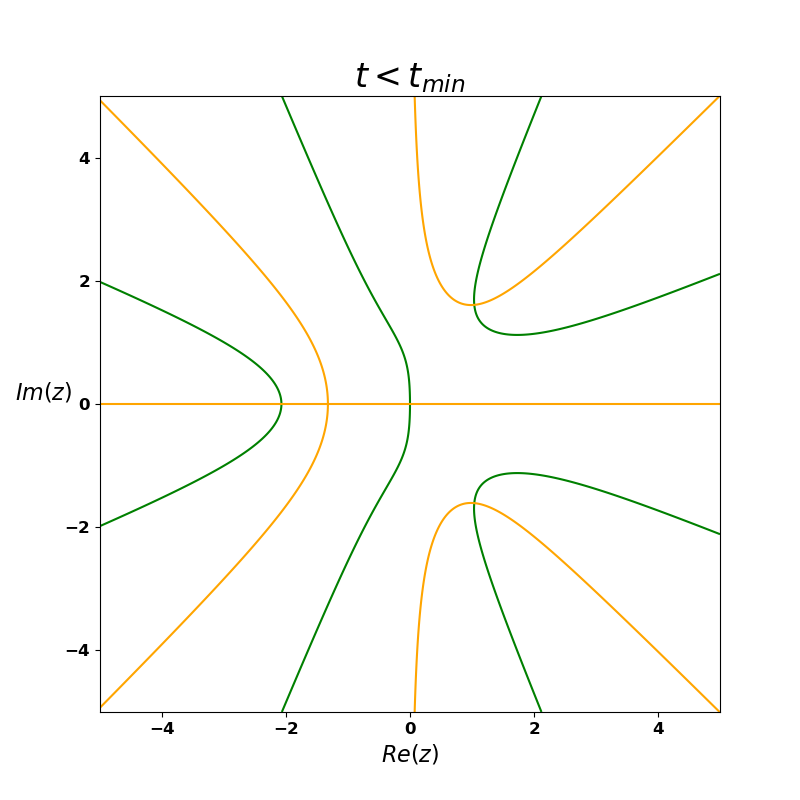}%\\
		\hspace{-0.4cm}
		\includegraphics[scale=.345]{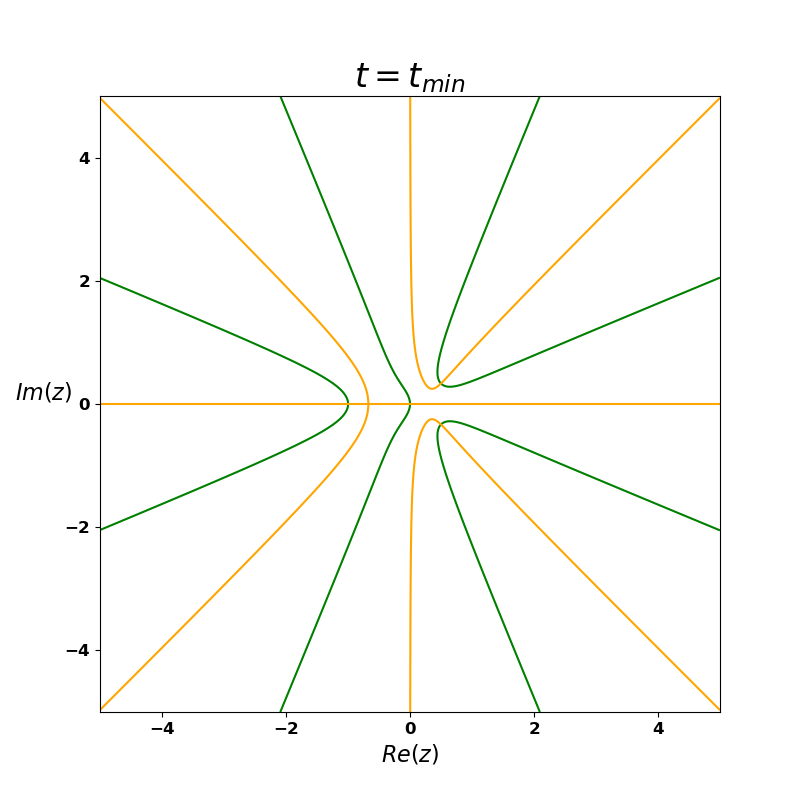}
	\end{tabbing}
	\vspace{-0.7cm}

\end{figure}
\vspace{-1.5cm}

\begin{figure}[H]
	\begin{tabbing}
		\hspace{-2.3cm}
		\centering
		%		\hspace{-2.3cm}%\=\kill
		\includegraphics[scale=.335]{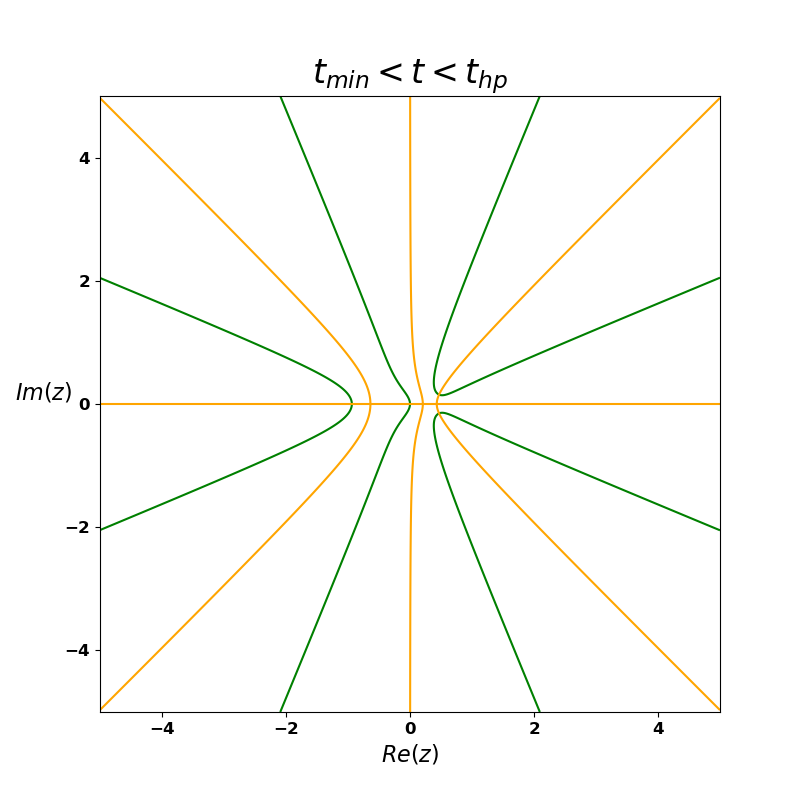}%\>
		\hspace{-0.4cm}
		\includegraphics[scale=.34]{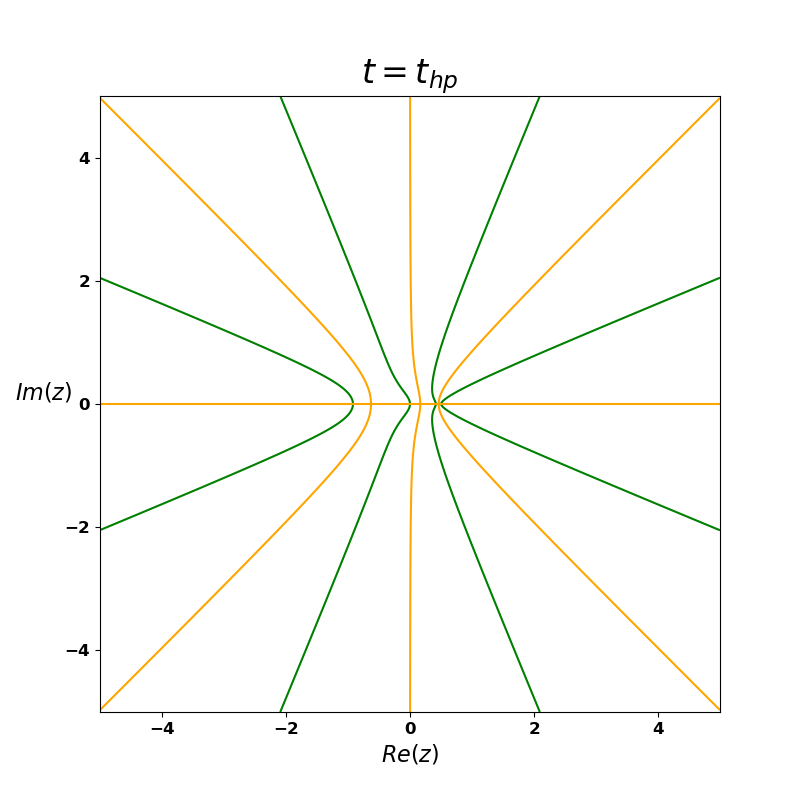}%\\
		\hspace{-0.4cm}
		\includegraphics[scale=.34]{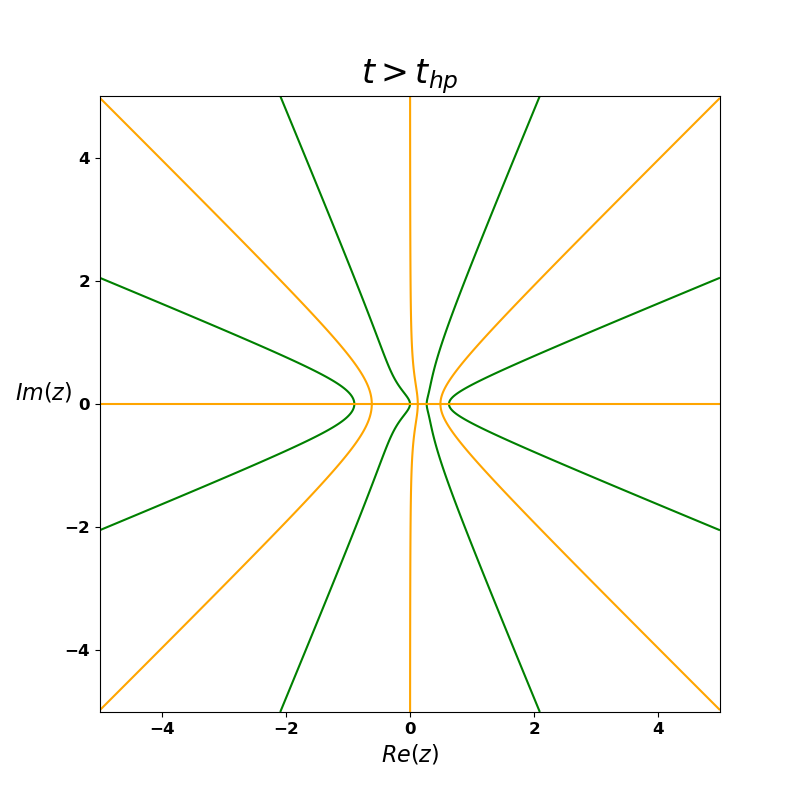}%\\
	\end{tabbing}
\end{figure}

\vspace{-1.8cm}

%textcolor{red}{comment}

%%%%%%%%%%%%%%%%%%%%%%%%%%%%%%%%%%%%%%
%\subsection{}
%%%%%%%%%%%%%%%%%%%%%%%%%%%%%%%%%%%%%%
%need t=0.01, t=0.15, t=1.1
%\begin{figure}[ H]
%	\centering
%	\begin{tabbing}
%		\centering
%		%			\hspace{-2.3cm}%\=\kill
%		\includegraphics[scale=.42]{uv_0_t_0.003}\label{}%\>
%		\hspace{-0.7cm}
%		\includegraphics[scale=.42]{uv_0_t_0.02}%\\
%	\end{tabbing}
%	\vspace{-0.7cm}
%	
%	\label{fig:prenyiv}
%\end{figure}
%
%\begin{figure}[ H]
%	\centering
%	\begin{tabbing}
%		\centering
%		%		\hspace{-2.3cm}%\=\kill
%		\includegraphics[scale=.42]{uv_0_t_0.05}\label{}%\>
%		\hspace{-0.7cm}
%		\includegraphics[scale=.42]{uv_0_t_0.1}%\\
%	\end{tabbing}
%	\vspace{-0.7cm}
%	
%	\label{fig:prenyiv}
%\end{figure}
%
%
%\begin{figure}[ H]
%	\centering
%	\begin{tabbing}
%		\centering
%		%		\hspace{-2.3cm}%\=\kill
%		\includegraphics[scale=.42]{uv_0_tmin}\label{}%\>
%		\hspace{-0.7cm}
%		\includegraphics[scale=.42]{uv_0_t_between}%\\
%	\end{tabbing}
%	\vspace{-0.7cm}
%	
%	\label{fig:prenyiv}
%\end{figure}
%
%\begin{figure}[ H]
%	\centering
%	\begin{tabbing}
%		\centering
%		%		\hspace{-2.3cm}%\=\kill
%		\includegraphics[scale=.42]{uv_0_thp}\label{}%\>
%		\hspace{-0.7cm}
%		\includegraphics[scale=.42]{uv_0_2thp}%\\
%	\end{tabbing}
%	\vspace{-0.7cm}
%	
%	\label{fig:prenyiv}
%\end{figure}
%
%
%\begin{figure}[ H]
%	\centering
%	\begin{tabbing}
%		\centering
%		%		\hspace{-2.3cm}%\=\kill
%		\includegraphics[scale=.42]{uv_0_5thp}\label{}%\>
%	\end{tabbing}
%	\vspace{-0.7cm}
%	\caption{}
%	\label{fig:prenyiv}
%\end{figure}
%
%

%%%%%%%%%%%%%%%%%%%%%%%%%%%%%%%%%%%%%%
\newpage
\begin{figure}[H]
	\centering
	\vspace{-2cm}
	\begin{tabbing}
		
		\centering
		\hspace{-2.5cm}%\=\kill
		\includegraphics[scale=.37]{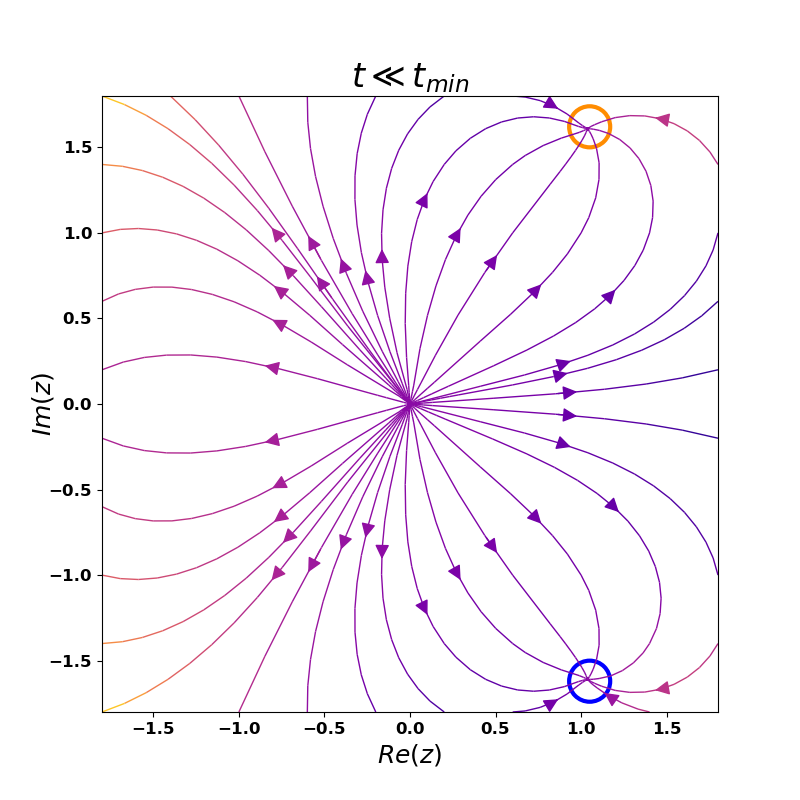}\label{}%\>
		\hspace{-0.85cm}
		\includegraphics[scale=.37]{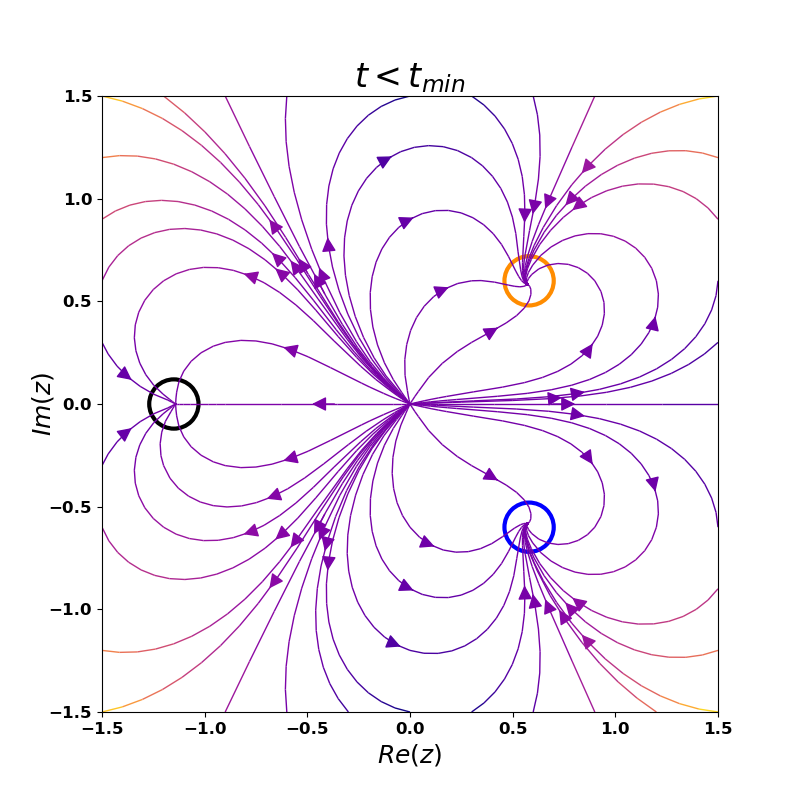}%\\
		\hspace{-0.85cm}
		\includegraphics[scale=.37]{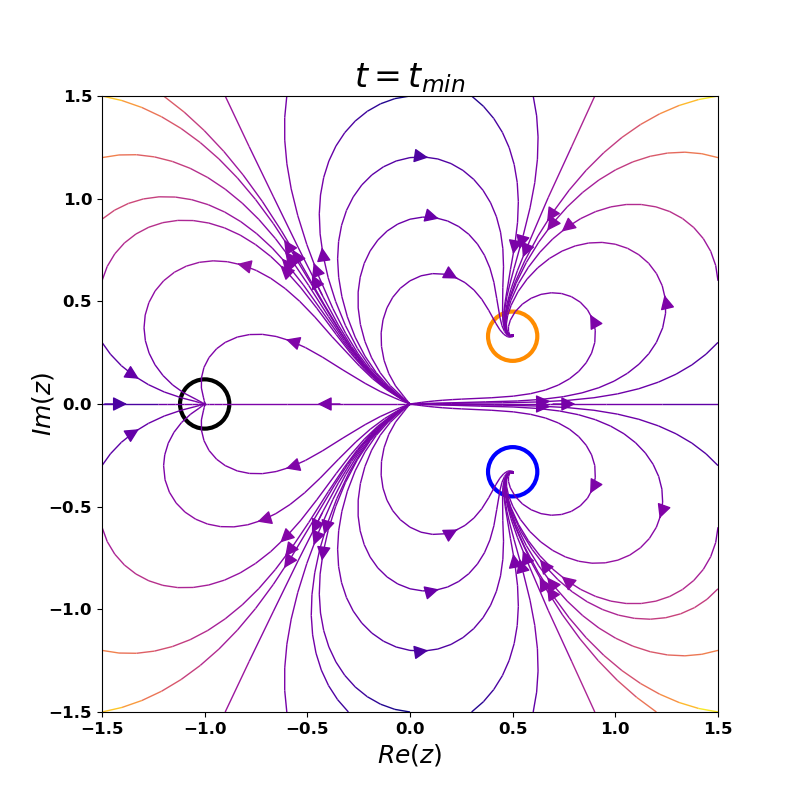}
	\end{tabbing}
	\vspace{-0.7cm}
\end{figure}
\vspace{-1cm}
\begin{figure}[H]
	\centering
	\begin{tabbing}
		\centering
		\hspace{-2.5cm}%\=\kill
		\includegraphics[scale=.37]{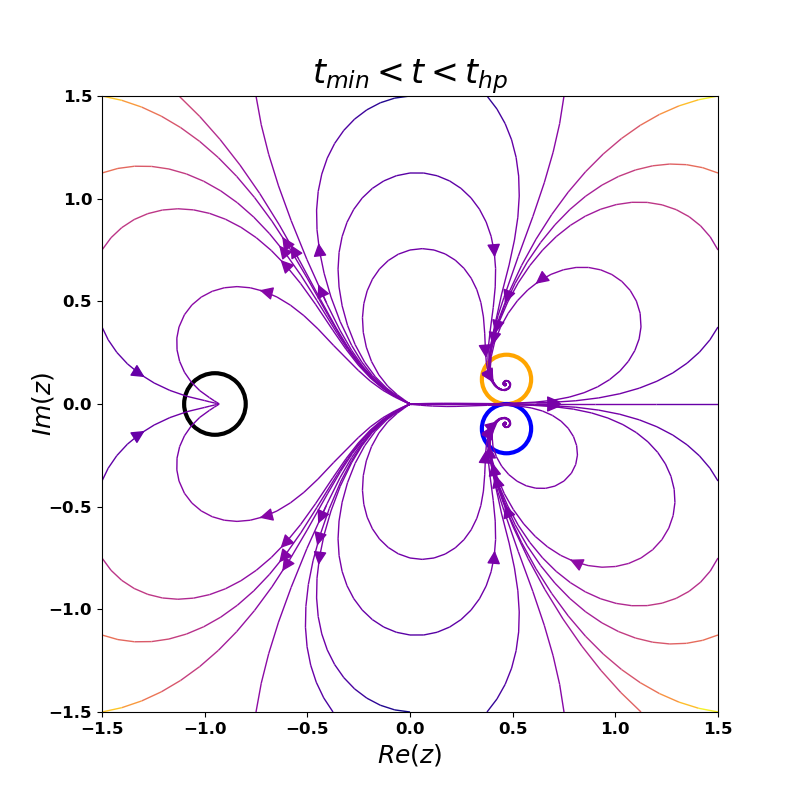}\label{}%\>
		\hspace{-0.85cm}
		\includegraphics[scale=.37]{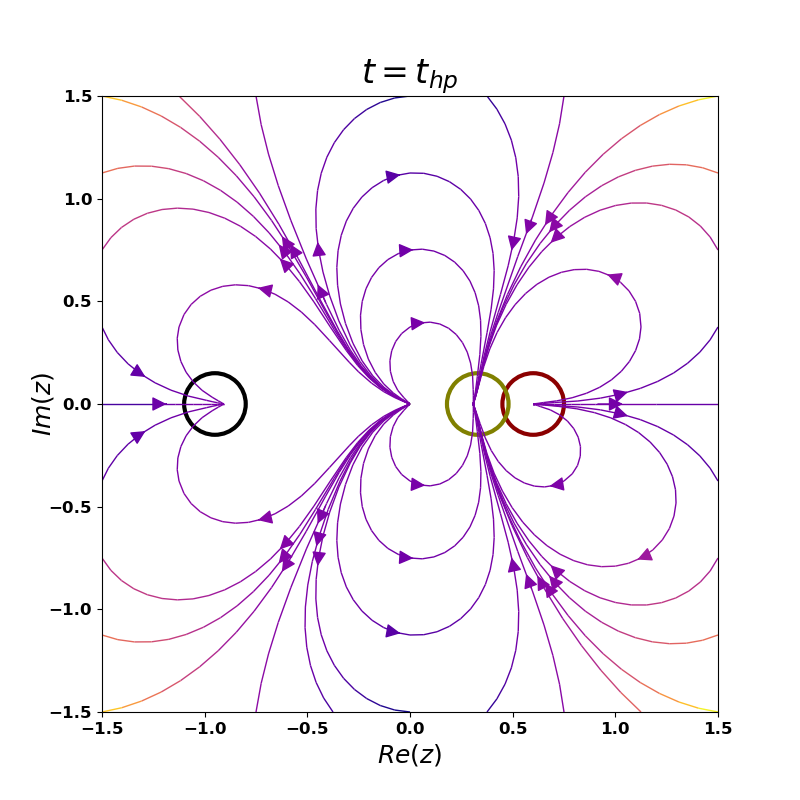},
		\hspace{-1.1cm}
		\includegraphics[scale=.37]{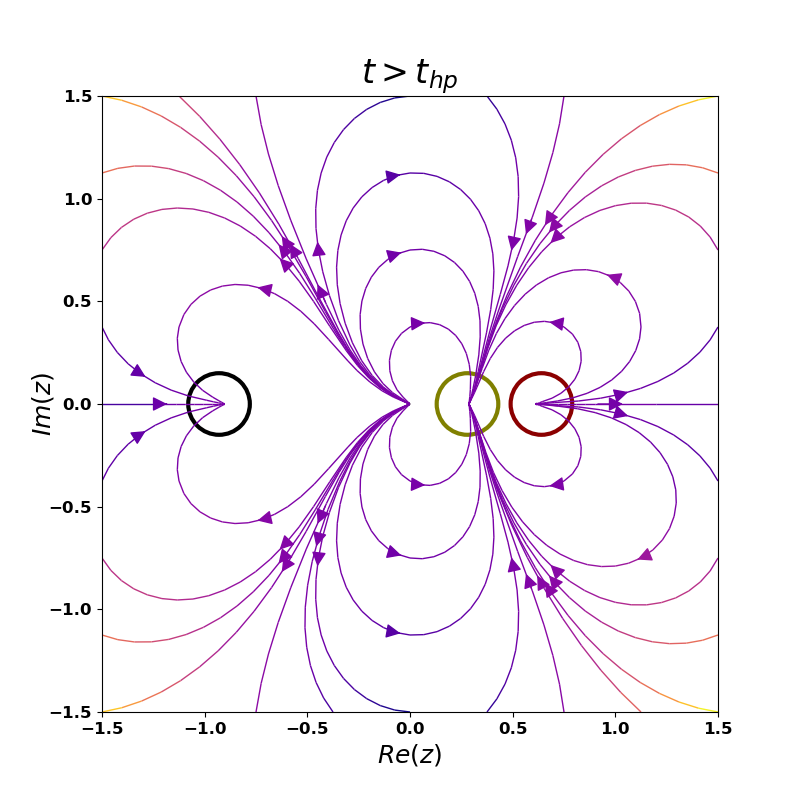}
	\end{tabbing}
	\vspace{-0.7cm}
\end{figure}
\vspace{-1cm}
%%%%%%%%%%%%%%%%%%%%%%%%%%%%%%%%%%%%%%

\subsection{The winding numbers in the grand canonical ensemble}
\begin{figure}[H]
	\vspace{-0.4cm}
	\centering
	\begin{tabbing}
		\centering
		\hspace{3cm}%\=\kill
		\includegraphics[scale=.6]{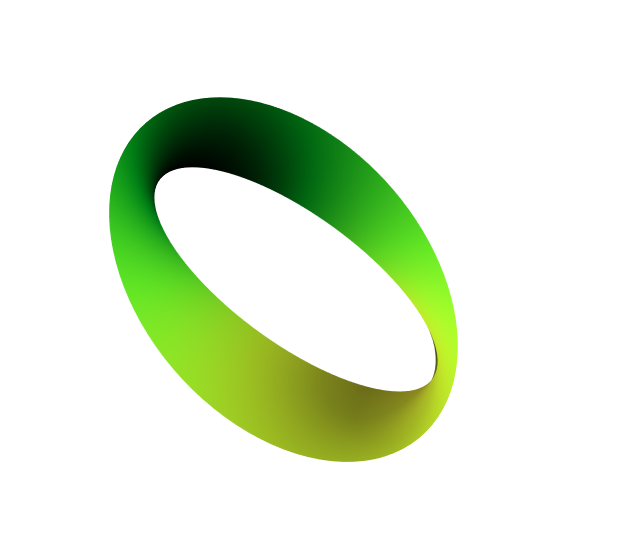}\label{}%\>
%		\hspace{-0.85cm}
%		\includegraphics[scale=.37]{flow_hp_8},
%		\hspace{-1.1cm}
%		\includegraphics[scale=.37]{flow_hp_9}
	\end{tabbing}
	\vspace{-1cm}	
	\caption{\footnotesize\it  Riemann surface associated with the R\'enyi-Hawking-Page phase transition in $d=4$.}\label{fig:fig_Rie_d4_1}
\end{figure}

In the grand canonical ensemble, we define the analytic function as the derivative of the complex Bragg-Williams function with respect to the Rényi entropy, as follows
\begin{equation}\label{def_analytic}
h_\phi(z)=\frac{df_\phi}{dS_R}=t_{R,eff}-t,
\end{equation}
here the $t_{R,eff}$ is the complex physical temperature corresponding to physical states. The analytic function of the four-dimensional R\'enyi charged-flat black hole is given by
\begin{equation}\label{exp_analytic}
h_\phi(z)=\displaystyle  \frac{\left(1-\phi^{2}\right) \left( \pi z^{2} + 1\right)-4 \pi tz}{4 \pi z}.
\end{equation}
\paragraph{} The zeros of the analytic function correspond to the extrema of the Braggs-Williams free energy. These zeros represent the stable (minima) and unstable (maxima) phase states of the black hole. 
This complex function is manifestly single-valued, as is the case for all analytic functions in arbitrary dimensions. Consequently, it does not exhibit branches or non-trivial Riemann surfaces. However, one can associate with it the Riemann surface of its inverse, $h^{-1}_\phi(z)$, which could be a multi-valued function.  \textit{Therefore, the proposed setup is to associate the Riemann surface of the multi-valued inverse analytic function with the profile of Hawking-Page phase transitions}. The inverse of the analytic function, Eq.\eqref{exp_analytic} admits a one-sheet trivial Riemann surface as shown in Fig.\ref{fig:fig_Rie_d4_1}. The reason is that $h_\phi$ has two zeros and one pole at the origin, which amounts to a winding number $\mathcal{W}(h_\phi)=2-1=1$ and hence a number of sheets, $\mathcal{S}(h^{-1}_\phi)=1$.

We identify the winding number, $\mathcal{W}_{HP}$ associated with the Hawking-Page phase transition of the R\'enyi charged black hole with the number of sheets of the Riemann surface of the inverse analytic function, $\mathcal{S}_{HP}$, thus we write
\begin{equation}
    \mathcal{W}_{HP}=\mathcal{S}_{HP}.
\end{equation}

Then, $\mathcal{W}_{HP} = 1$ for the four-dimensional case. Since the Hawking-Page phase transition is a second-order phase transition, this establishes a reference for winding numbers in four-dimensional spacetime: a single second-order phase transition requires a winding number of at least $\mathcal{W} \geq 1$. For a general four-dimensional asymptotically flat Rényi black hole with a given winding number $\mathcal{W} \geq 1$, the maximum number of second-order phase transitions, $N_2$, is given by
\begin{equation}\label{N2}
    N_2=\mathcal{W}.
\end{equation}
In what follows, we focus on using complex analysis formalism and the complexified Braggs-Williams free energy to investigate Riemann surfaces and their winding numbers within the canonical ensemble. This ensemble, characterized by a fixed charge, allows us to delve into the Rényi statistical mechanics of charged-flat black holes exhibiting a Van der Waals-like phase structure.

\section{Van der Waals phase transition of the 4-dimensional R\'enyi charged-flat  black hole from complex analysis perception}

\subsection{Complexification of the R\'eniy Bragg-Williams free energy in the canonical ensemble}

Whitin a fixed charge $Q$, the R\'eniy Gibbs-Williams free energy of the charged R\'enyi-flat black hole in the canonical ensemble is found to be,
\begin{align}
f_Q(r)&=M-t S_R-\phi Q\\
 &=\displaystyle \frac{16 p r \left(- Q^{2} + r^{2}\right) + 3 t \left(Q^{2} - r^{2}\right) \log{\displaystyle\left(\frac{3 Q^{2} - 32 \pi p r^{4} - 3 r^{2}}{3(Q^{2} - r^{2})} \right)} }{32 p r^{2}}.
\end{align}
Here $p$ designates the R\'enyi pressure \cite{Promsiri:2020jga,Barzi:2023mit}.  Following the same process of the previous sections, the analytic continuation, $r\longrightarrow z$, gives,
\begin{equation}\label{key}
f_Q(z)=  \displaystyle \frac{16 p z \left(- Q^{2} + z^{2}\right) + 3 t \left(Q^{2} - z^{2}\right) \log{\displaystyle\left(\frac{3 Q^{2} - 32 \pi p z^{4} - 3 z^{2}}{3(Q^{2} - z^{2})} \right)} }{32 p z^{2}}.
\end{equation}
In the limit $0<p<<1$, and performing the following scaling definitions, $z \rightarrow p^{-1/2}z$, $t\rightarrow p^{1/2} t$, $Q\rightarrow p^{-1/2}Q$, and $ f_Q\rightarrow  p^{-1/2}f_Q$, we obtain,
% \begin{equation}\label{key}
% \bar{f}_R(r,t,Q)=\displaystyle \frac{- Q^{2} + r^{2} \left(\pi^{2} r^{3} t - 2 \pi r t + 1\right)}{2 r}
% \end{equation}

\begin{equation}\label{f_Q_4}
f_Q(z)=\displaystyle \frac{3 \left(Q^2-z^2\right)^2+2 \pi  z^3 t \left(3 Q^2+16 \pi z^4-3 z^2\right)}{6 z\left(z^2-Q^2\right)}.
\end{equation}
The real and imaginary parts of $f_Q(z)$ are given by,
\begin{equation}
\begin{cases}
\begin{split}
\chi(x,y)=&\frac{1}{2}
x \left(1-\frac{ Q^2}{x^2+y^2}\right)+ \pi t \left(y^2-x^2\right)\\\\&-\frac{16 \pi ^2 t \left[-x^8+4 x^6 y^2+10 x^4 y^4+4 x^2 y^6-y^8+Q^2 \left(x^6-15 x^4 y^2+15 x^2 y^4-y^6\right)\right]}{3\left[Q^4+2 Q^2 \left(y^2-x^2\right)+\left(x^2+y^2\right)^2\right]}
\end{split}\\\\
\begin{split}
\psi(x,y)=&\frac{3 yQ^2}{2(x^2+y^2)}+\frac{y}{2}(1-4 \pi t) \\\\&+\frac{8}{3} y \pi ^2 t \left[\frac{Q^5}{(Q+x)^2+y^2}-\frac{Q^5}{(Q-x)^2+y^2}+4 Q^2 x+8 x (x^2-y^2)\right]
\end{split}
\end{cases}
\end{equation}

The complex function $f_Q(z)$, Eq.\eqref{f_Q_4}, presents a great similarity with the complex potential flow of a doublet. Indeed, it can be readily verified that $\nabla^2\chi=0$ and $\nabla^2\psi=0$, asserting the potential character of the energy flow in the canonical ensemble.

\begin{figure}[H]
	\begin{tabbing}
		\hspace{-2.3cm}
		\centering
		%		\hspace{-2.3cm}%\=\kill
		\includegraphics[scale=.34]{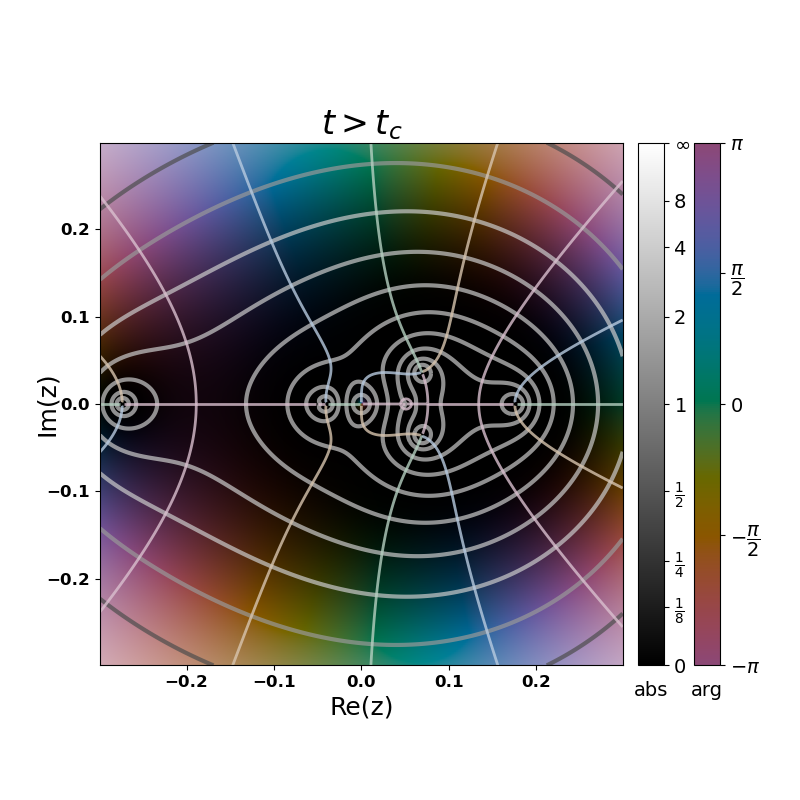}%\>
		\hspace{-0.4cm}
		\includegraphics[scale=.34]{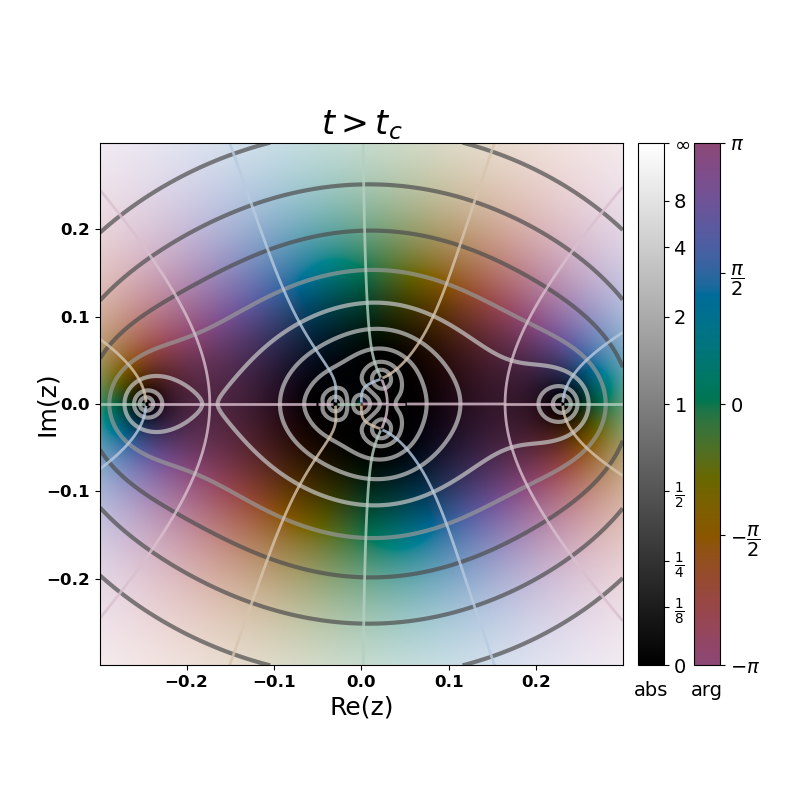}%\\
		\hspace{-0.4cm}
		\includegraphics[scale=.34]{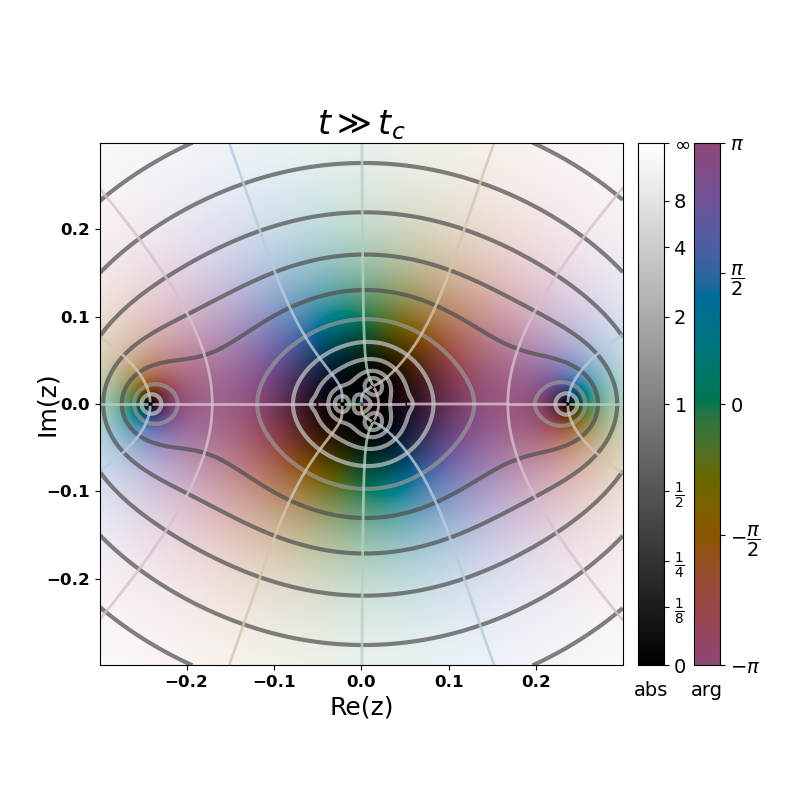}%\\
	\end{tabbing}
	\vspace{-0.7cm}
	\caption{\footnotesize \it R\'enyi-Van-der-Waals phase transition: The absolute value and phase of the complex Bragg-Williams free energy for the 4d-charged-flat black hole as the temperature gradually increases. The electric charge has been set to
  $Q=0.05$.}	\label{fig:fig_d4_4}
\end{figure}

\paragraph{}Figure \ref{fig:fig_d4_4} illustrates the thermal evolution of both the absolute value and the phase of the complex Bragg-Williams free energy in the canonical ensemble of the four-dimensional Rényi charged-flat black hole and within the subcritical regime where the scaled electric charge of the black hole is below the scaled critical charge $Q_c=0.08726$\cite{Barzi:2023msl}. Similarly, a continuous process drives three points, two complex conjugate points with a positive real part ($Re(z)>0$) towards the real axis and a third real point which shifts its position to meet the complex pair. Upon merging at the temperature \( t_c \), they form the Van der Waals transition point. This point represents the transition SBH$\longleftrightarrow$LBH which proceeds through the intermediate black hole state (IBH). 

\begin{figure}[H]
	\centering
	\begin{tabbing}
		\centering
		\hspace{-2.5cm}%\=\kill
		\includegraphics[scale=.37]{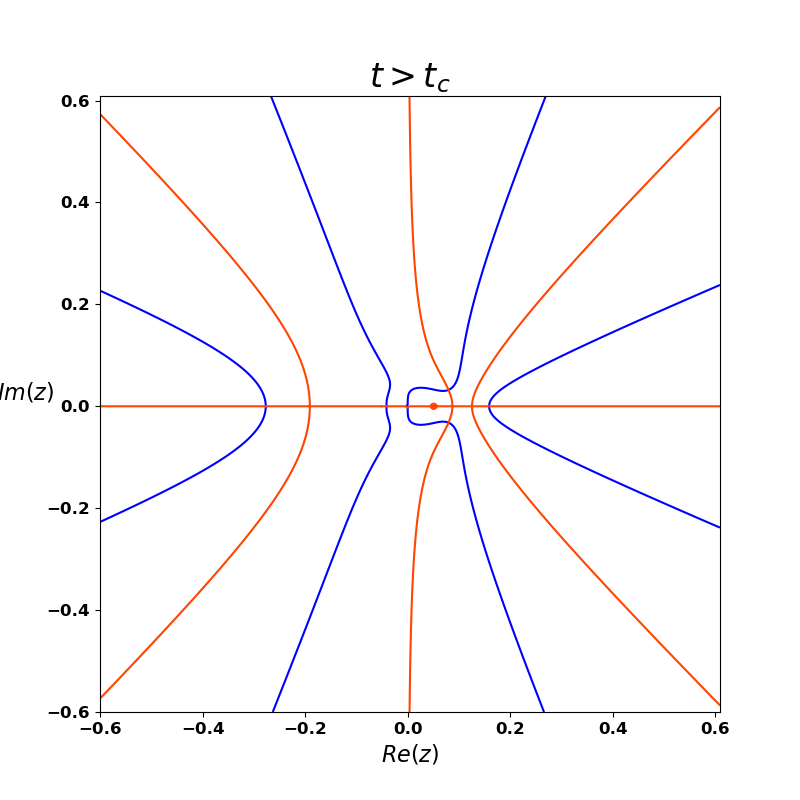}\label{}%\>
		\hspace{-0.85cm}
		\includegraphics[scale=.37]{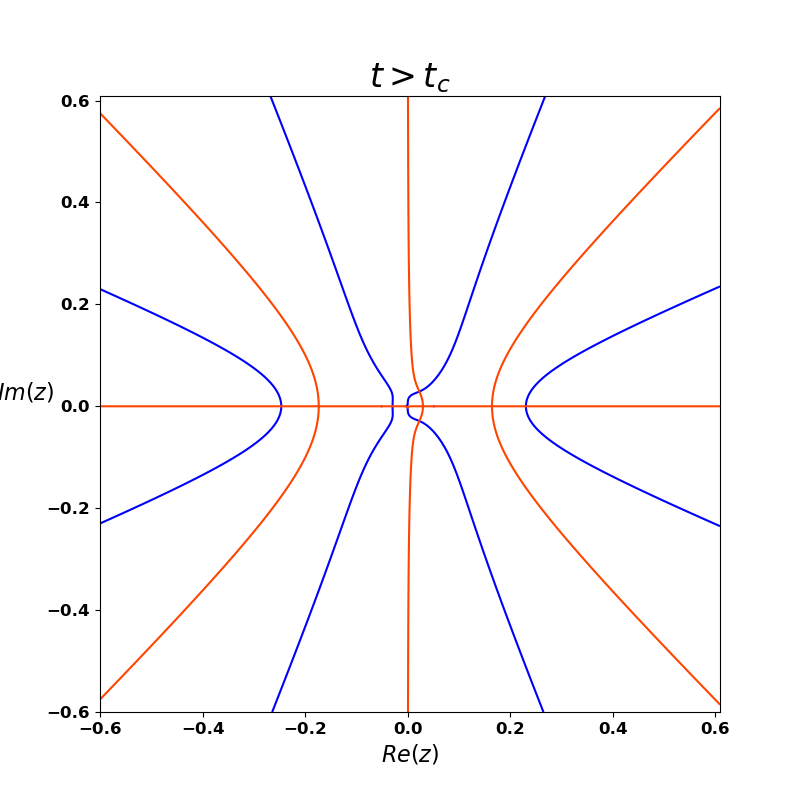},
		\hspace{-1.1cm}
		\includegraphics[scale=.37]{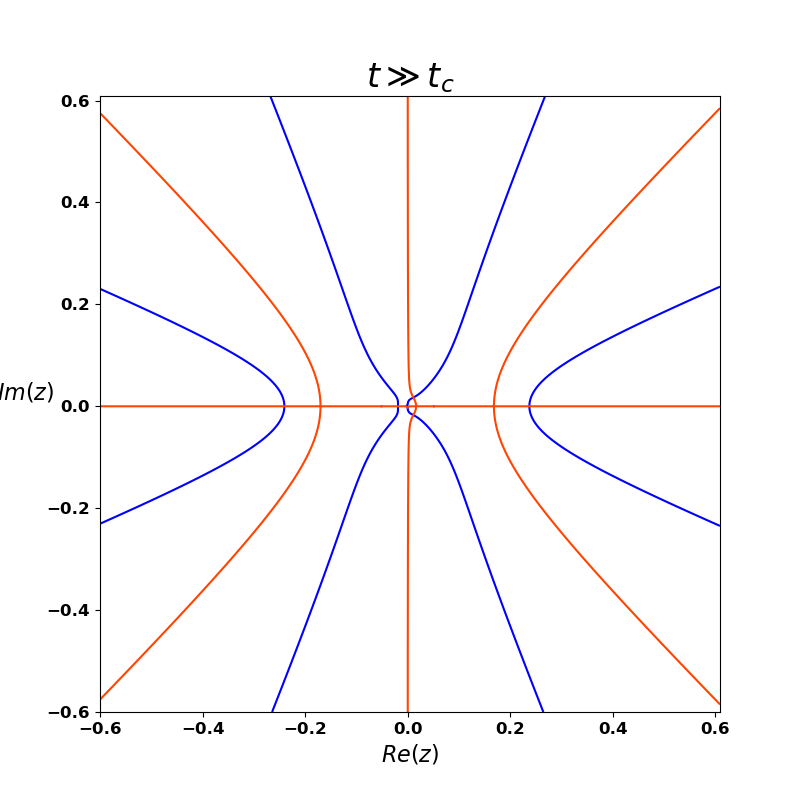}
	\end{tabbing}
	\vspace{-0.7cm}
	\caption{\footnotesize\it  R\'enyi-Van-der-Waals phase transition: The evolution of the zeros of the real part $\chi=0$ (\textbf{blue lines}) 
 and the zeros of the imaginary part $\psi=0$ (\textbf{orange-red lines}) of the complex free energy for the 4d-charged-flat black hole as the temperature gradually increases. The electric charge has been set to $Q=0.05$.}	\label{fig:fig_d4_5}
\end{figure}

To further elucidate the complex structure of the Van der Waals transition, Fig.\ref{fig:fig_d4_5} illustrates the evolution of the zeros of the real and imaginary parts of the complex Braggs-Williams free energy. The blue lines correspond to the solutions of $\chi(x, y)=0$, while the orange lines represent $\psi(x,y)=0$. In this framework, the intersections of the blue and orange lines mark the zeros of the complex free energy. As the temperature increases, a real zero near the small values of the scaled black hole horizon radius, $r$, indicative of the SBH$\longleftrightarrow$IBH transition, gradually shifts due to its interaction with a nearby complex conjugate pair of zeros. This transformation leads to a new real zero at a higher horizon radius, signaling the IBH$\longleftrightarrow$LBH phase transition.

% \textcolor{red}{HERE.   more text about figs 8 and 9}
\newpage

\begin{figure}[H]
	
	\vspace{-3cm}
	\centering
	\begin{tabbing}
		\centering
		\hspace{-2.3cm}%\=\kill
		\includegraphics[scale=.34]{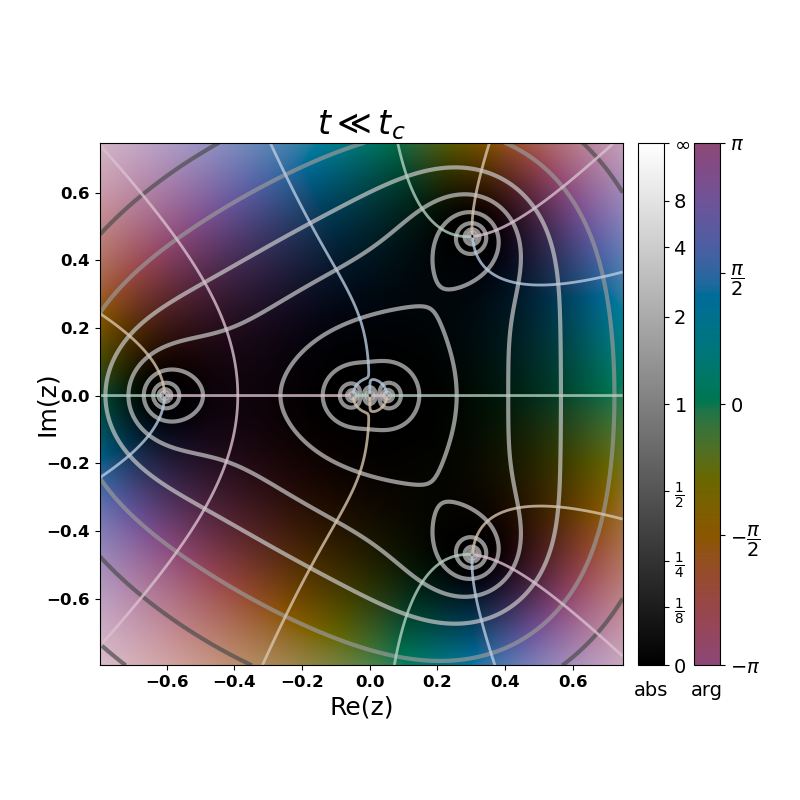}%\>
		\hspace{-0.4cm}
		\includegraphics[scale=.345]{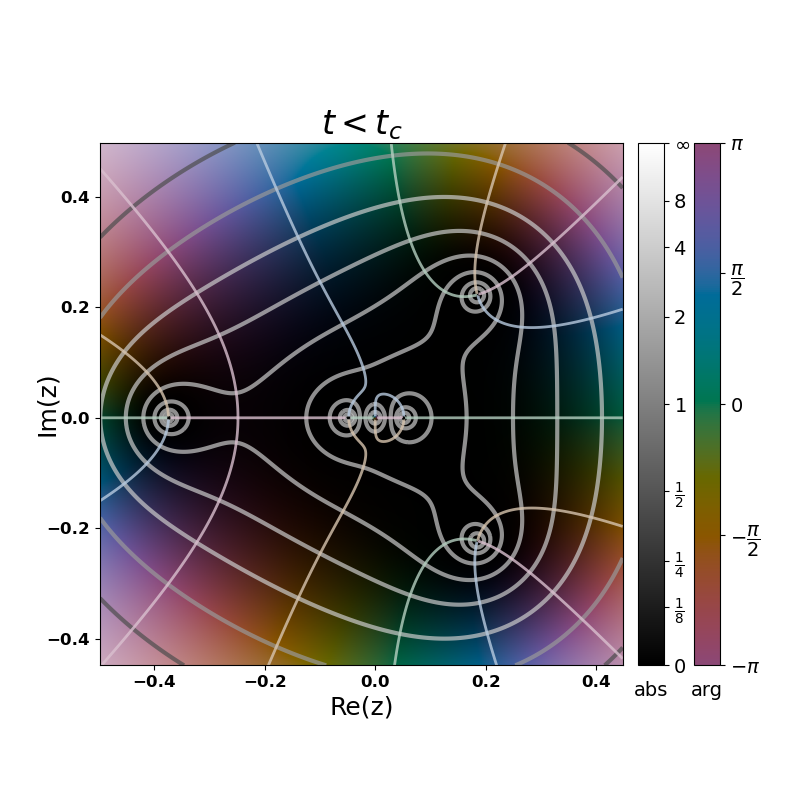}%\\
		\hspace{-0.4cm}
		\includegraphics[scale=.34]{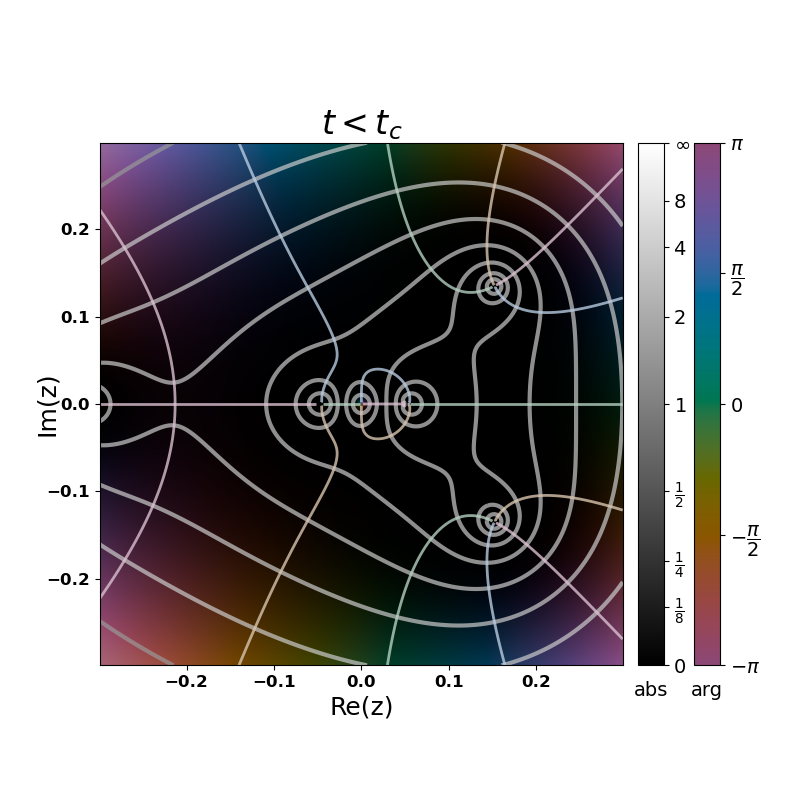}
	\end{tabbing}
	\vspace{-0.7cm}
	
\end{figure}
\vspace{-1.5cm}

\begin{figure}[H]
	\begin{tabbing}
		\hspace{-2.3cm}
		\centering
		%		\hspace{-2.3cm}%\=\kill
		\includegraphics[scale=.34]{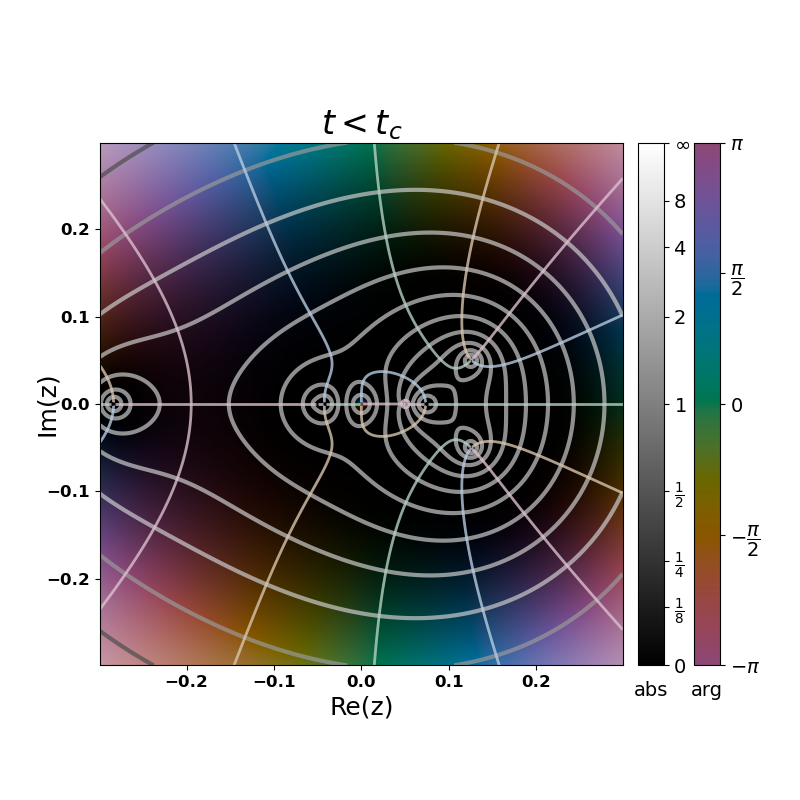}%\>
		\hspace{-0.4cm}
		\includegraphics[scale=.34]{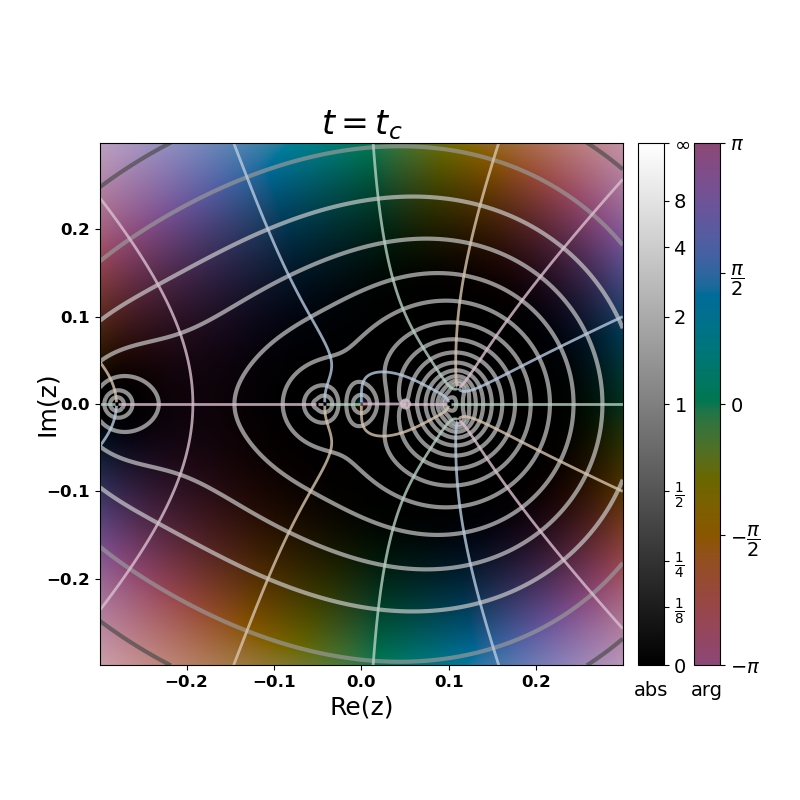}%\\
		\hspace{-0.4cm}
		\includegraphics[scale=.34]{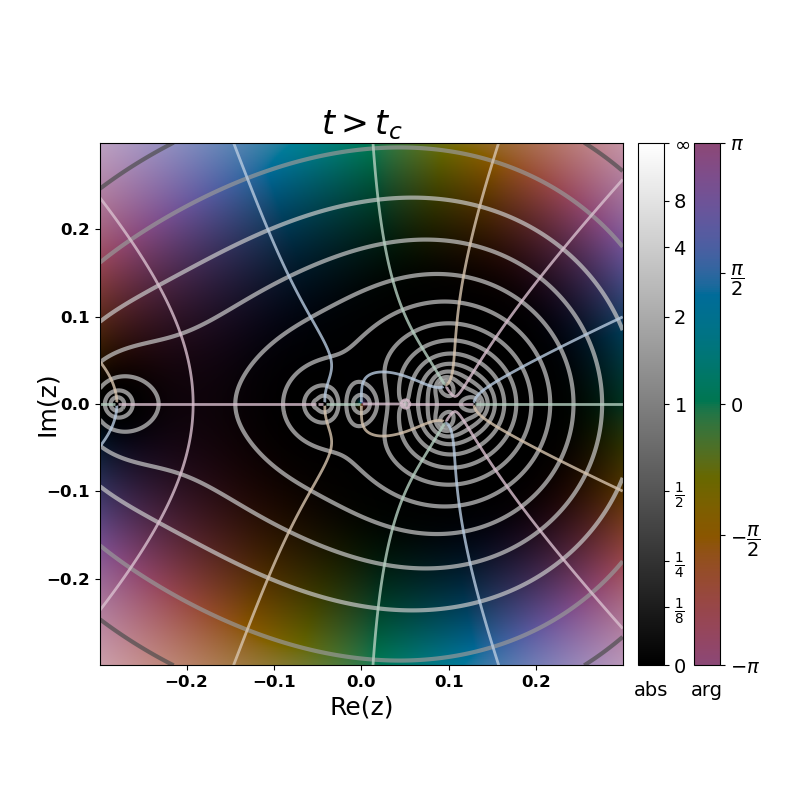}%\\
	\end{tabbing}
	\vspace{-0.7cm}
\end{figure}

\vspace{-1.5cm}

%%%%%%%%%%%%%%%%%%%%%%%%
\newpage
\begin{figure}[H]
	\centering
	\vspace{-2cm}
	\begin{tabbing}
		\centering
		\hspace{-2.5cm}%\=\kill
		\includegraphics[scale=.37]{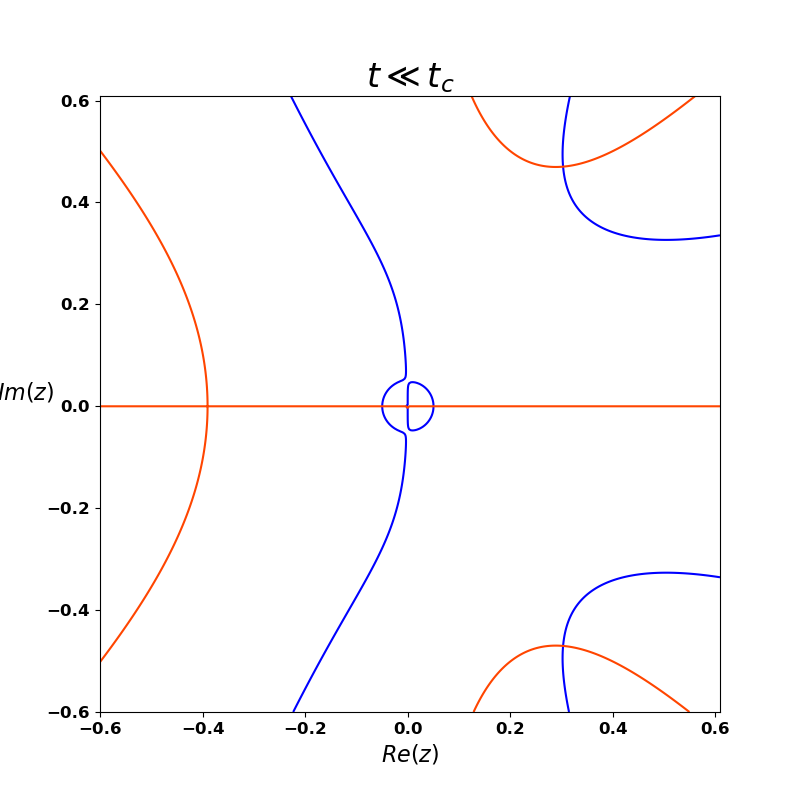}\label{}%\>
		\hspace{-0.85cm}
		\includegraphics[scale=.37]{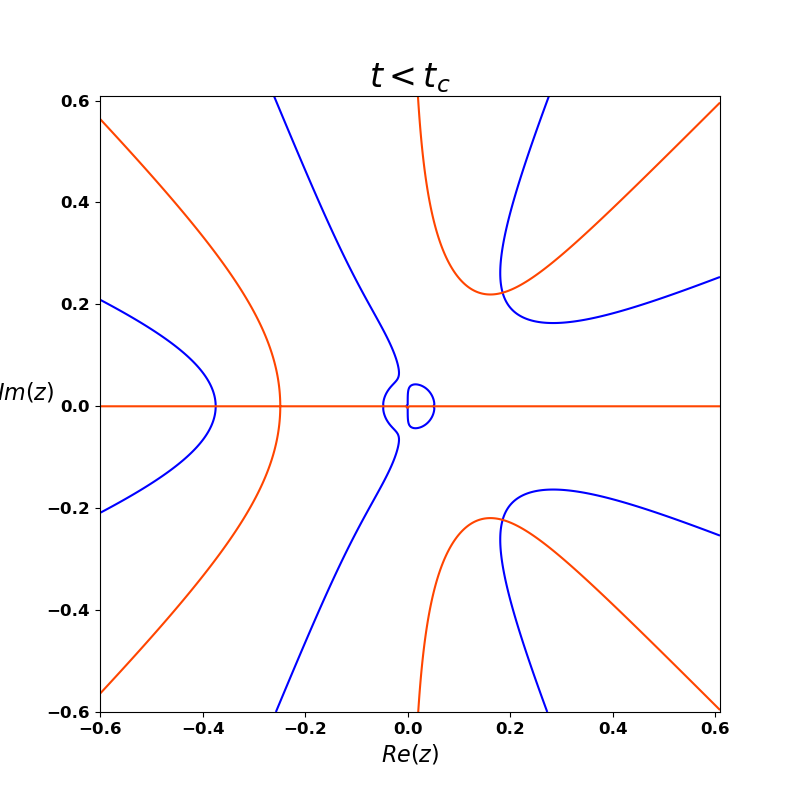}%\\
		\hspace{-0.85cm}
		\includegraphics[scale=.37]{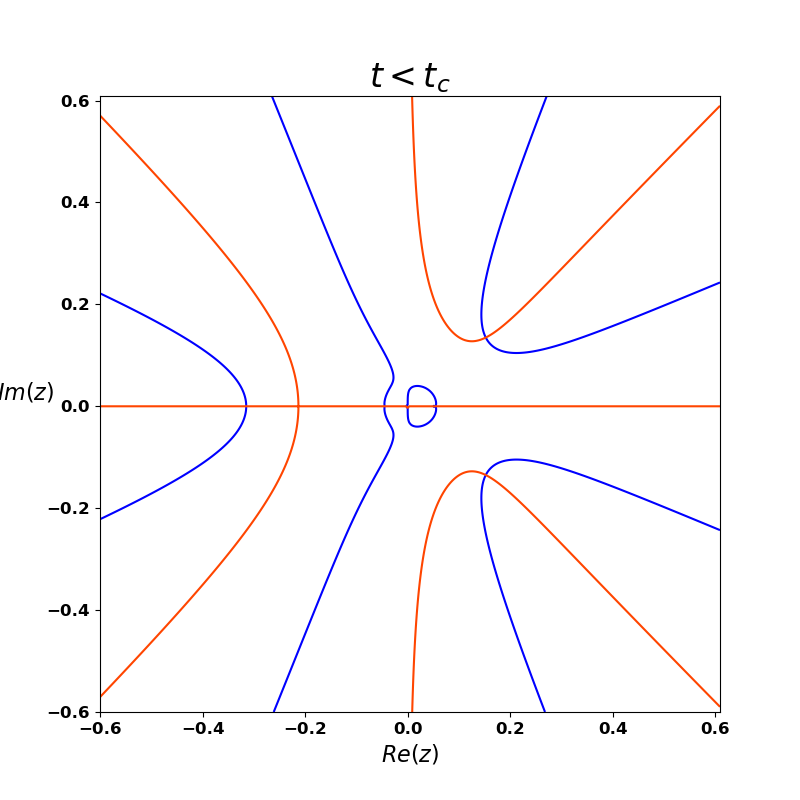}
	\end{tabbing}
	\vspace{-0.7cm}
\end{figure}
\vspace{-1cm}
\begin{figure}[H]
	\centering
	\begin{tabbing}
		\centering
		\hspace{-2.5cm}%\=\kill
		\includegraphics[scale=.37]{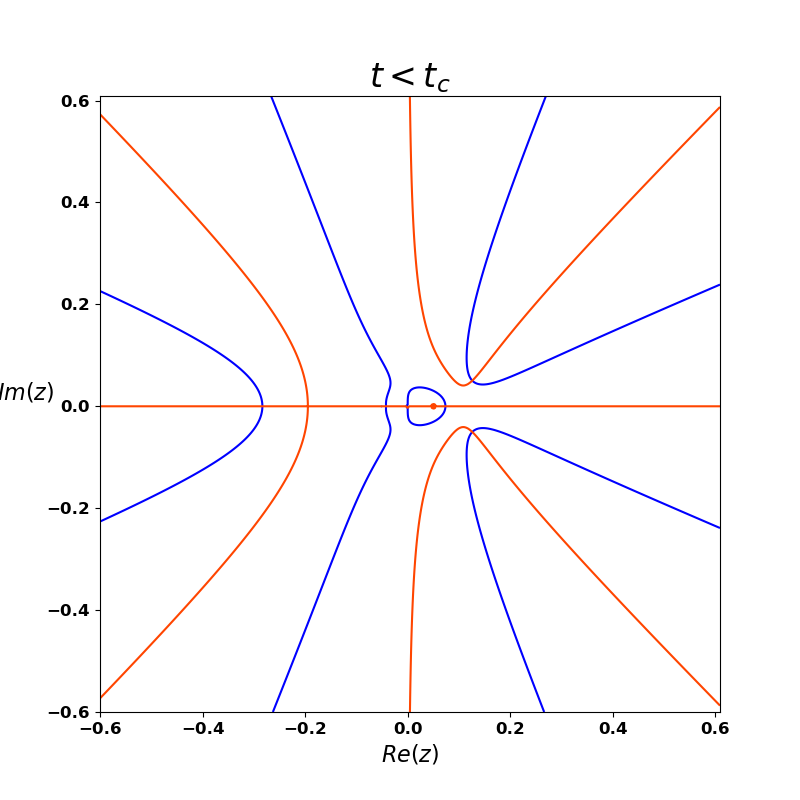}\label{}%\>
		\hspace{-0.85cm}
		\includegraphics[scale=.37]{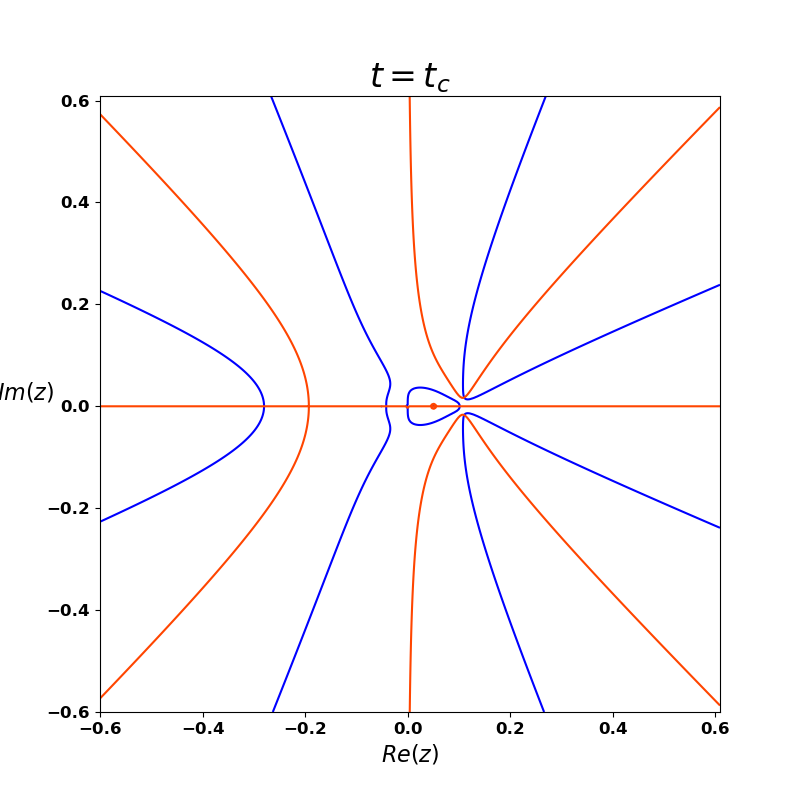},
		\hspace{-1.1cm}
		\includegraphics[scale=.37]{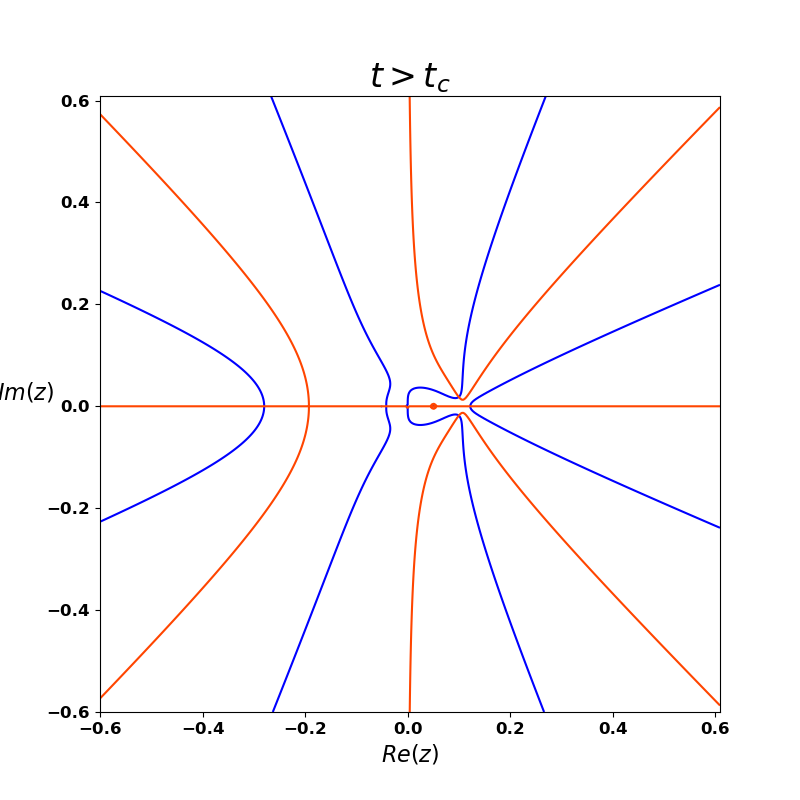}
	\end{tabbing}
	\vspace{-0.7cm}
\end{figure}
\vspace{-1cm}

\begin{figure}[H]
	\centering
	\begin{tabbing}
		\centering
		\hspace{-2.5cm}%\=\kill
		\includegraphics[scale=.37]{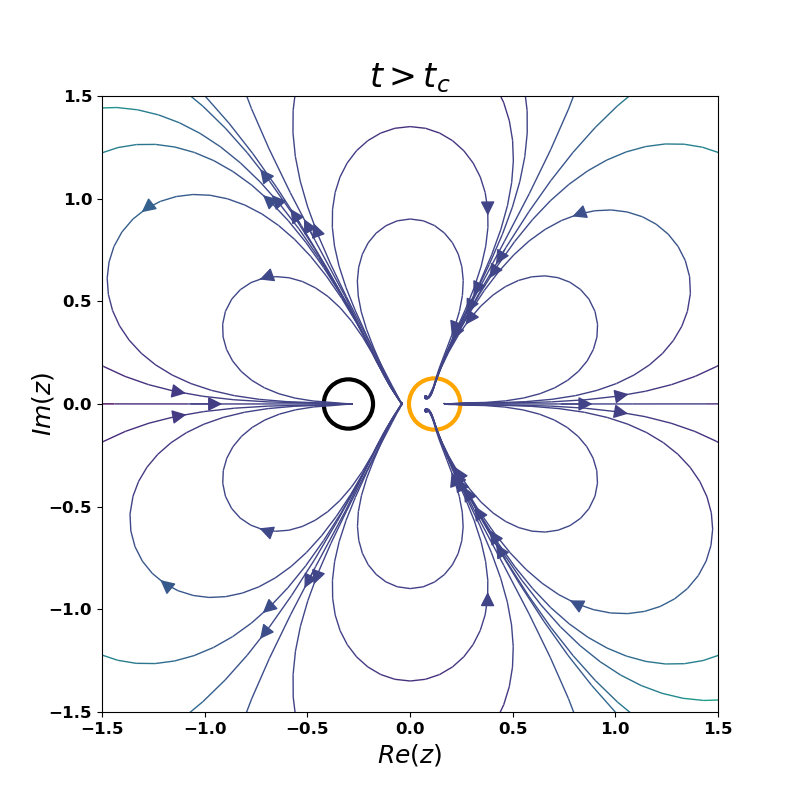}\label{}%\>
		\hspace{-0.85cm}
		\includegraphics[scale=.37]{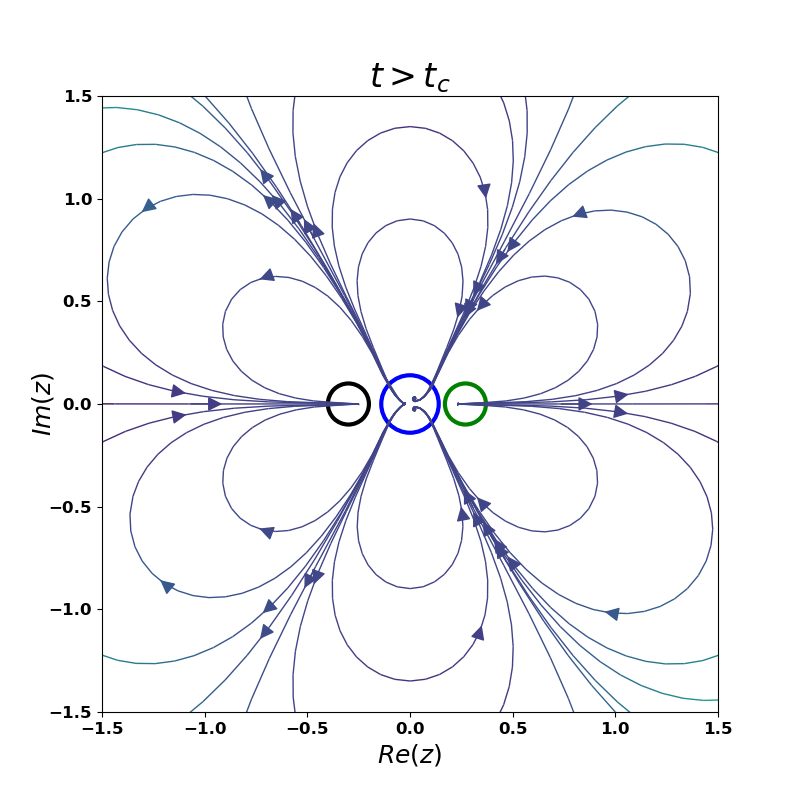},
		\hspace{-1.1cm}
		\includegraphics[scale=.37]{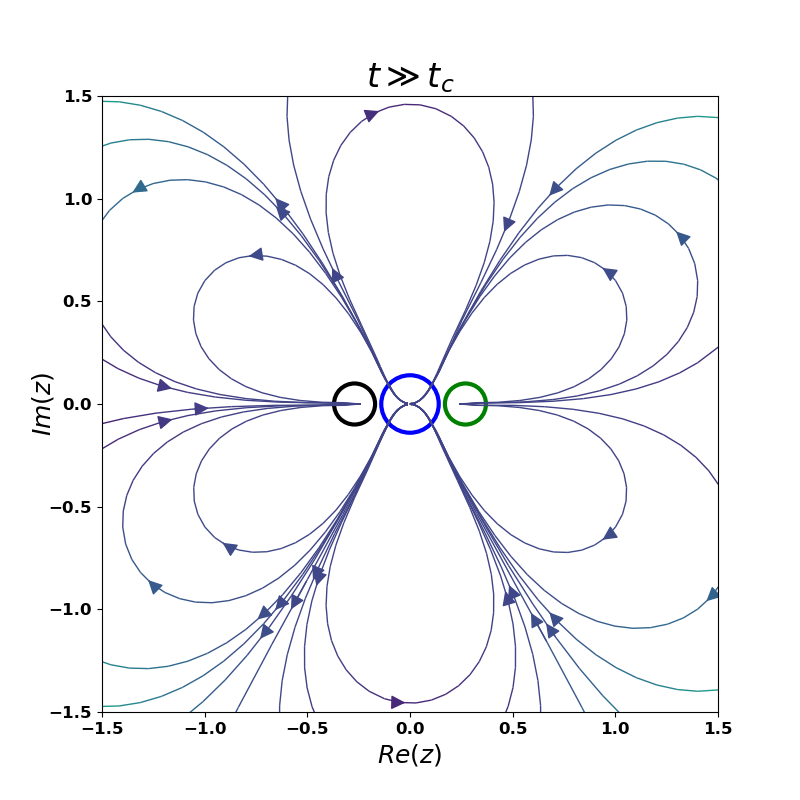}
	\end{tabbing}
	\vspace{-0.7cm}
	\caption{\footnotesize\it  R\'enyi-Van-der-Waals phase transition: The energy flow vector illustrations for the 4d-charged-flat black hole as the temperature gradually increases. Colored circles indicate relevant sources and sinks in the energy flow. The electric charge has been set to
  $Q=0.05$.}	\label{fig:fig_d4_6}
\end{figure}

The evolution of the complex Braggs-Williams energy flow in the canonical ensemble within the subcritical regime is portrayed in Fig.\ref{fig:fig_d4_6}. At low temperatures, the flow is characterized by a set of source and sink points, notably a pair of complex conjugated sinks with a positive real part approaching the real line in addition to two real and positive sources at and close to $z=0$. The energy flow from the sources to the pair of sinks pushes them to merge with non-zero source points at $t=t_c$ which constitutes thus the distinguishing feature of the Van der Waals phase transition. As temperature exceeds $t_c$, the complex conjugated pair detaches from the non-zero source point and gradually reconnects with the source at $z=0$. 

It is important to highlight a significant distinction: unlike the energy flow observed in the Hawking-Page phase transition depicted in Fig.\ref{fig:fig_d4_3}, the energy flow for the Van der Waals transition originates from two distinct sources. The first is the source at $z=0$, representing the thermal radiation phase without a black hole, and the second is the non-zero source corresponding to the Small Black Hole (SBH) phase. This difference marks a clear distinction between the complex structures of the Hawking-Page and Van der Waals phase transitions in the charged R\'enyi-flat black hole.
\newpage
\begin{figure}[H]
	\centering
	\vspace{-2cm}
	\begin{tabbing}
		\centering
		\hspace{-2.5cm}%\=\kill
		\includegraphics[scale=.37]{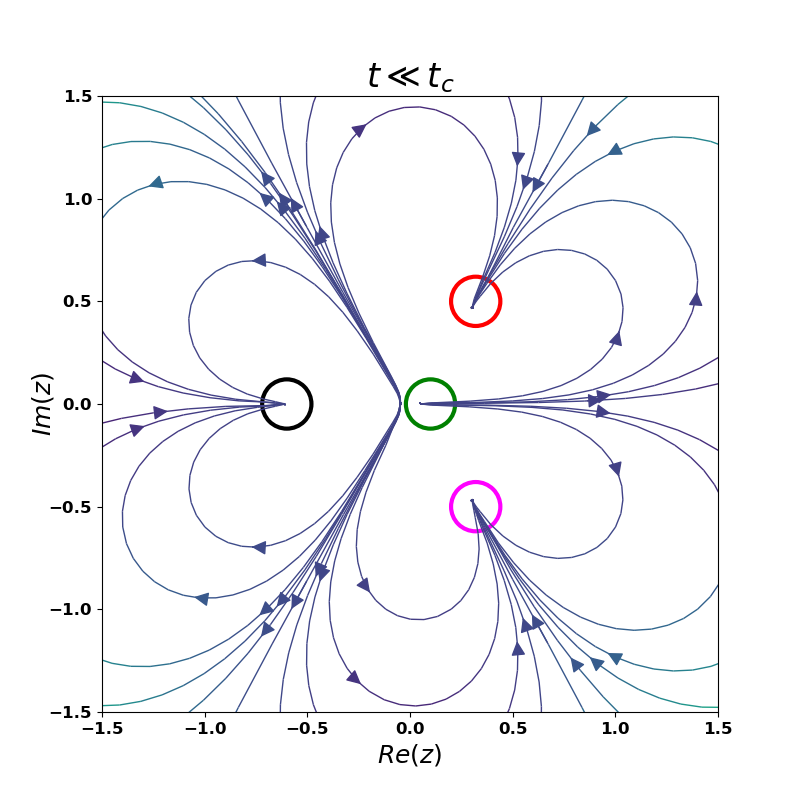}\label{}%\>
		\hspace{-0.85cm}
		\includegraphics[scale=.37]{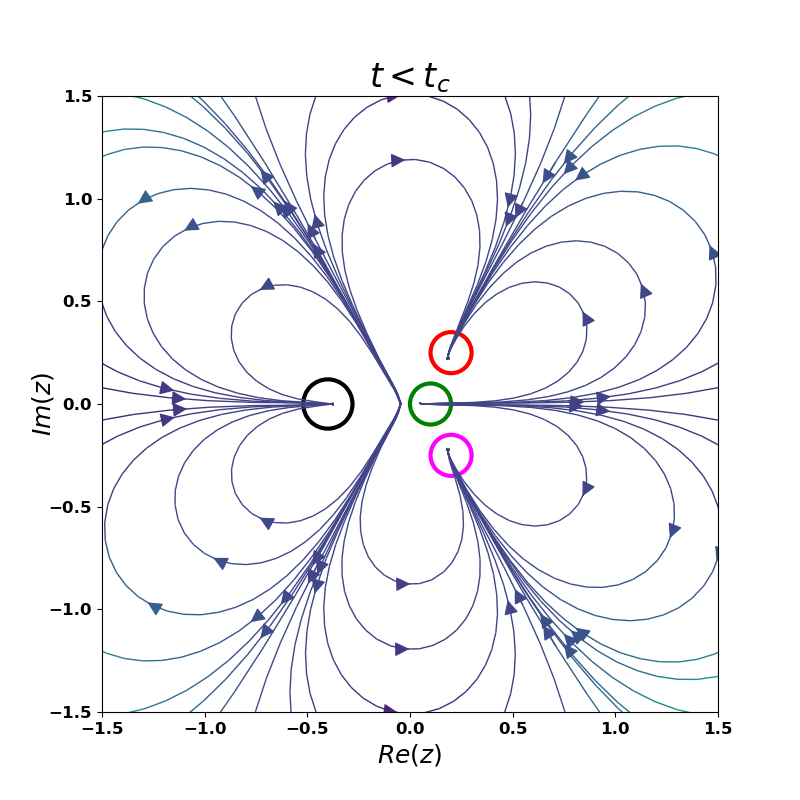}%\\
		\hspace{-0.85cm}
		\includegraphics[scale=.37]{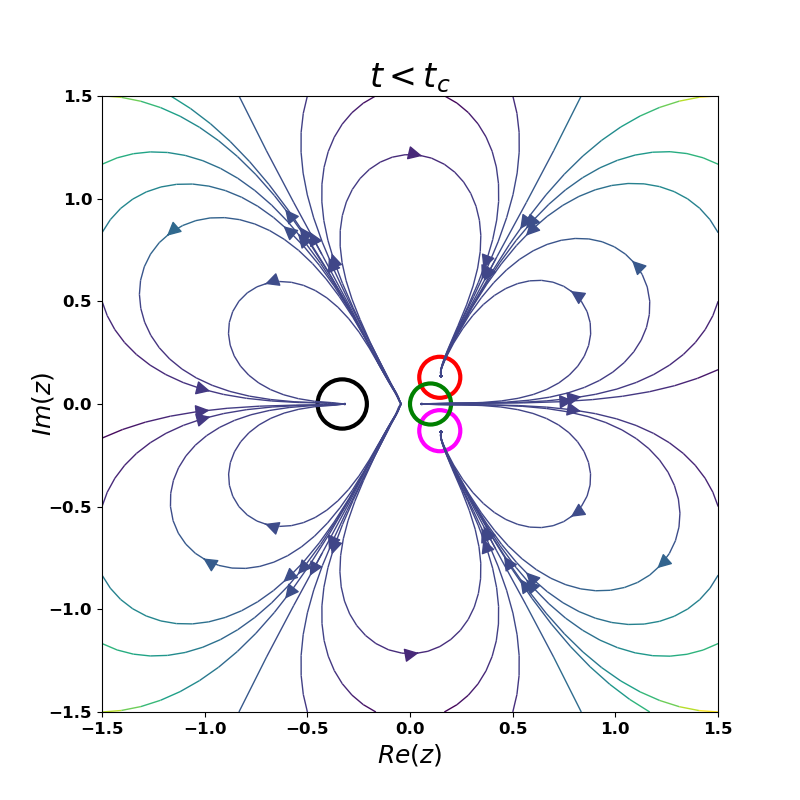}
	\end{tabbing}
	\vspace{-0.7cm}
\end{figure}
\vspace{-1cm}
\begin{figure}[H]
	\centering
	\begin{tabbing}
		\centering
		\hspace{-2.5cm}%\=\kill
		\includegraphics[scale=.37]{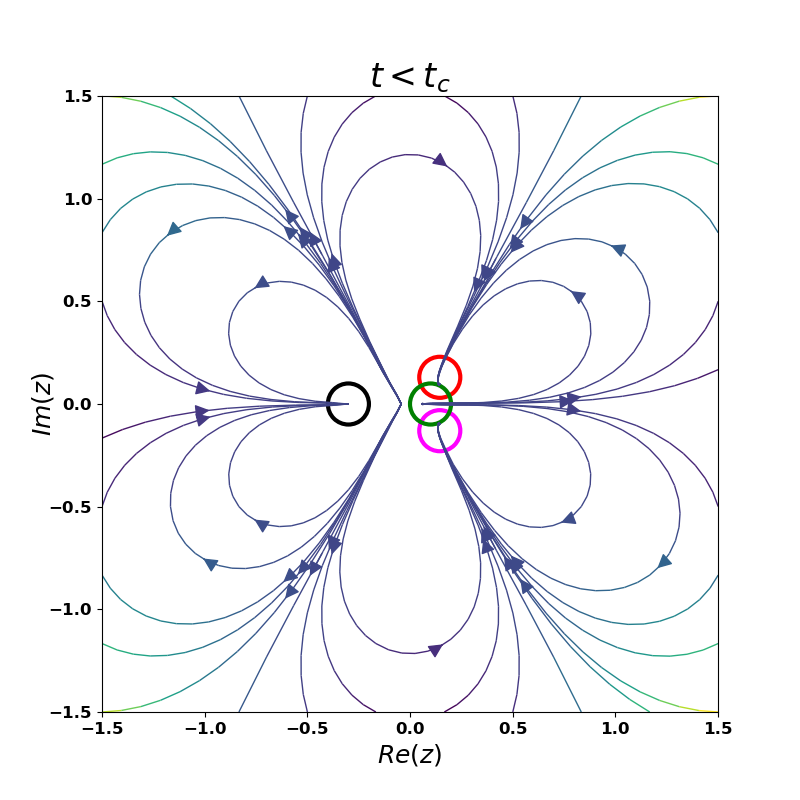}\label{}%\>
		\hspace{-0.85cm}
		\includegraphics[scale=.37]{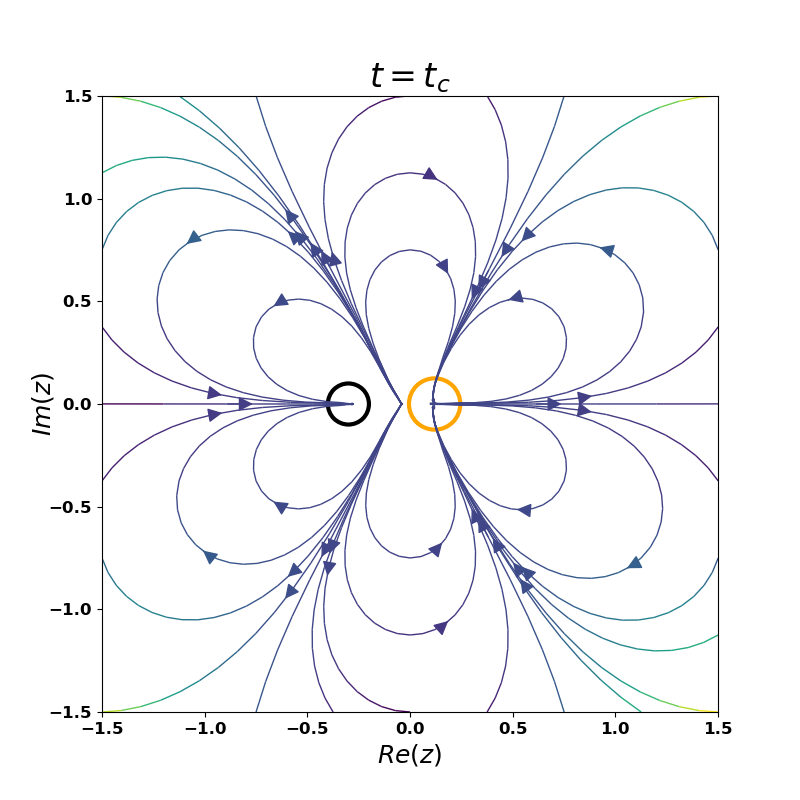},
		\hspace{-1.1cm}
		\includegraphics[scale=.37]{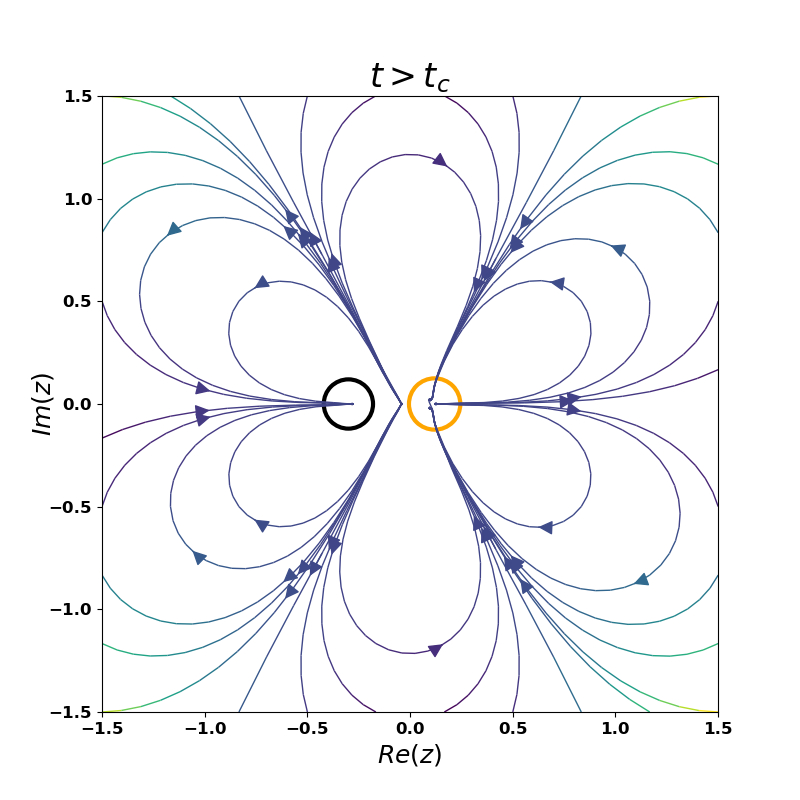}
	\end{tabbing}
	\vspace{-0.7cm}
\end{figure}
\vspace{-1cm}

\newpage

%% in documenet
%\includemovie{10cm}{10cm}{hp.gif}

\subsection{Analytic function and winding numbers in the canonical ensemble}

\begin{figure}[H]
	\vspace{-0.5cm}
	\centering
	\begin{tabbing}
		\centering
		\hspace{2.5cm}%\=\kill
		\includegraphics[scale=.42]{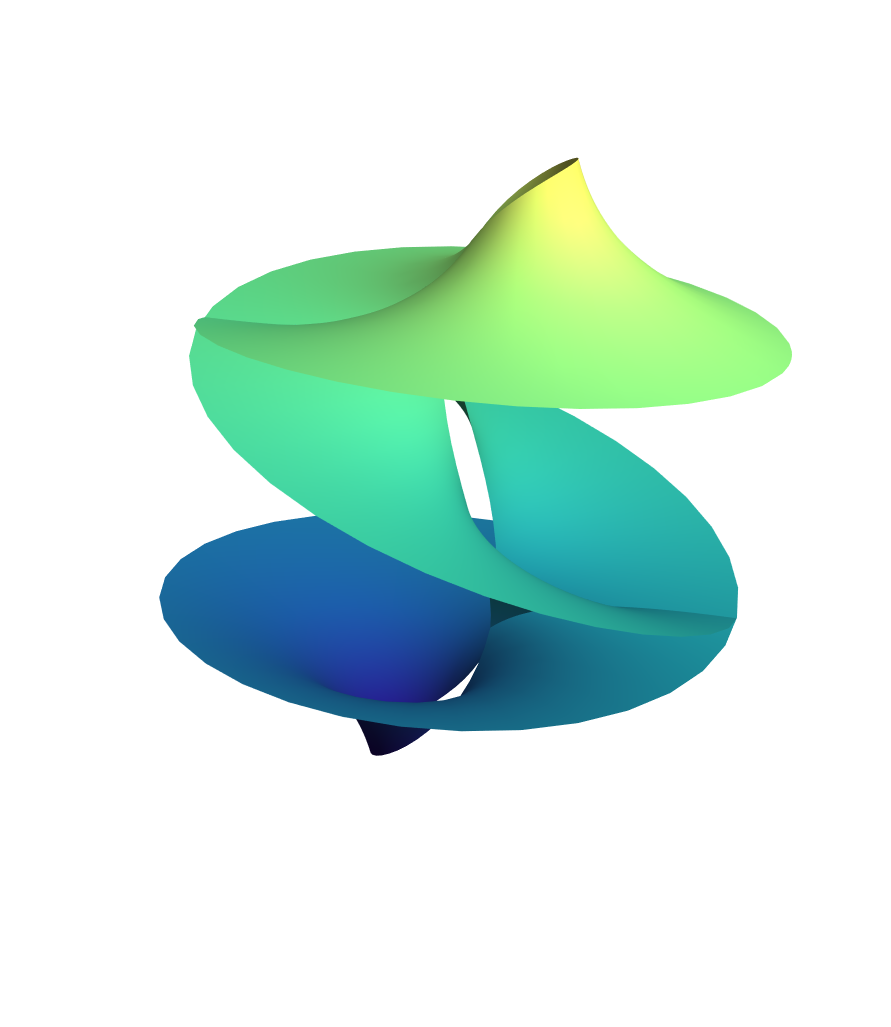}%\>
		%		\hspace{-0.85cm}
		%		\includegraphics[scale=.37]{flow_hp_8},
		%		\hspace{-1.1cm}
		%		\includegraphics[scale=.37]{flow_hp_9}
	\end{tabbing}
	\vspace{-3cm}	
	\caption{\footnotesize Riemann surface associated with the Van-der-Waals phase transition in $d=4$.}\label{fig:fig_Rie_d4_Q}
\end{figure}

\paragraph{}The analytic function of the four dimensional R\'enyi-flat black hole in the canonical ensemble is given by
\begin{equation}\label{analytic_Q_4}
h_Q(z)=\displaystyle  - \frac{Q^{2}}{4 \pi z^{3}}+ \frac{1}{4 \pi z} + \frac{8 z}{3} - t.
\end{equation}
This is a single-valued function and has four zeros $\mathcal{N}=4$, and one pole at the origin $\mathcal{P}=1$, thus its winding number is $\mathcal{W}=4-1=3$, which corresponds exactly to the number of sheets of the Riemann surface of its multi-valued inverse function as illustrated in Fig.\ref{fig:fig_Rie_d4_Q}.
Similar to the grand canonical ensemble, we associate the Riemann surface of the inverse analytic function with the first-order Van der Waals phase transition. We identify the winding number associated with the Van der Waals phase transition, $\mathcal{W}_{VdW}$, with the number of sheets of this Riemann surface, $\mathcal{S}_{VdW}$.
\begin{equation}
   \mathcal{W}_{VdW}=\mathcal{S}_{VdW}.
\end{equation}

This establishes a reference for first-order phase transitions: To occur, a first-order phase transition requires a winding number at least equal to $3$, $\mathcal{W}\geq 3$. For a general four-dimensional asymptotically flat R\'enyi black hole with a given winding number $\mathcal{W}\geq 3$, the maximum number of first-order phase transitions, $N_1$, is determined as,
\begin{equation}\label{N1}
    N_1=\left\lfloor\frac{\mathcal{W}}{3}\right\rfloor,
\end{equation}
where $\lfloor .\rfloor$ denotes the floor function. Moreover, since the winding number $\mathcal{W}_{VdW}$ is greater than $1$, a second-order phase transition is also possible according to Eq.\eqref{N2}. This aligns with the existence of a critical second-order phase transition in Van der Waals systems.

\section{Phase transitions of R\'enyi charged-flat black hole in arbitrary dimensions}

\begin{figure}[H]
	\begin{tabbing}
		\hspace{-2.3cm}
		\centering
		%		\hspace{-2.3cm}%\=\kill
		\includegraphics[scale=.34]{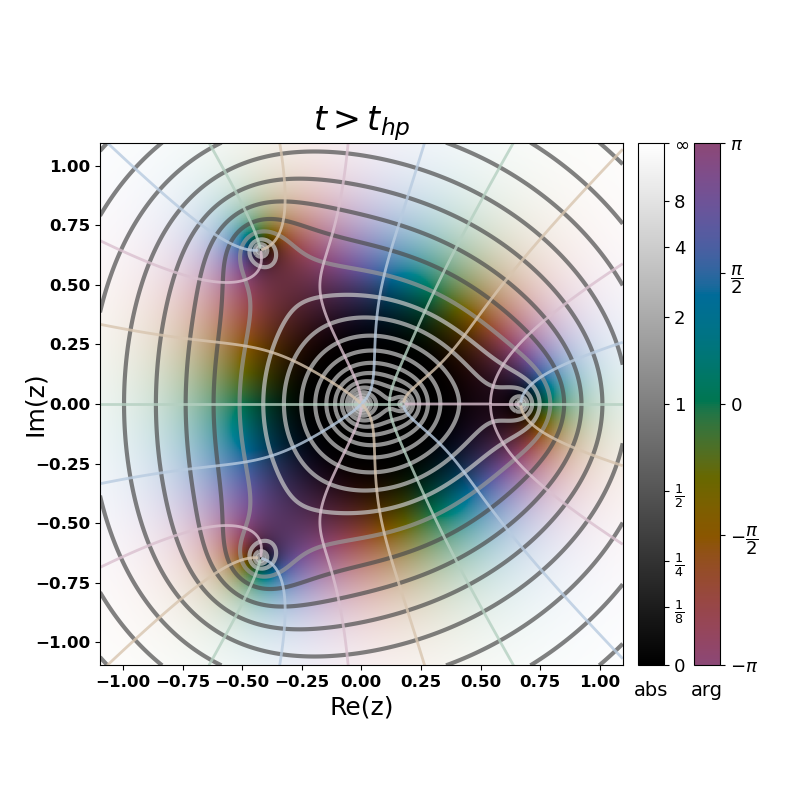}%\>
		\hspace{-0.4cm}
		\includegraphics[scale=.34]{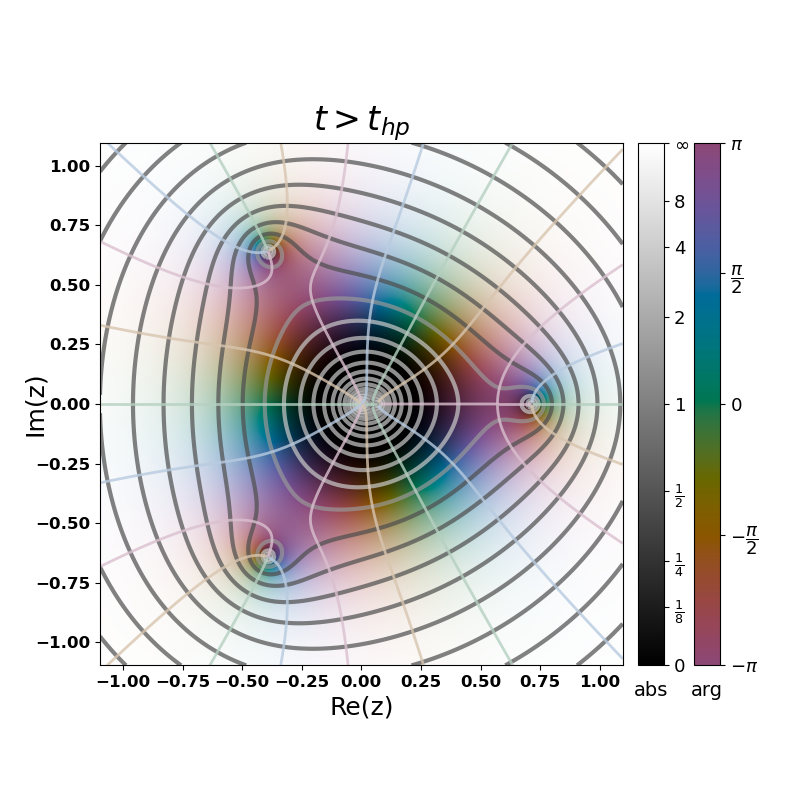}%\\
		\hspace{-0.4cm}
		\includegraphics[scale=.34]{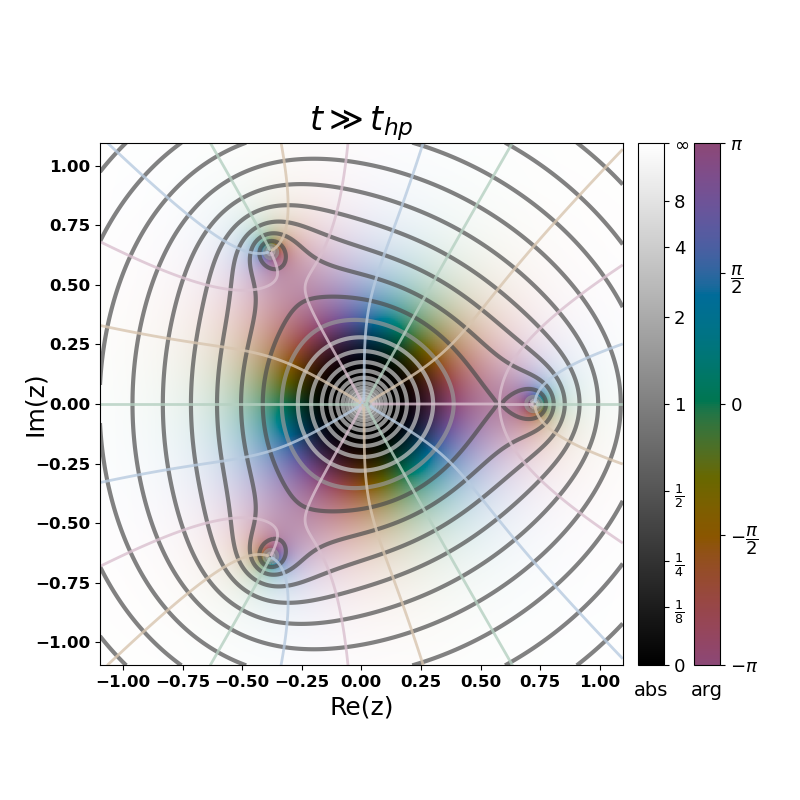}%\\
	\end{tabbing}
	\vspace{-0.7cm}
	\caption{\footnotesize \it R\'enyi-Hawking-Page phase transition in five dimensions spacetimes: The absolute value and phase of the complex Bragg-Williams free energy for the charged-flat black hole as the temperature gradually increases. The electric potential has been set to
  $\phi=0.5$.}		\label{fig:fig_d5_1}
\end{figure}

\begin{figure}[H]
	\begin{tabbing}
		\hspace{-2.3cm}
		\centering
		%		\hspace{-2.3cm}%\=\kill
		\includegraphics[scale=.34]{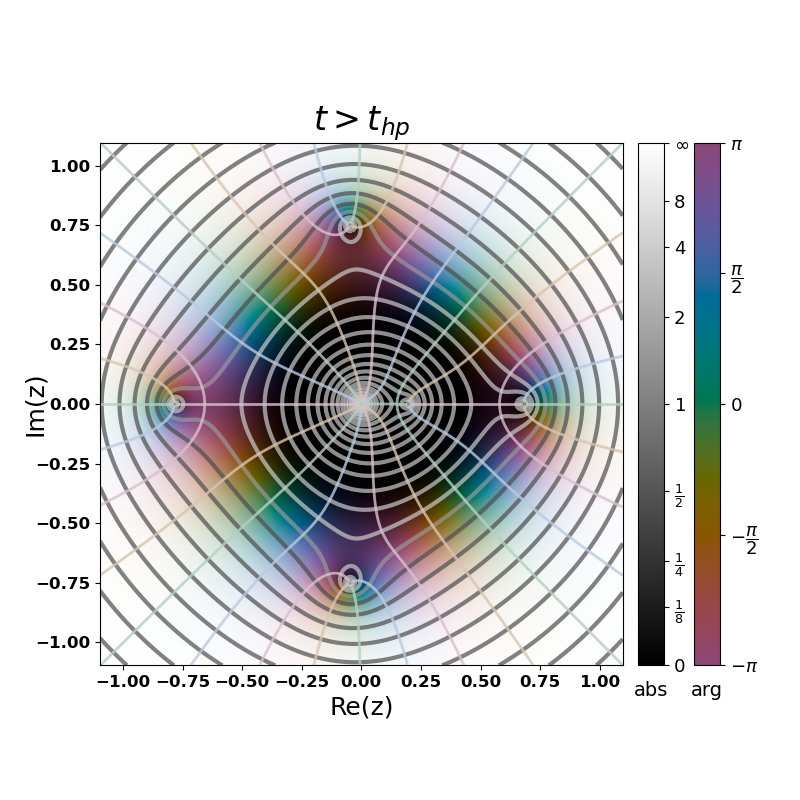}%\>
		\hspace{-0.4cm}
		\includegraphics[scale=.34]{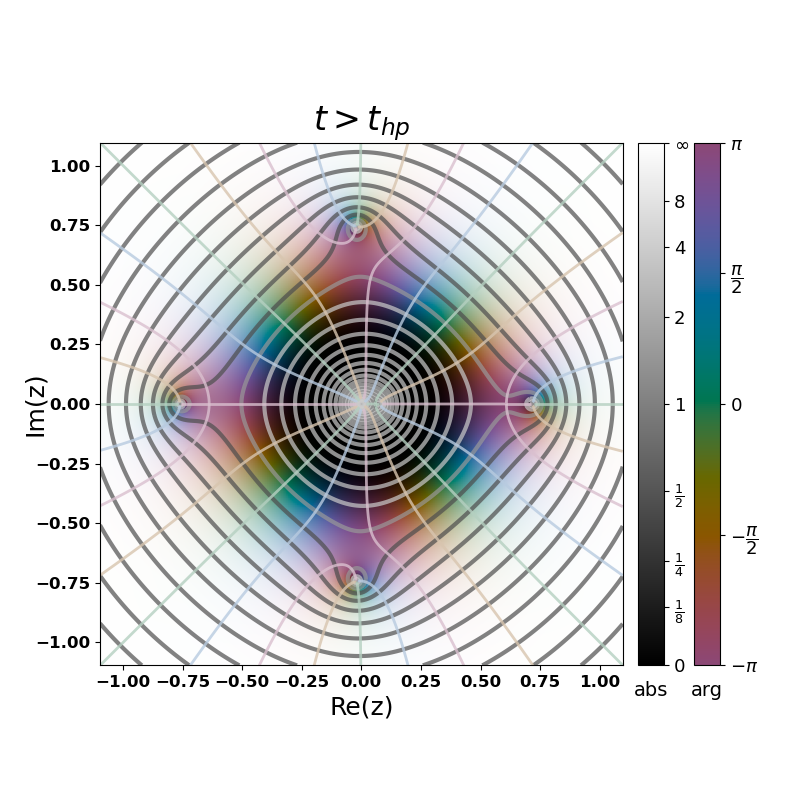}%\\
		\hspace{-0.4cm}
		\includegraphics[scale=.34]{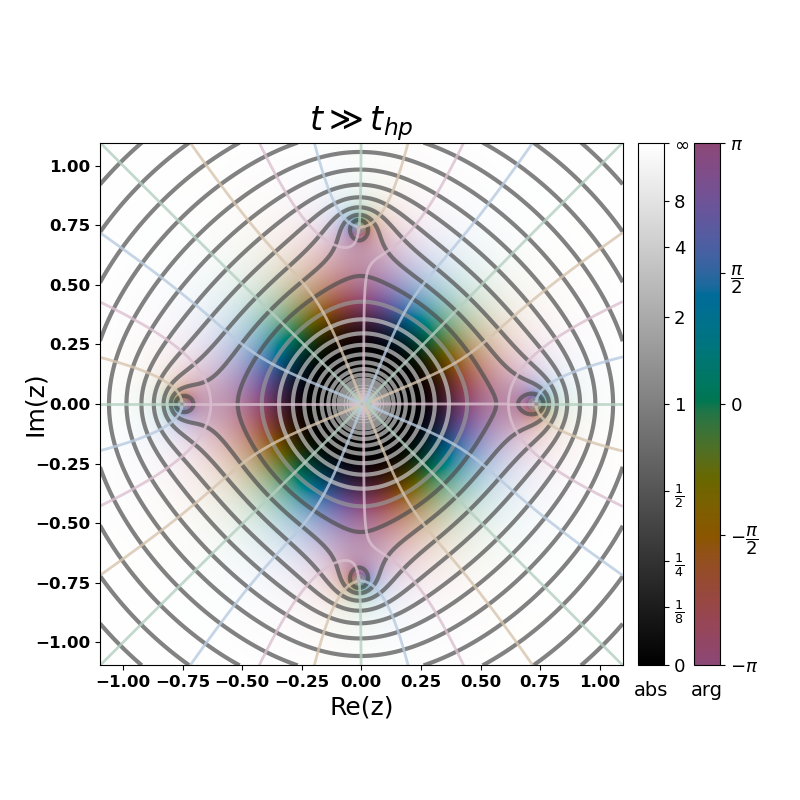}%\\
	\end{tabbing}
	\vspace{-0.7cm}
	\caption{\footnotesize \it R\'enyi-Hawking-Page phase transition in six dimensions spacetimes: The absolute value and phase of the complex Bragg-Williams free energy for the charged-flat black hole as the temperature gradually increases. The electric potential has been set to
  $\phi=0.5$.}		\label{fig:fig_d6_1}
\end{figure}

In this section, we extend our investigation of the Hawking-Page and Van der Waals phase transitions to an arbitrary dimensional R\'enyi charged black hole in asymptotically flat spacetime. A straightforward generalization of the scaled complex Braggs-Williams free energy is given by \cite{Barzi:2022ygr},
\begin{equation}\label{f_R_phi}
f_\phi(z)=\displaystyle \frac{\lambda z^{d - 3} \left[d-2 - 2 \phi^{2} \left(d - 3\right)\right]- \Omega_{d} t \left(d - 2\right) \ln{\displaystyle \left(1+\frac{ 4 \pi \lambda z^{d - 2}}{\Omega_{d} \left(d - 2\right)} \right)}}{\Omega_{d} \lambda \left(d - 2\right)},
\end{equation}
where $\Omega_d$ is by definition,
\begin{equation}\label{omega_d}
\Omega_{d}=\frac{8 \Gamma \left(\frac{d - 1}{2}\right)}{(d - 2) \pi^{\frac{d-3}{2}}},
\end{equation}
 For $0<\lambda<<1$, and the scaling of the free parameters, $z\longrightarrow \lambda^{\frac{1}{2-d}}z$, $t\longrightarrow \lambda^{\frac{1}{d-2}}t$, $\phi\longrightarrow \phi$ and $f_\phi\longrightarrow \lambda^{\frac{1}{2-d}}f_\phi$, we obtain
\begin{equation}\label{f_R_d}
f_\phi(z)=\frac{z^{d-3}\left[8 \pi ^2 t z^{d-1}+(d-2) \Omega_d\left[d-2-4 \pi z t-2 (d-3) \phi^2\right]\right]}{(d-2)^2 \Omega _d^2}.
\end{equation}
In Fig\ref{fig:fig_d5_1} and Fig.\ref{fig:fig_d6_1} we depict the absolute value and phase of the complex Bragg-Williams free energy in 5-and 6-dimensional spacetimes, respectively. The Hawking-Page phase transition results from a continuous process where, similar to the four-dimensional case, two complex conjugate points approach each other and merge on the real line when the temperature reaches $t_{hp}$.
%
%\begin{equation}\label{eq3}
%f_\phi(z)=\frac{\pi ^{\frac{d-3}{2}} z^{d-3} \left(-2 (d-3) \phi ^2+d-2\right)}{8 \Gamma \left(\frac{d-1}{2}\right)}-t \log \left(1+\frac{\pi ^{\frac{d-1}{2}} z^{d-2}}{2 \Gamma \left(\frac{d-1}{2}\right)}\right),
%\end{equation}
%which gives in the limit of small nonextensivity parameter,
%\begin{equation}\label{eq4}
%f_\phi(z)=\displaystyle - \frac{z^{d - 4} }{8 \pi^{3} \Gamma\left(\frac{d}{2} - \frac{1}{2}\right)} \left(
%\begin{split}
%&\pi^{\frac{d}{2} + \frac{3}{2}} d \phi^{2} z  - \pi^{\frac{d}{2} + \frac{3}{2}} d z  - 6 \pi^{\frac{d}{2} + \frac{3}{2}} \phi^{2} z  \\&+  2 \pi^{\frac{d}{2} + \frac{3}{2}} z  + 4 \pi^{\frac{d}{2} + \frac{5}{2}} t z^{2}  - \frac{\pi^{d + 2} t z^{d}}{\Gamma\left(\frac{d}{2} - \frac{1}{2}\right)}
%\end{split}
%\right)
%\end{equation}

\newpage

\begin{figure}[H]
	
	\vspace{-3cm}
	\centering
	\begin{tabbing}
		\centering
		\hspace{-2cm}%\=\kill
		\includegraphics[scale=.34]{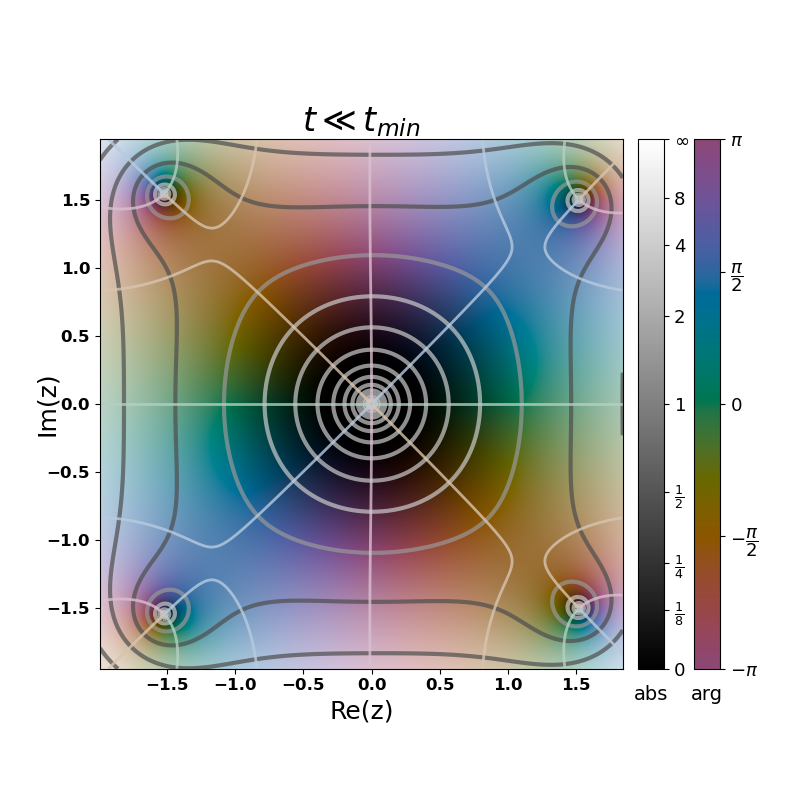}%\>
		\hspace{-0.4cm}
		\includegraphics[scale=.345]{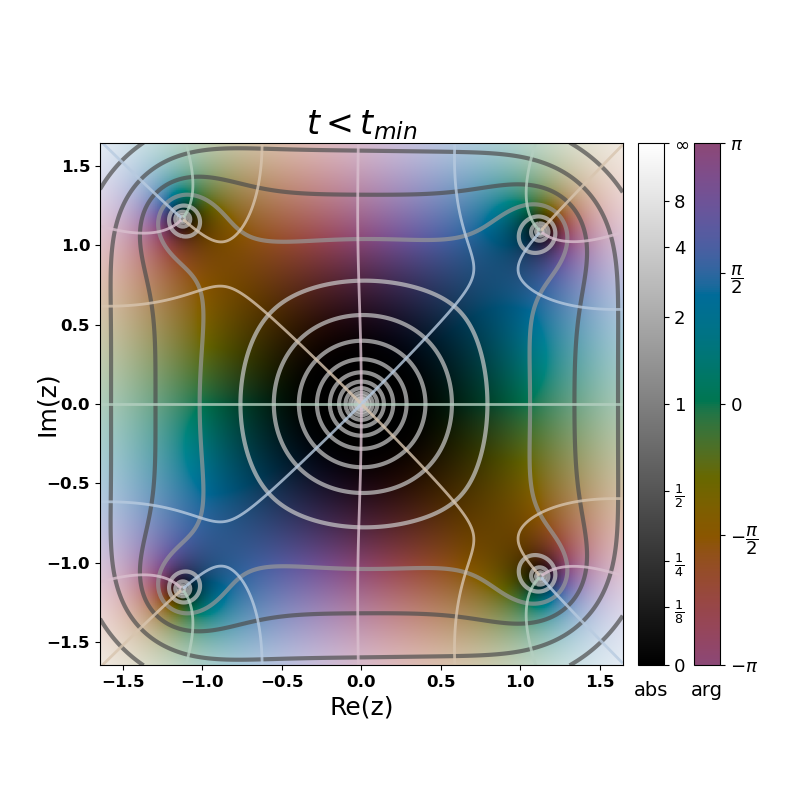}%\\
		\hspace{-0.4cm}
		\includegraphics[scale=.34]{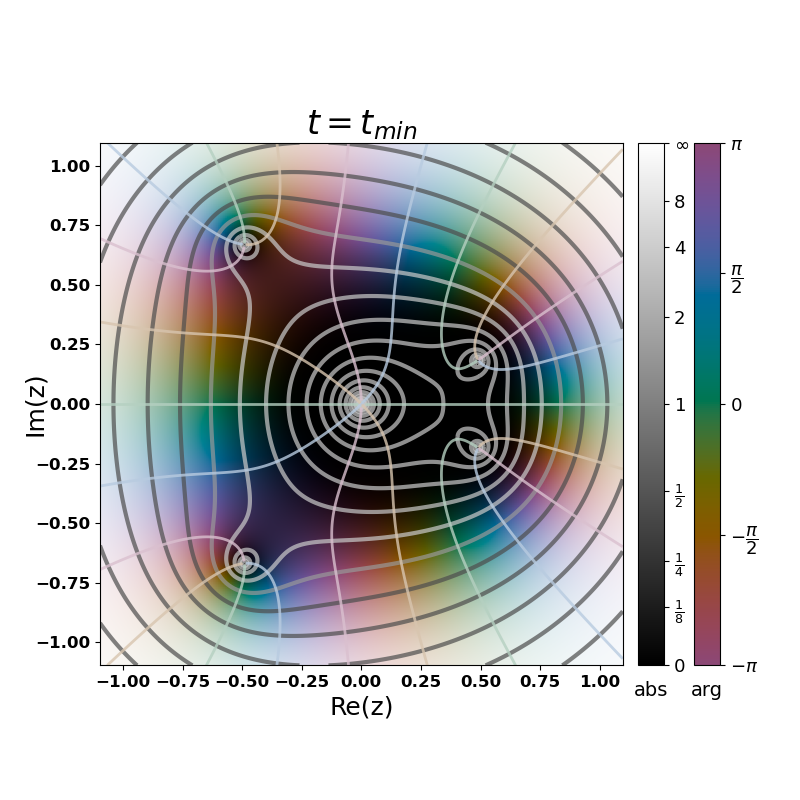}
	\end{tabbing}
	\vspace{-0.7cm}
	
\end{figure}
\vspace{-1.5cm}

\begin{figure}[H]
	\begin{tabbing}
		\hspace{-2.3cm}
		\centering
		%		\hspace{-2.3cm}%\=\kill
		\includegraphics[scale=.34]{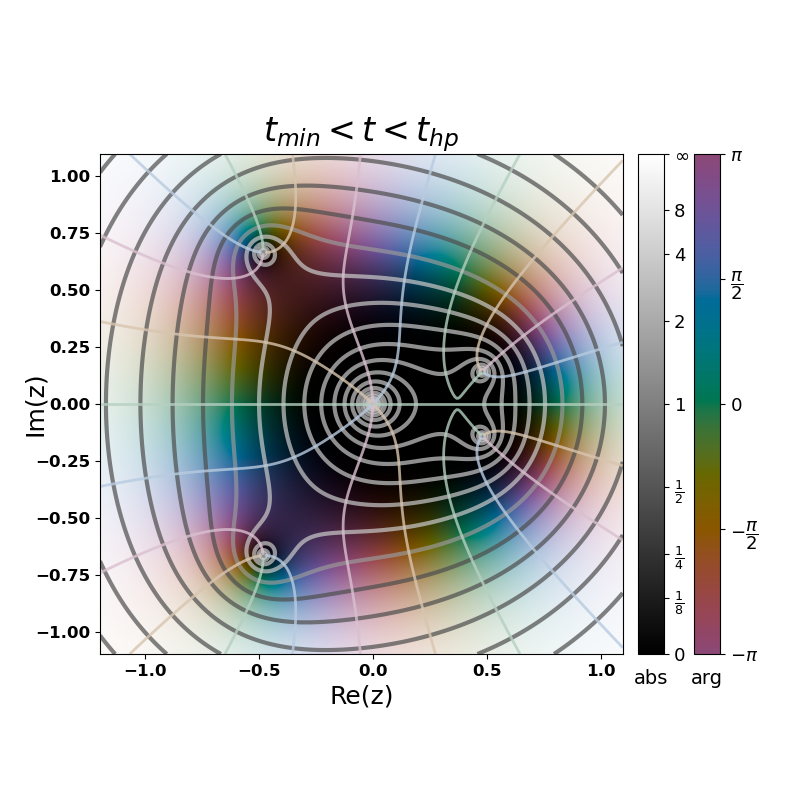}%\>
		\hspace{-0.4cm}
		\includegraphics[scale=.34]{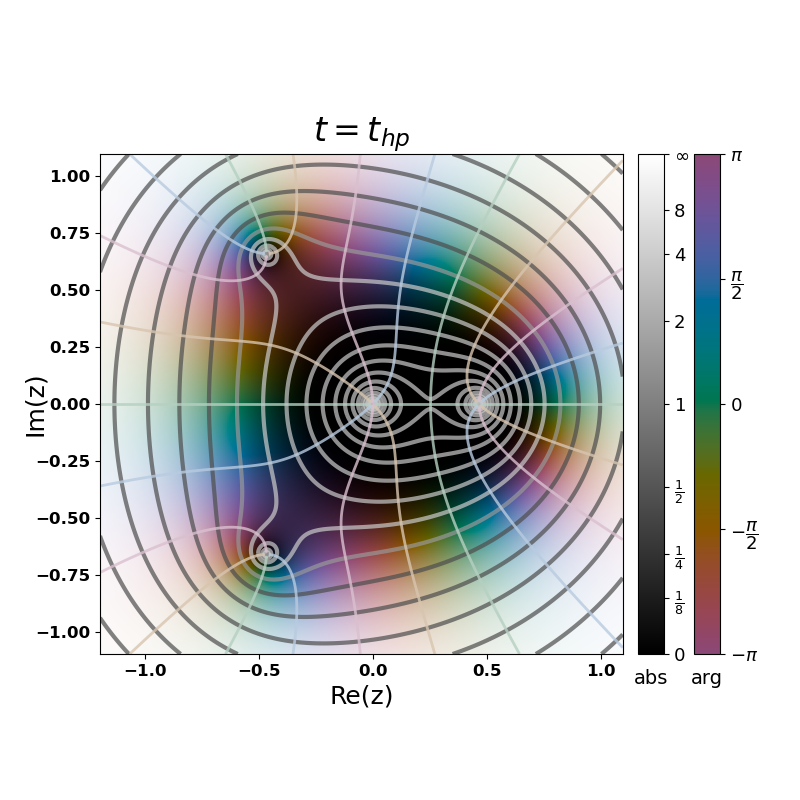}%\\
		\hspace{-0.4cm}
		\includegraphics[scale=.34]{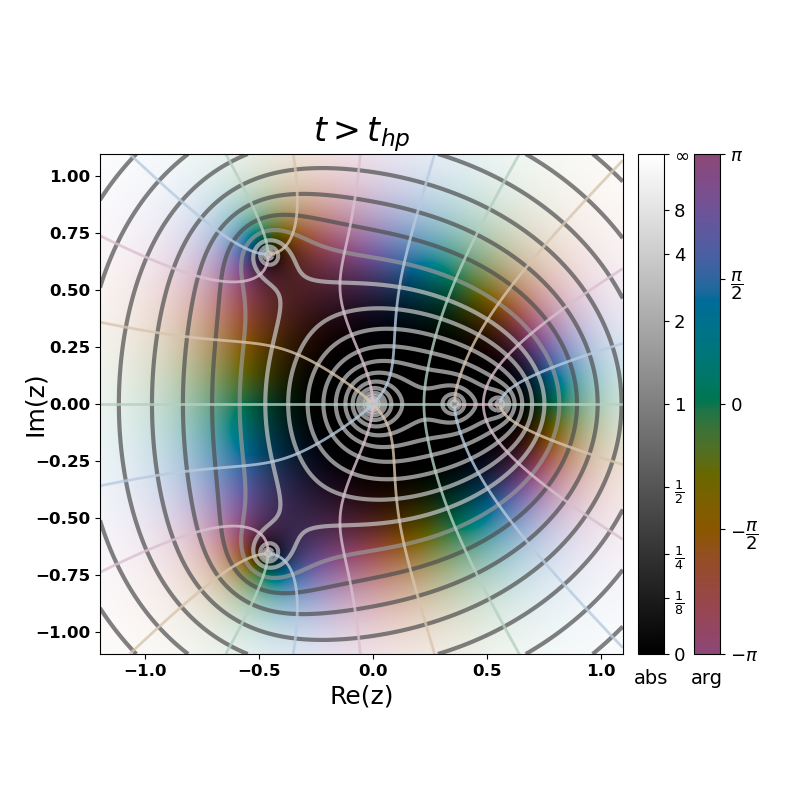}%\\
	\end{tabbing}
	\vspace{-0.7cm}
\end{figure}

\vspace{-1.5cm}

\newpage

\begin{figure}[H]
	
	\vspace{-3cm}
	\centering
	\begin{tabbing}
		\centering
		\hspace{-2.3cm}%\=\kill
		\includegraphics[scale=.34]{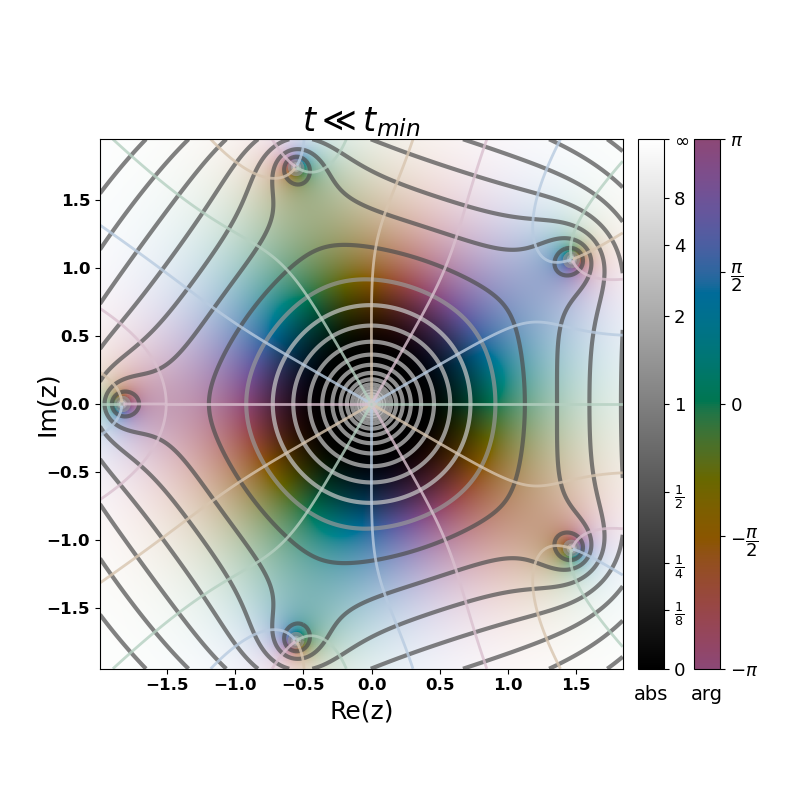}%\>
		\hspace{-0.4cm}
		\includegraphics[scale=.345]{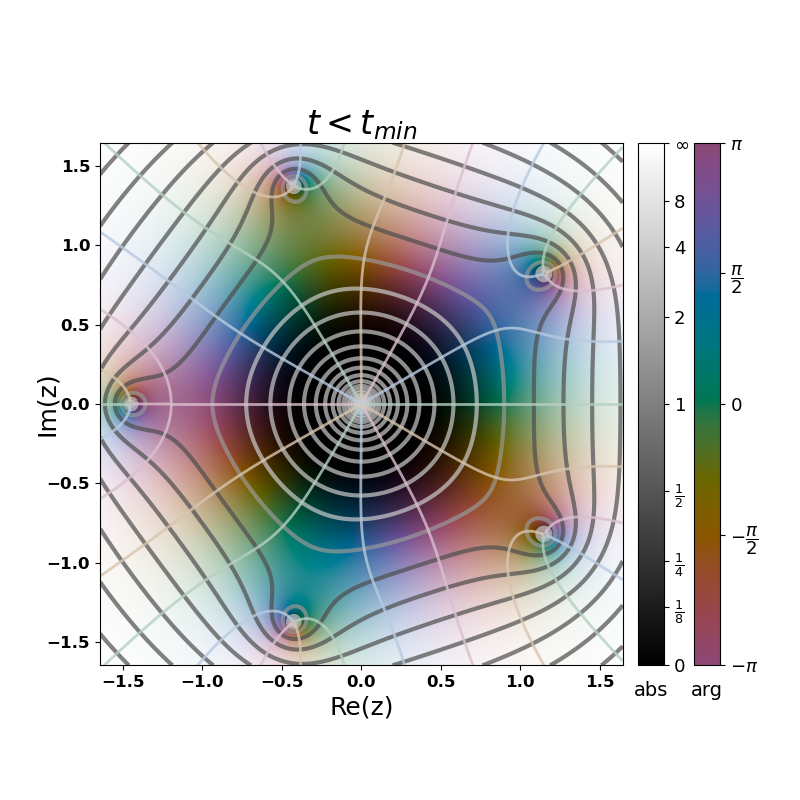}%\\
		\hspace{-0.4cm}
		\includegraphics[scale=.34]{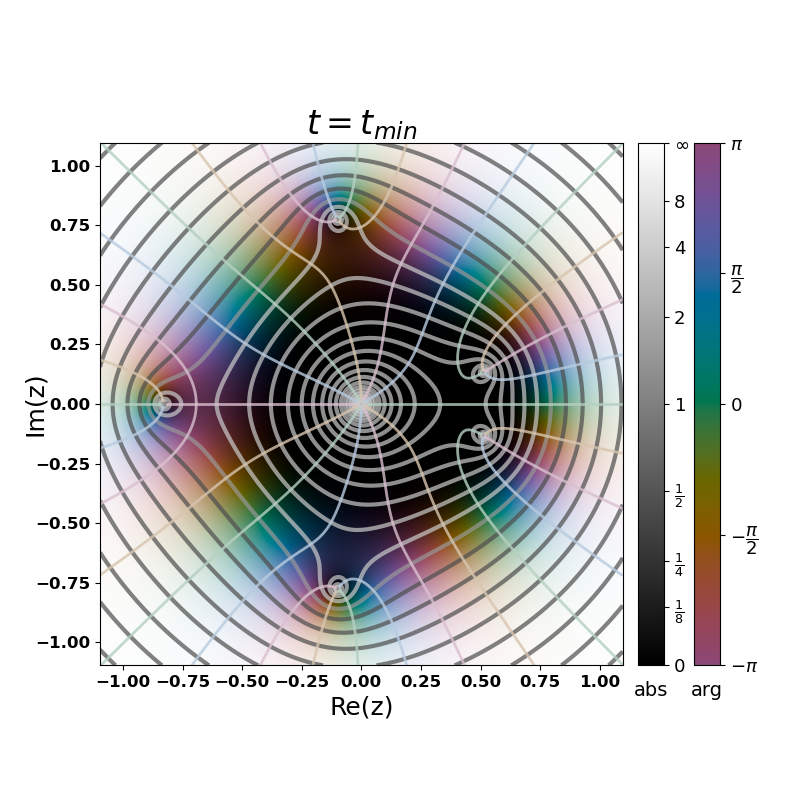}
	\end{tabbing}
	\vspace{-0.7cm}
	
\end{figure}
\vspace{-1.5cm}

\begin{figure}[H]
	\begin{tabbing}
		\hspace{-2.3cm}
		\centering
		%		\hspace{-2.3cm}%\=\kill
		\includegraphics[scale=.34]{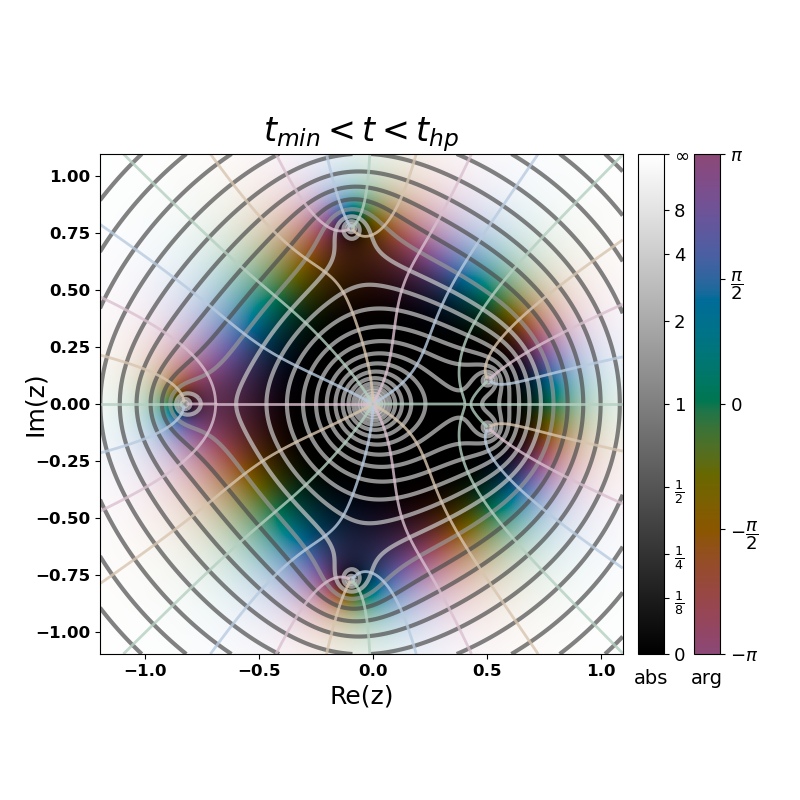}%\>
		\hspace{-0.4cm}
		\includegraphics[scale=.34]{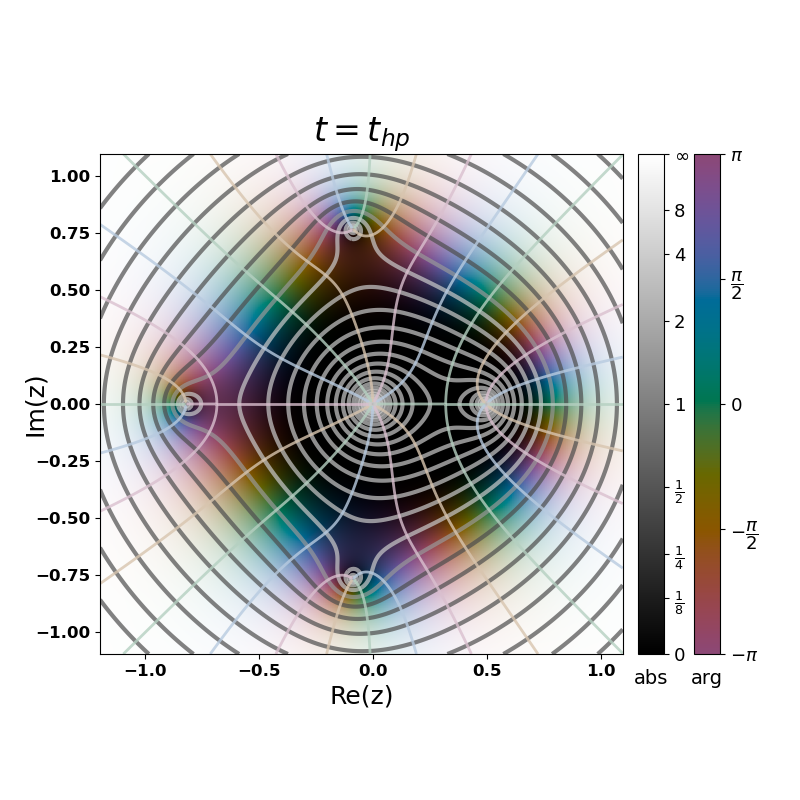}%\\
		\hspace{-0.4cm}
		\includegraphics[scale=.34]{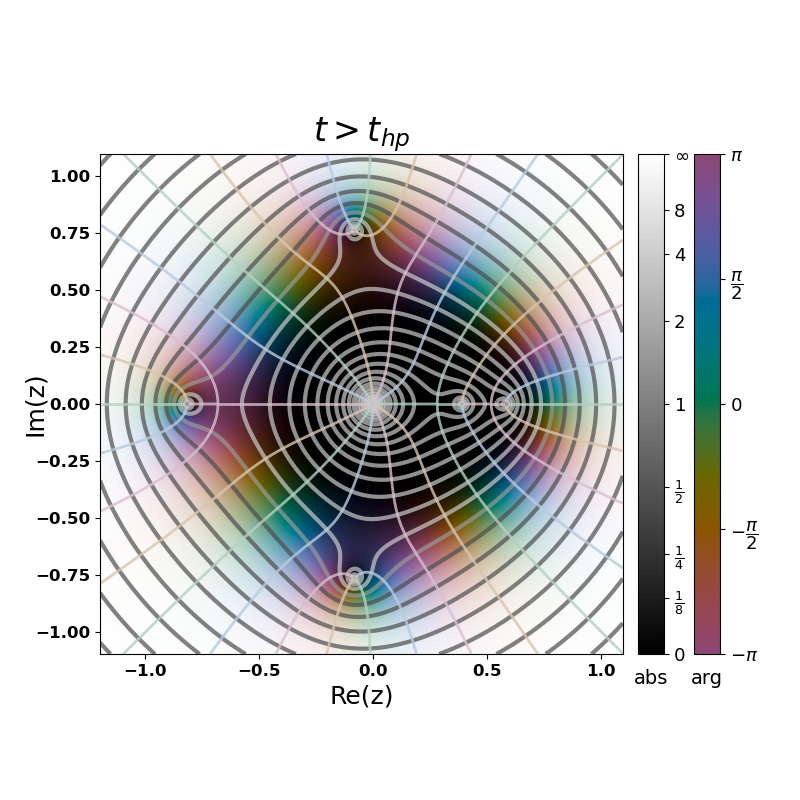}%\\
	\end{tabbing}
	\vspace{-0.7cm}
\end{figure}

\vspace{-1.5cm}

%*******Here*****

\begin{figure}[!htb]
        \vspace{-0.5cm}
		\centering
		\begin{tabbing}
		  \hspace{-.4cm}
		\includegraphics[scale=0.4]{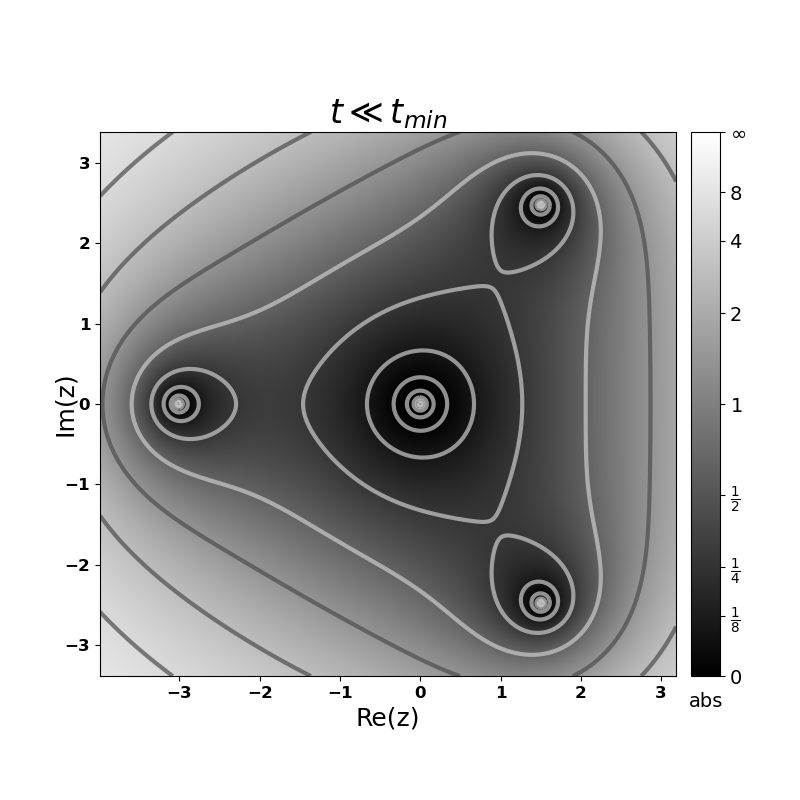}
		\hspace{0.6cm}
		\includegraphics[scale=0.4]{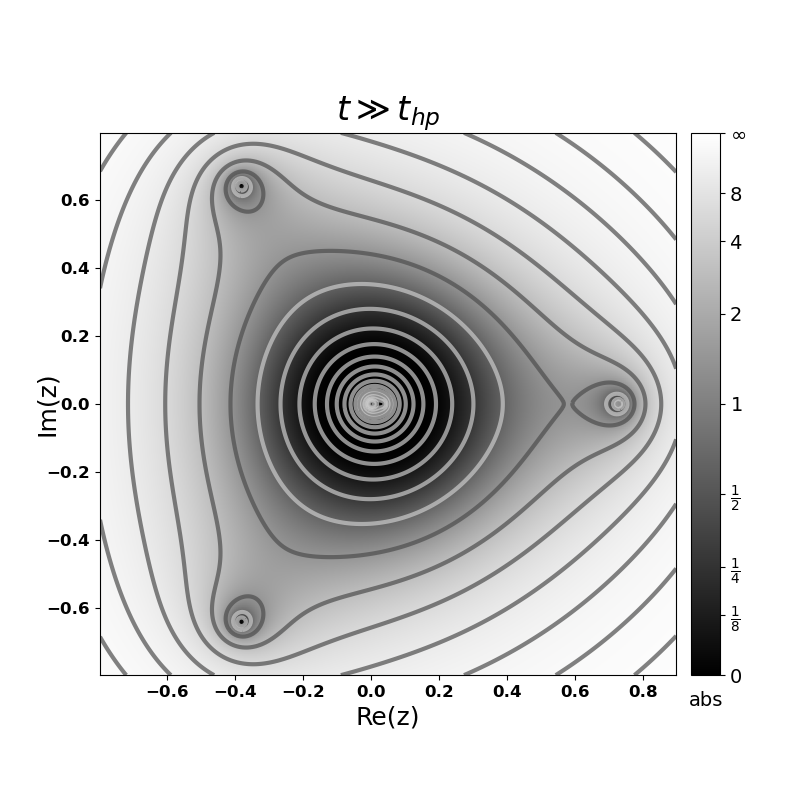}
		\end{tabbing}
		\vspace{-1.2cm}
		\caption{\footnotesize Existence of a holographic correspondence manifested by the grand canonical complex Braggs-Williams free energy in 4-dimensional at $t<<t_{min}$ (\textbf{Left}) and 5-dimensional $t>>t_{hp}$ (\textbf{Right}), R\'enyi charged-flat black holes. }
		\label{fig:holography_hp}
	\end{figure}

\begin{figure}[!ht]
	% \vspace{-0.5cm}
	\centering
	\begin{tabbing}
		\centering
		\hspace{-1.6cm}%\=\kill
		\includegraphics[scale=.3]{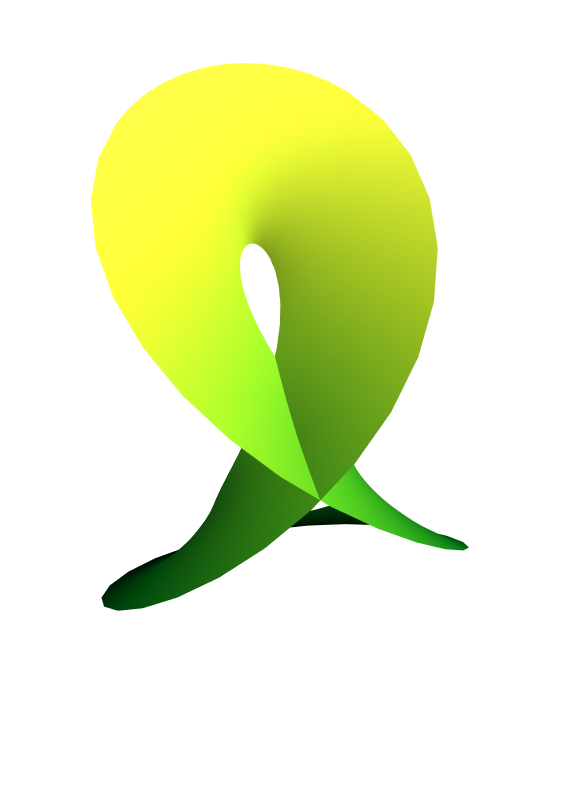}%\>
		\hspace{-1.1cm}
		\includegraphics[scale=.3]{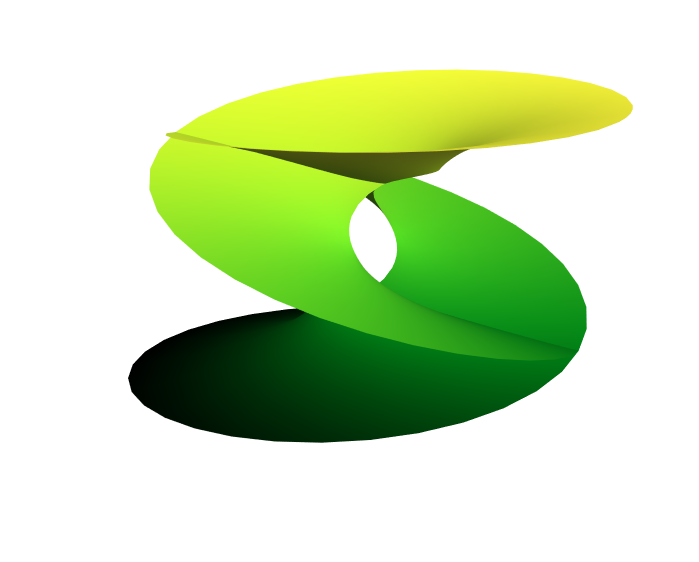},
		\hspace{-0.9cm}
		\includegraphics[scale=.3]{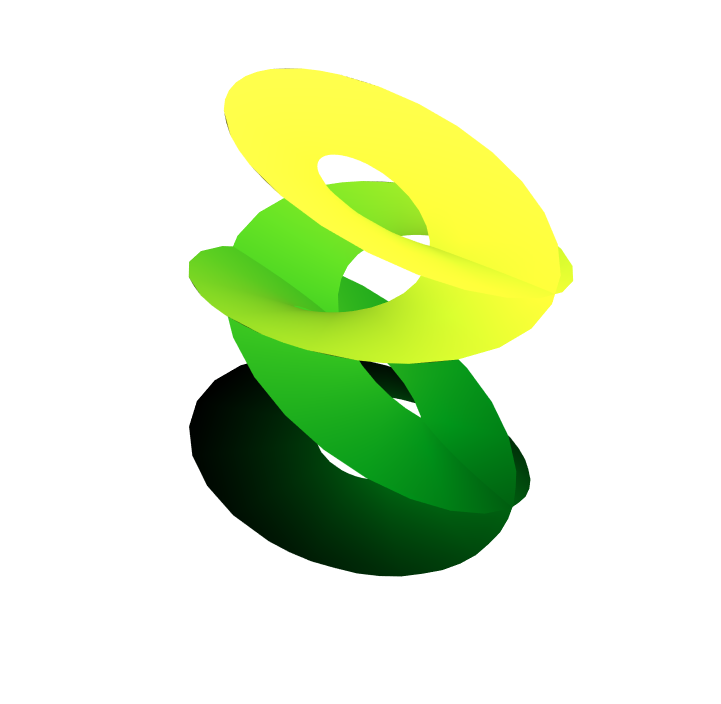}\hspace{-1.2cm}
		\includegraphics[scale=.3]{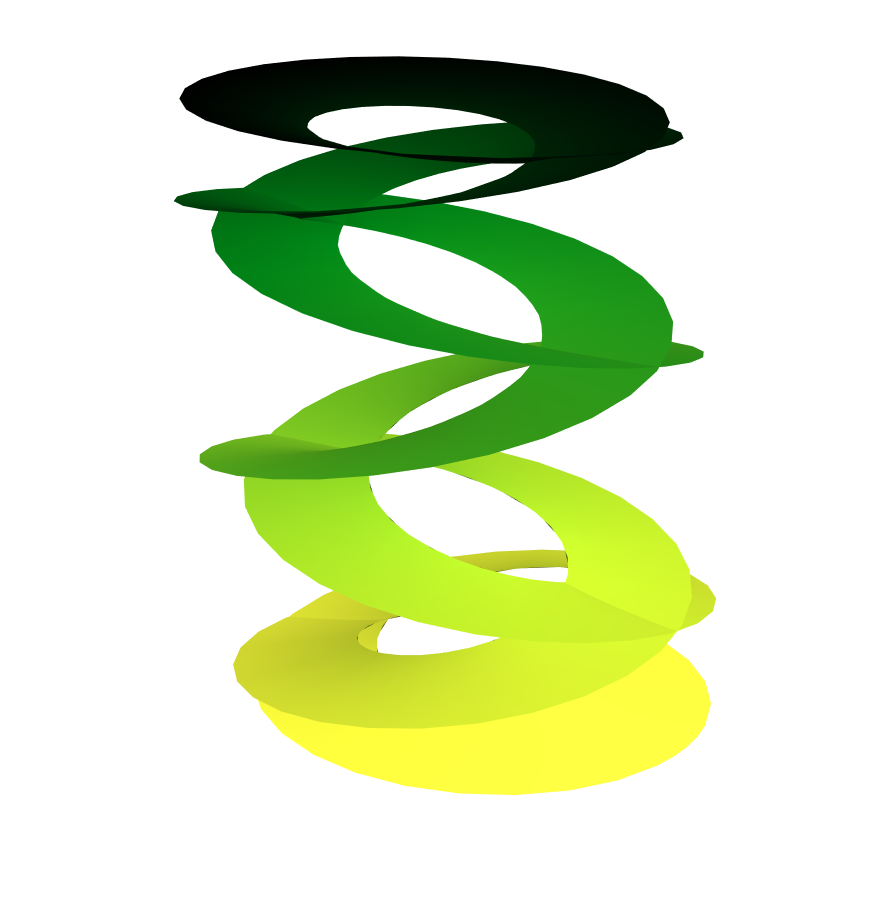}
		
	\end{tabbing}
	\vspace{-0.7cm}
	\caption{\footnotesize Riemann surfaces associated with the Hawking-Page phase transitions of $d=5,\;6,\;7$ and $10$  dimensional R\'enyi charged-flat black hole. The winding numbers are $2$, $3$, $4$ and $7$, respectively.}
	\label{fig:fig_Rie_higher_hp}
\end{figure}

A surprising feature revealed by Figs.\ref{fig:fig_d4_1}, \ref{fig:fig_d5_1} and \ref{fig:fig_d6_1} is the remarkable similarity between the behavior of the complex Rényi Braggs-Williams free energy far below $t \ll t_{min}$ in $d$-dimensional spacetime and its behavior far above $t \gg t_{hp}$ in $(d+1)$-dimensional spacetime. Closer investigation shows that a scaling transformation combined with a phase rotation precisely maps one behavior to the other. This connection suggests an intriguing holographic-like relationship, where the intermediate charged Rényi black hole state far below the scaled minimum temperature $t_{min}(d)$ in $d$-dimensional spacetime exhibits the same behavior as the stable Rényi black hole state far above the scaled Hawking-Page temperature $t_{hp}(d+1)$ in $(d+1)$-dimensional spacetime. This relationship can be expressed through the following transformation
\begin{equation}\label{key}
f_\phi(z,d)_{t\ll t_{min}}=\displaystyle \alpha f_\phi(\beta z,d+1)_{t\gg t_{hp}},
\end{equation}
where $\alpha$ and $\beta$ are complex numbers. Fig.\ref{fig:holography_hp}  illustrates this holographic-like relationship between the complex Bragg-Williams free energy in four- and five-dimensional spacetimes. The triangular complex structure in four dimensions at $t \ll t_{min}$ is a rotated and distorted image of its triangular counterpart in five dimensions at $t \gg t_{hp}$. From a topological point of view, these structures are identical because they differ only by a continuous transformation. This intriguing relationship requires further investigation.
\paragraph{}Continuing our analysis, we express the analytic function for the $d$-dimensional Rényi charged-flat black hole in the grand canonical ensemble as follows
%\begin{equation}\label{key}
%h_\phi(z)=\frac{\pi ^{-\frac{d}{2}-1} \pi ^{\frac{d-3}{2}}(d-3) \left(-2 \phi ^2(d-3)+d-2\right) \left(\pi ^{d/2} \pi \lambda  z^d+2 \pi ^{3/2} z^2 \Gamma \left(\frac{d-1}{2}\right)\right)}{8 z^2 z(d-2) \Gamma \left(\frac{d-1}{2}\right)}-t
%\end{equation}
\begin{equation}
h_\phi(z)=\frac{\left(d - 3\right) \left[\Omega_{d} \left(d - 2\right) + 4 \pi \lambda z^{d - 2}\right] \left[2- d + 2 \phi^{2} \left(d - 3\right)\right]-4 \pi \Omega_{d}\left(d - 2\right) t z}{4 \pi \Omega_{d} \left(d - 2\right)z }.
\end{equation}
 Fig.\ref{fig:fig_Rie_higher_hp} displays the Riemann surfaces associated with the Hawking-Page phase transitions of the Rényi-flat black hole in higher dimensions. The winding number $\mathcal{W}_{HP}$ corresponds to the number of sheets comprising the Riemann surface. Accordingly, we obtain
\begin{equation}\label{winding_num}
\mathcal{W}_{HP}=d-3.
\end{equation}
%\begin{figure}[H]
%	\centering
%	\begin{tabbing}
%		\centering
%		\hspace{3cm}%\=\kill
%		\includegraphics[scale=.3]{Rie-d10-hp}%\>
%%		\hspace{-0.85cm}
%%		\includegraphics[scale=.42]{Rie-d6-hp},
%%		\hspace{-1.1cm}
%%		\includegraphics[scale=.42]{Rie-d7-hp}
%	\end{tabbing}
%	\vspace{-0.7cm}	
%	\caption{}\label{fig:fig_higher_hp}
%\end{figure}

\paragraph{} A similar analysis in $d$-dimensional asymptotically Anti-de Sitter (AdS) spacetime yields the same winding number. Specifically, for the charged-AdS black hole in four dimensions, the Hawking-Page transition exhibits a winding number of $\mathcal{W}_{HP}=1$. This result provides solid consolidation to the potential and significant connection between the nonextensivity Rényi parameter $\lambda$ and the cosmological constant $\Lambda$ observed previously in \cite{Promsiri:2020jga,Promsiri:2021hhv,Barzi:2023mit,Barzi:2023msl}. 
\paragraph{}In the canonical ensemble, we deal with Van der Waals phase transitions. To this end, we re-express the nonextensivity parameter $\lambda$ as a function of the Rényi pressure $p$\cite{Barzi:2022ygr} as follows
\begin{equation}
\lambda=\frac{16(d-2)\Omega_{d}\:r_{h}^{d - 2} \:p}{(d-1)(d-3)\left[\displaystyle (d-2)r_h^{2(d-3)}-2(d-3) Q^{2}\right]}.
\end{equation}
The complex Braggs-Williams free energy of  the $d$-dimensional R\'enyi-flat black hole in the canonical ensemble reads as
\begin{equation}\label{key}
\begin{split}
f_Q(z)&=\frac{z^{d-4}\Delta_d}{128 p(d-3) \Omega _d}\times\\&\left(\begin{split}&16 p z
-(d-1)(d-3)t\\\\& \times\left[\begin{split}&\log \left[\frac{(d-2)(d-1)}{\Delta_d}\right]+\\\\&\log \left[8\left(d-3\right)\left(d-2\right)\left(d-1\right)+\left(d-1\right)\left[\Delta_d -8 (d-3)\right]+512 \pi p z^2\right]\end{split}\right]
\end{split}\right).
\end{split}
\end{equation}
Where $\Delta_d=8 (d-3)-z^{6-2 d}(d-2)\Omega_d^2Q^2$. As before we take the limit of small $p$ and perform the scaling of the free parameters, $z\longrightarrow p^{\frac{1}{2-d}}z$, $t\longrightarrow p^{\frac{1}{d-2}}t$, $Q\longrightarrow p^{\frac{1}{2-d}}Q$ and $f_Q\longrightarrow p^{\frac{1}{2-d}}f_Q$. 
 The analytic function of the charged R\'enyi-flat black hole in the canonical ensemble has the form
\begin{equation}\label{key}
h_Q(z)=\displaystyle \frac{ 512 \pi p z}{32 \pi  \left(d - 2\right)(d-1) }  -  \frac{ \left[\Omega_{d}^{2} Q^{2} z^{6 - 2 d}(d-2) - 8 (d-3)\right]}{32 \pi z}-t.
\end{equation}

\begin{figure}[H]
	\begin{tabbing}
		\hspace{-2cm}
		\centering
		%		\hspace{-2.3cm}%\=\kill
		\includegraphics[scale=.33]{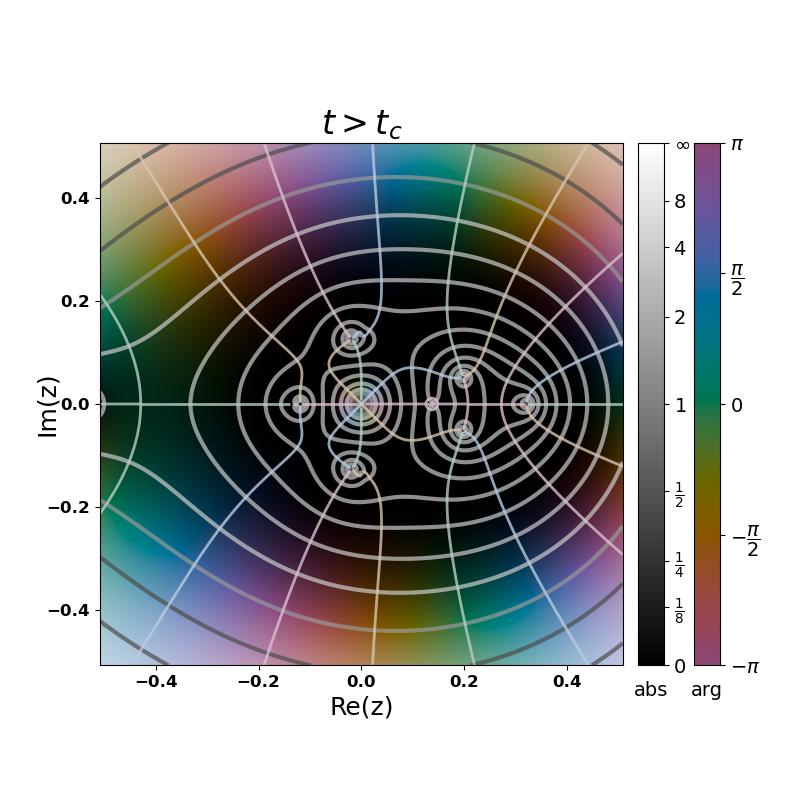}%\>
		\hspace{-0.4cm}
		\includegraphics[scale=.33]{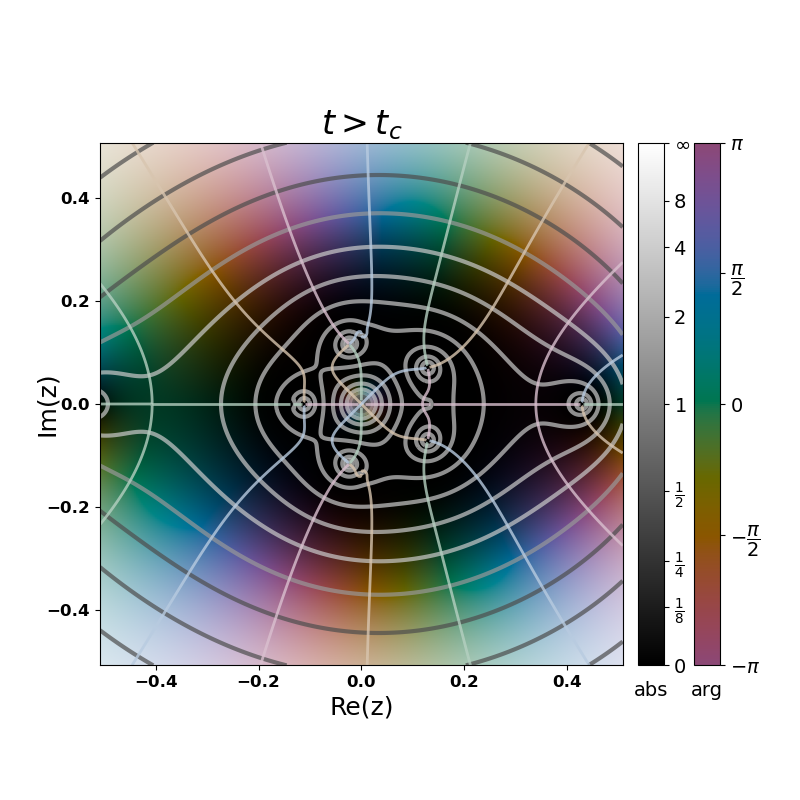}%\\
		\hspace{-0.4cm}
		\includegraphics[scale=.33]{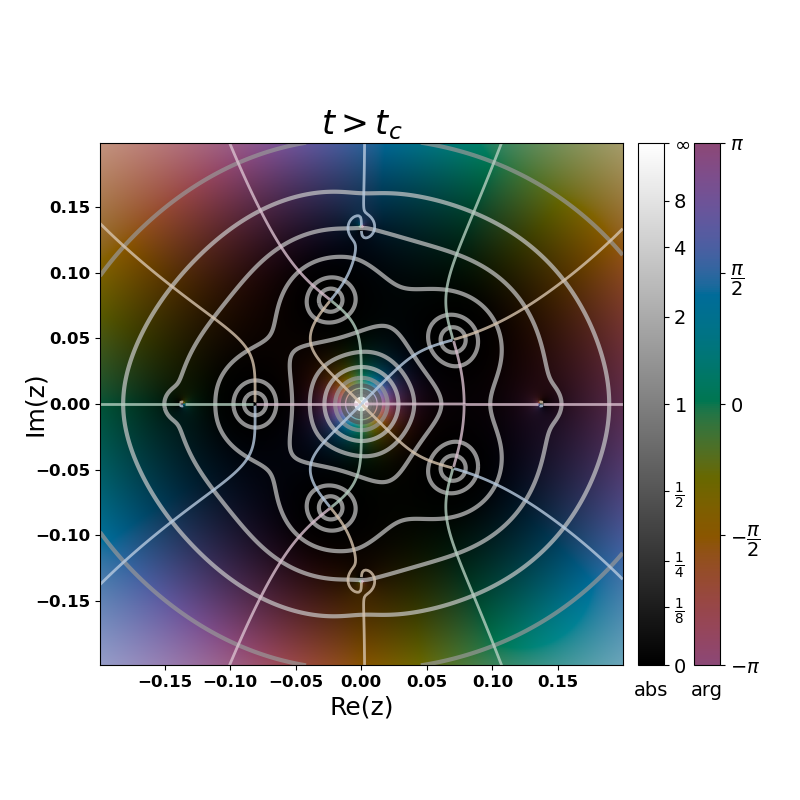}%\\
	\end{tabbing}
	\vspace{-1.2cm}	
	\caption{\footnotesize Van-der-Waals complex structure associated with the five-dimensional charged R\'enyi-flat black hole. The electric charge has been set to $Q=0.05$.}
	\label{fig:fig_d5_Q}
\end{figure}

Portrayed in Fig.\ref{fig:fig_d5_Q} are the absolute value and phase of the complex Bragg-Williams free energy in five dimensions. The Van-der-Waals phase transition is likewise the product of a continuous process where two complex conjugate points coalesce on the real line at the critical temperature $t_{c}$.

%*****Here******

\begin{figure}[!ht]
        \vspace{-0.45cm}
		\centering
		\begin{tabbing}
		  \hspace{-.4cm}
		\includegraphics[scale=0.4]{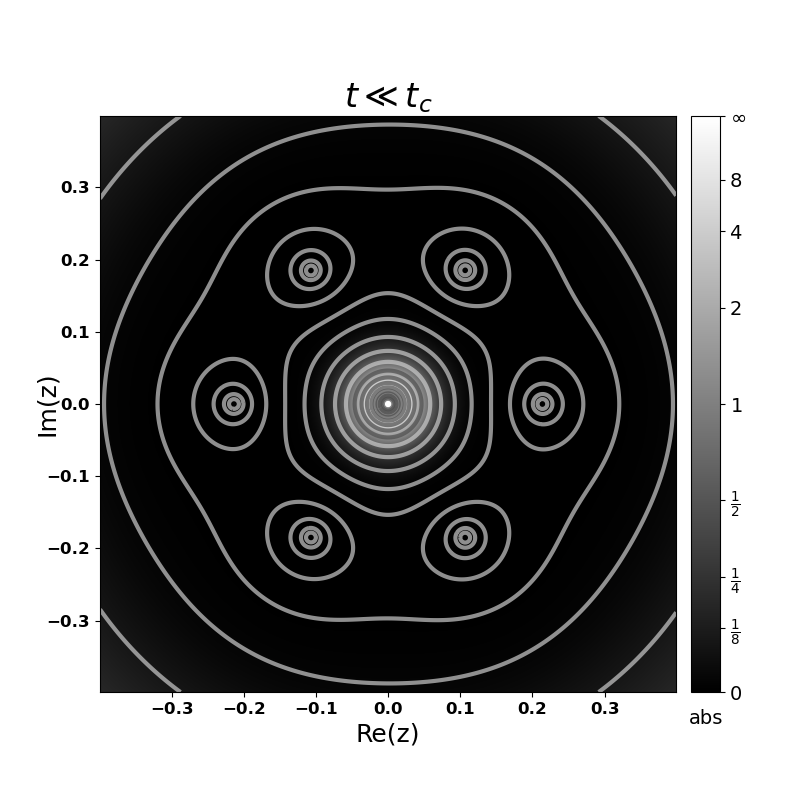}
		\hspace{0.6cm}
		\includegraphics[scale=0.4]{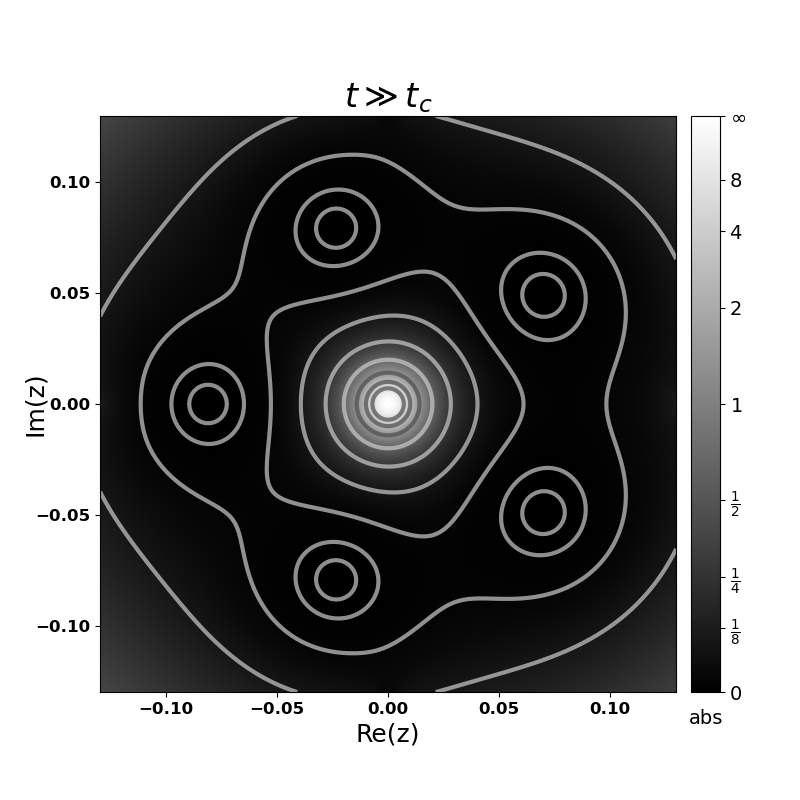}
		\end{tabbing}
		\vspace{-1.2cm}
		\caption{\footnotesize Illustration of the absence of holographic correspondence in the canonical complex Braggs-Williams free energy in 6-dimensional at $t<<t_c$ (\textbf{Left}) and 5-dimensional  at $t>>t_c$ (\textbf{Right}) R\'enyi charged-flat black holes. }
		\label{fig:holography_vdw}
	\end{figure}

\paragraph{}A close inspection of Fig.\ref{fig:fig_d4_4} and Fig.\ref{fig:fig_d5_Q} reveals an absence, as manifested in the Hawking-Page case, of a direct holographic correspondence between the fluctuating black hole state at $t<<t_c$ in four-dimensional spacetime and the stable black hole state in five-dimensional spacetime at $t>>t_c$. Notably, in the vicinity of the origin, $z=0$, the patterns are different as depicted in Fig.\ref{fig:holography_vdw}, where we plotted the absolute value of the complex free energy in five and six dimensions. This suggests that a fundamental difference between Van-der-Waals phase transitions in different dimensions.
\newpage
%\begin{landscape}
\begin{figure}[H]
	
	\vspace{-3cm}
	\centering
	\begin{tabbing}
		\centering
		\hspace{-2cm}%\=\kill
		\includegraphics[scale=.33]{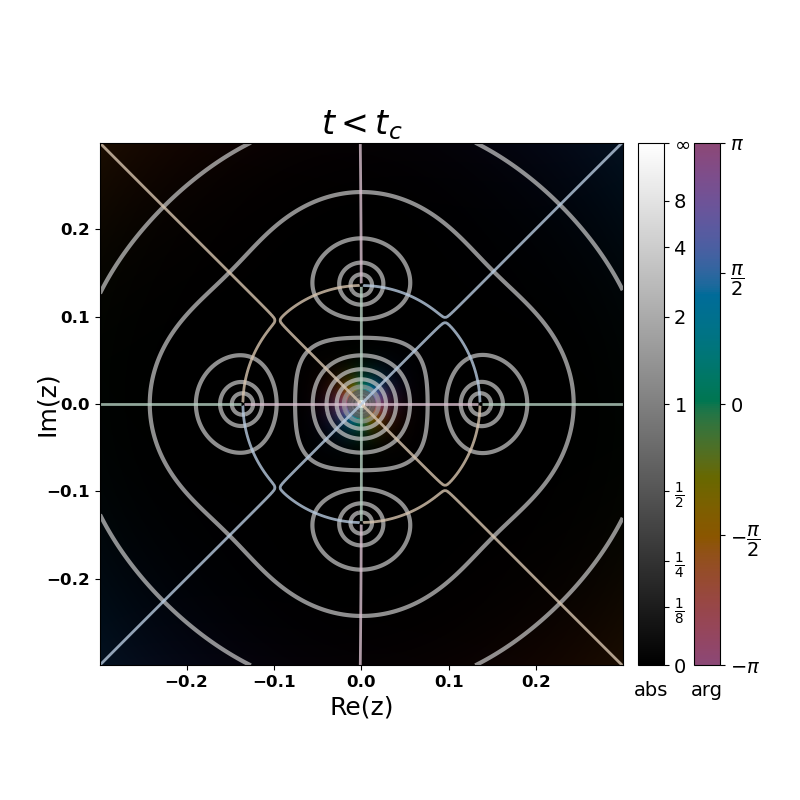}%\>
		\hspace{-0.4cm}
		\includegraphics[scale=.33]{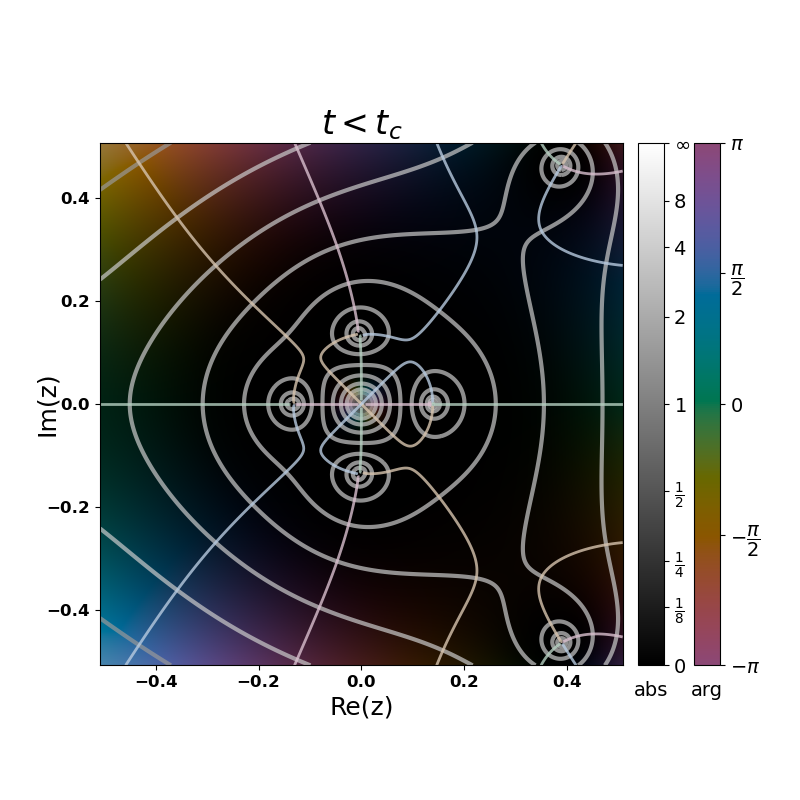}%\\
		\hspace{-0.4cm}
		\includegraphics[scale=.33]{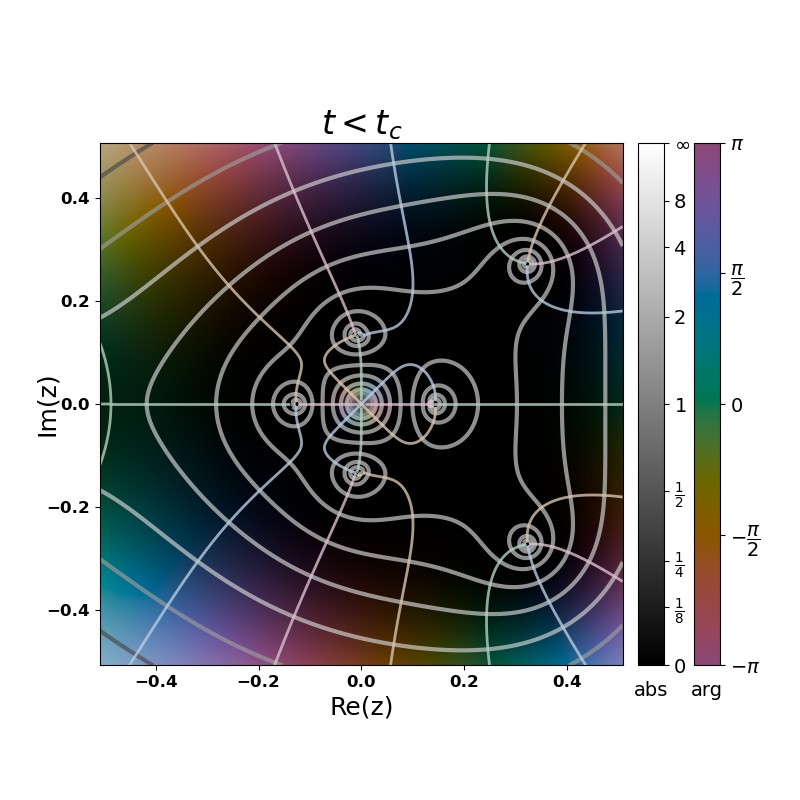}
	\end{tabbing}
	\vspace{-0.7cm}
	
\end{figure}
\vspace{-1.5cm}

\begin{figure}[H]
	\begin{tabbing}
		\hspace{-2cm}
		\centering
		%		\hspace{-2.3cm}%\=\kill
		\includegraphics[scale=.33]{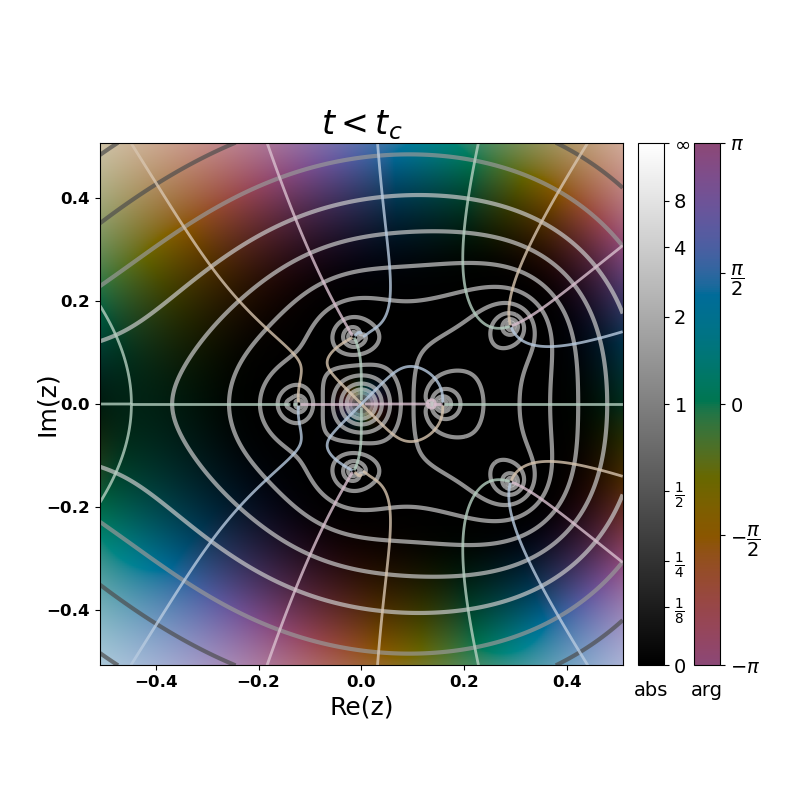}%\>
				\hspace{-0.4cm}
		\includegraphics[scale=.33]{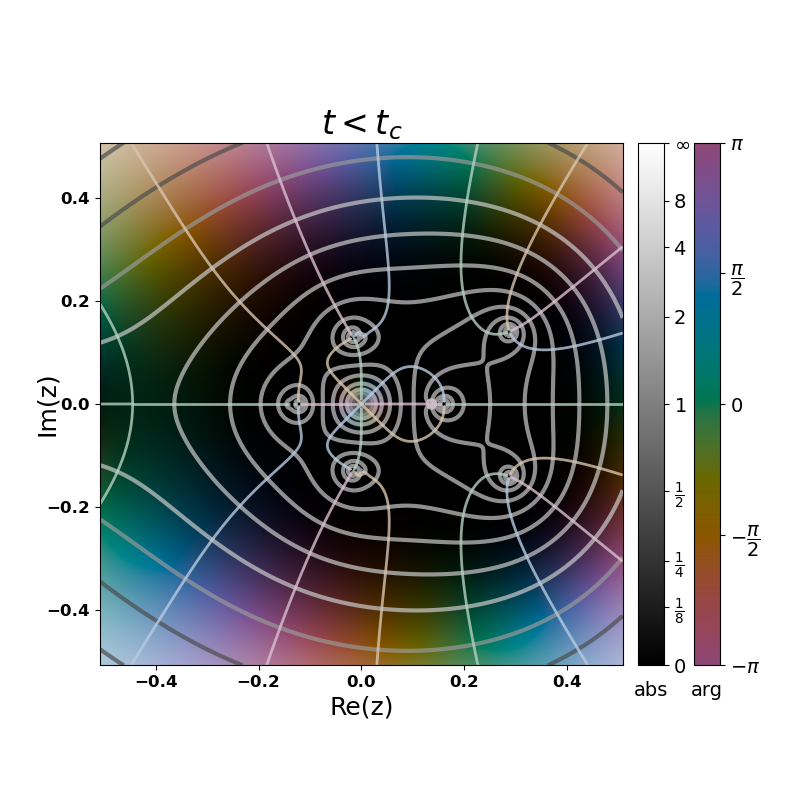}%\\
		\hspace{-0.4cm}
		\includegraphics[scale=.33]{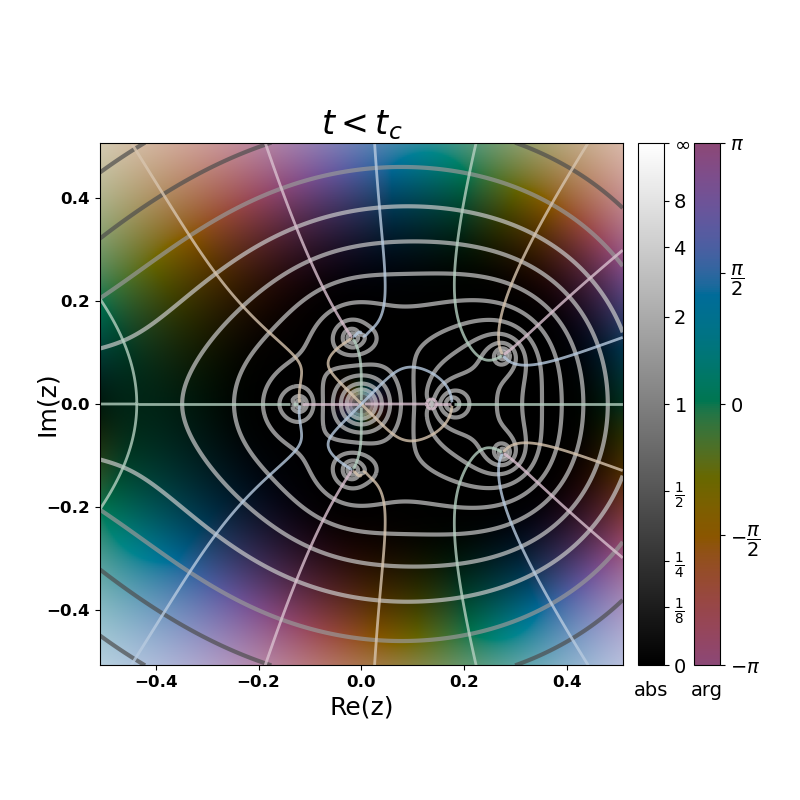}%\\
	\end{tabbing}
	\vspace{-0.7cm}
\end{figure}

\vspace{-1.5cm}

\begin{figure}[H]
	\begin{tabbing}
		\hspace{-2cm}
		\centering
		%		\hspace{-2.3cm}%\=\kill
		\includegraphics[scale=.33]{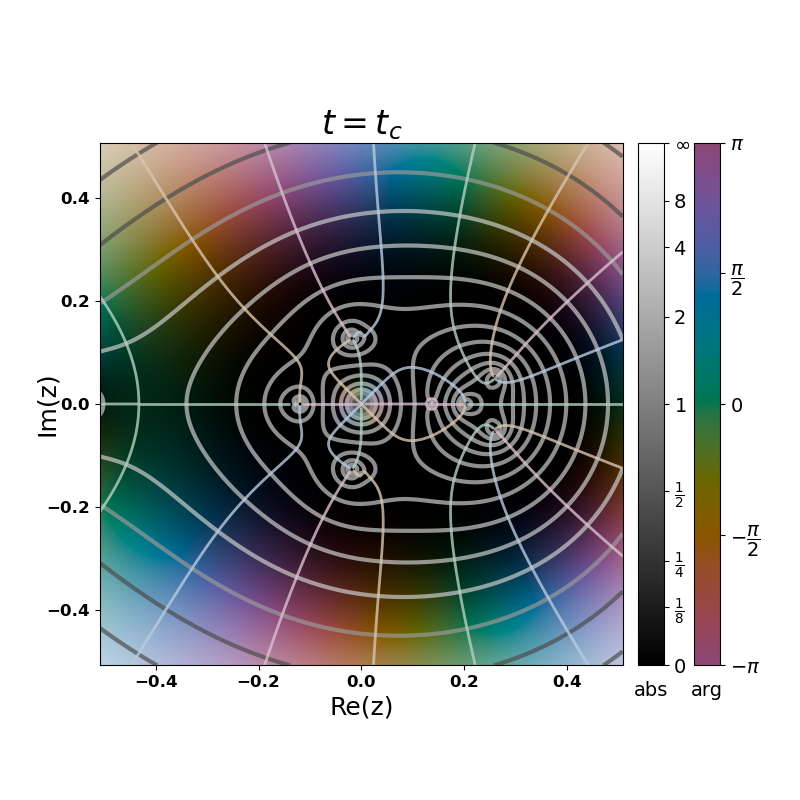}%\>
		\hspace{-0.4cm}
		\includegraphics[scale=.33]{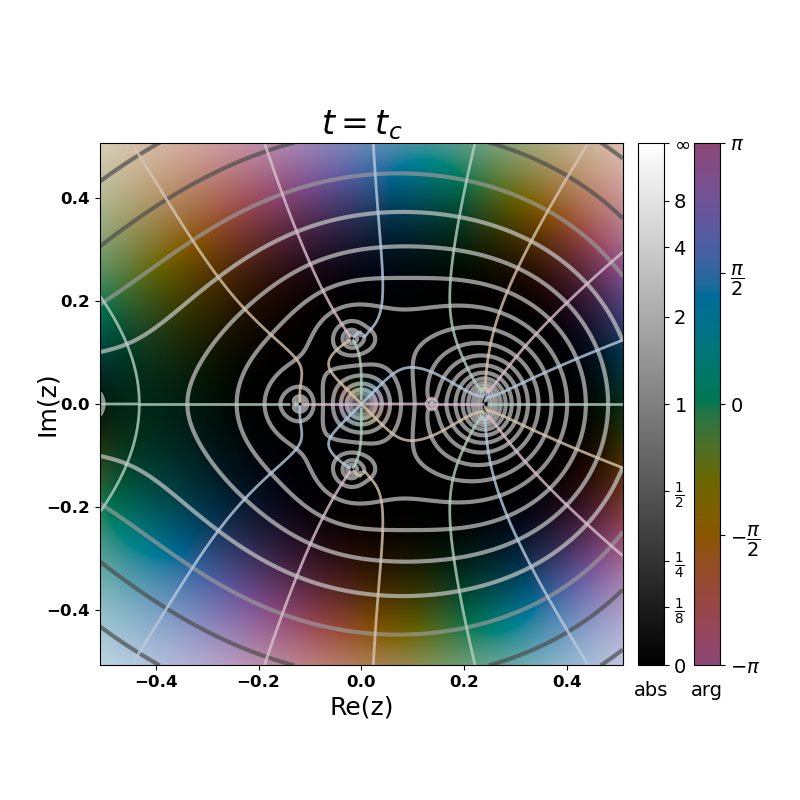}%\\
		\hspace{-0.4cm}
		\includegraphics[scale=.33]{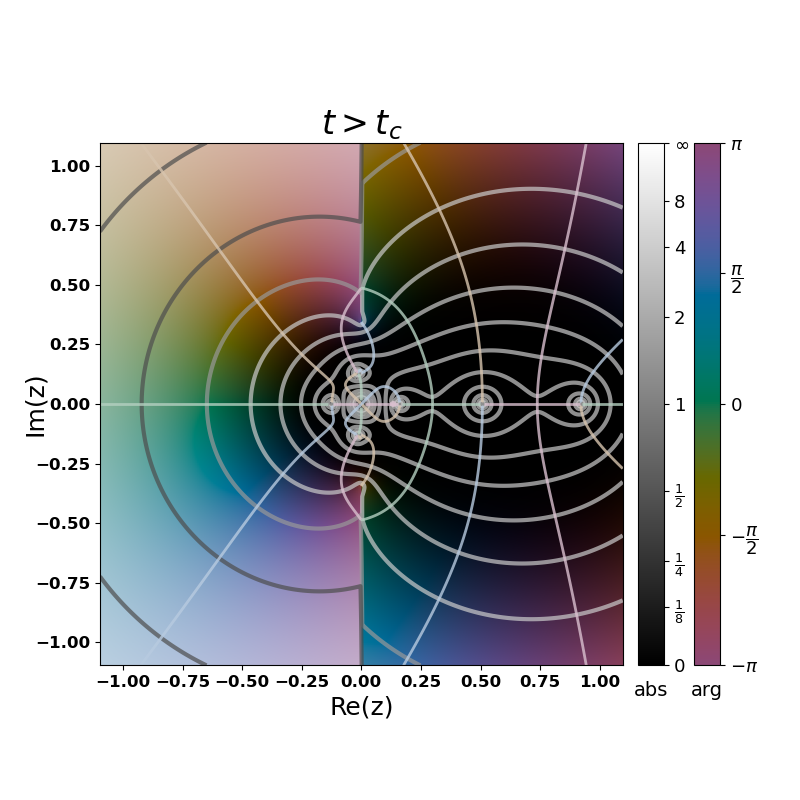}%\\
	\end{tabbing}
	\vspace{-1cm}	
\end{figure}
\vspace{-1.8cm}
%\end{landscape}
%
%\begin{equation}
%\begin{split}
%h_Q(z)&=\frac{1}{256 (d-2) \pi \Gamma \left(\frac{d-1}{2}\right)}\\\\&\times \frac{\pi ^{-\frac{d+3}{2}}}{z^{d}} \left[(d-2) \pi^d z^{2d-2}\left(\frac{64 \pi ^{3-d} Q^2 \Gamma ^2\left(\frac{d-1}{2}\right) z^{6-2 d}}{d-2}+8d-24\right)-128 \pi ^3 Q^2 z^4 \Gamma ^2\left(\frac{d-1}{2}\right)\right]\\\\&\times z^{1-d}\Gamma \left(\frac{d-3}{2}\right) \left[\frac{256 p \pi \pi ^{\frac{d+3}{2}} z^{2 d+2}}{(d-1) \left((d-2) \pi ^d z^{2 d}-2 \pi ^3 (d-3) Q^2 z^6 \Gamma ^2\left(\frac{d-3}{2}\right)\right)}+4 (d-3) \pi ^{\frac{3}{2}-\frac{d}{2}}\right]-t
%\end{split}
%\end{equation}

\begin{figure}[H]
	\centering
	\begin{tabbing}
		\centering
	\hspace{1.8cm}%\=\kill
		\includegraphics[scale=.3]{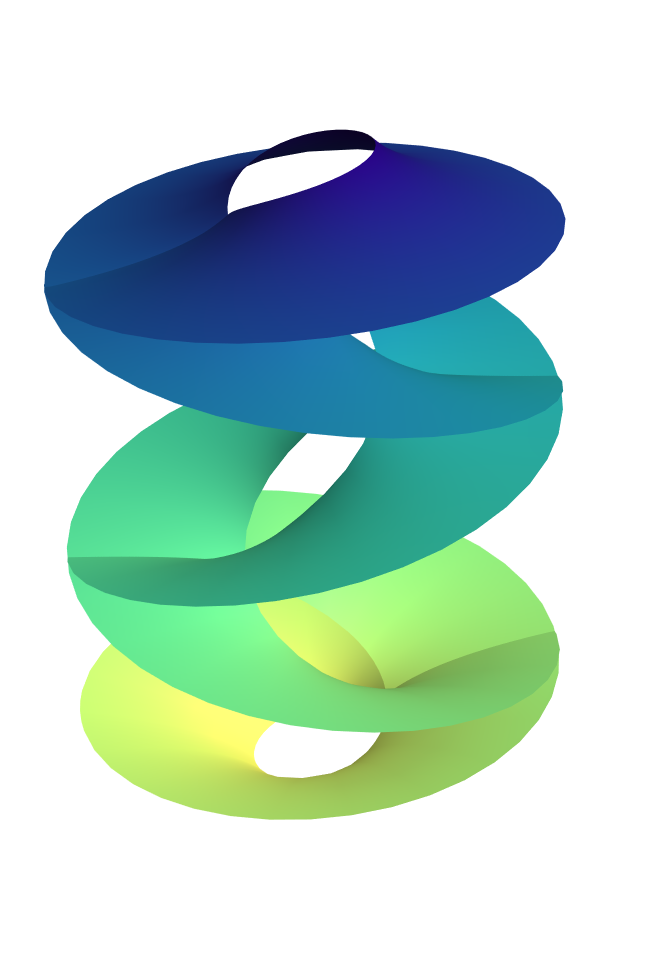}%\>
		\hspace{2.cm}
		\includegraphics[scale=.3]{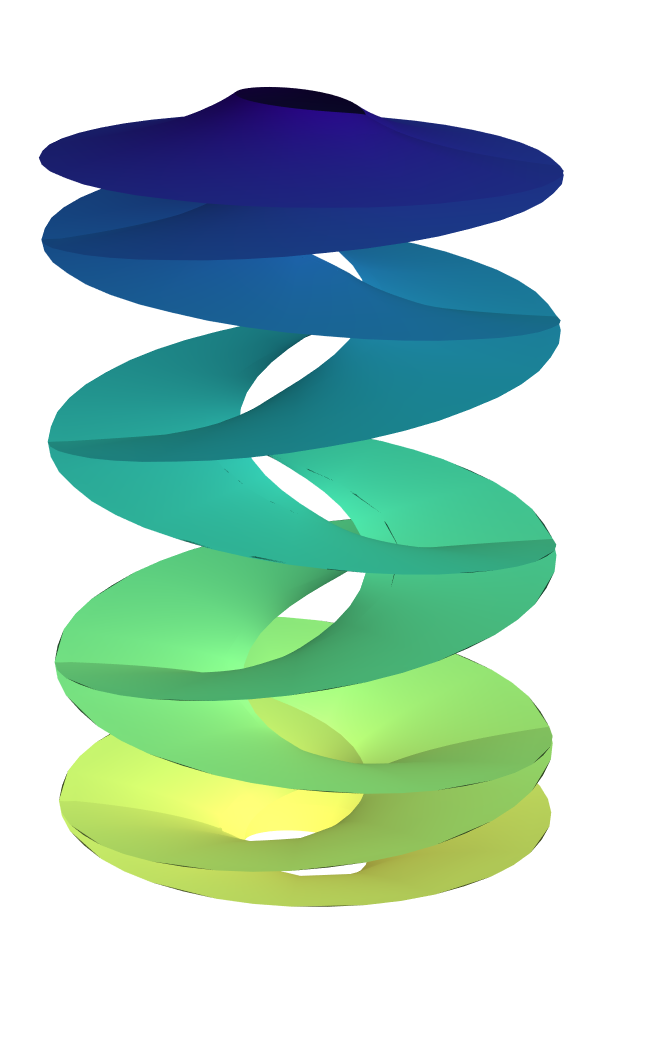},		
	\end{tabbing}
	\vspace{-1.1cm}
	\caption{\footnotesize Riemann surfaces associated with the Van-der-Waals phase transitions of $\;5-$ and $7-$dimensional R\'enyi charged-flat black holes. The winding numbers are $5$ and $9$, respectively.}\label{fig:fig_Rie_higher_Q}
\end{figure}

In Fig.\ref{fig:fig_Rie_higher_Q}, we display the Riemann surfaces related to the Van der Waals phase transitions of the Rényi-flat black hole in higher dimensions. The winding number $\mathcal{W}_{VdW}$ matches the number of sheets that form the Riemann surface, therefore
\begin{equation}\label{winding_num}
\mathcal{W}_{VdW}=2d-5
\end{equation}

\paragraph{}  We point out that for the $d$-dimensional asymptotically AdS charged black hole, an equal winding number is found, in particular, in four dimensions, the Van-der-Waals phase transition possesses a winding number equal to $\mathcal{W}_{VdW}=3$.

\paragraph{} Here, we propose a generalization of the formulas for the maximum number of Hawking-Page and Van der Waals phase transitions, Eqs.\eqref{N2} and \eqref{N1}, to a general $d$-dimensional asymptotically flat Rényi charged black hole with a given winding number $\mathcal{W}$, such that
\begin{eqnarray}
    N_2=\left\lfloor\frac{\mathcal{W}}{d-3}\right\rfloor, \quad N_1=\left\lfloor\frac{\mathcal{W}}{2d-5}\right\rfloor  
\end{eqnarray}

By comparing the Riemann surfaces of the Hawking-Page phase transitions in different dimensions one can infer that they are topologically unequivalent primarily due the their different number of sheets, the same is true for the Van der Waals phase transitions. Moreover, the sequences of winding numbers for the Hawking-Page and Van der Waals transitions have no common numbers, that is the equation $\mathcal{W}_{HP}$=$\mathcal{W}_{VdW}$ has no solution for $d>3$. This fact establishes the profound difference between the two types of phase transitions from a complex analysis and topological perspectives.

\section{Conclusion}

Understanding phase transitions is crucial for comprehending the nature and evolution of black hole thermodynamic systems. In this study, we explore the profound connection between black hole phase transitions and winding numbers derived through complex analysis within the framework of Rényi statistics. By incorporating the nonextensive nature of black hole entropy, we have predicted the specific types of phase transitions that black holes can undergo. Indeed, we have mapped out the behavior of black holes during various phase transitions, such as those analogous to Hawking-Page and Van der Waals transitions, within different thermodynamic ensembles and their association with the number of sheets comprising the Riemann surface.

%\textcolor{red}{Summary of results}

We began by introducing the complexification process of the Bragg-Williams free energy within the Rényi formalism. Such a process elevates this thermodynamic quantity to a well-defined differential complex function, enabling its interpretation as a complex energy flow potential. In this framework, the real part of the function is considered analogous to the velocity potential, while the imaginary part corresponds to the stream function of the energy flow. %Within this analogy, we indicated that the complex Braggs-Williams free energy is a superposition of well-known complex flow potentials representing a flow in a sector with well-defined velocities at infinity. 
Next, we examined the Hawking-Page phase structure that emerges in the grand canonical ensemble. Indeed, the illustrations of the absolute value and phase of the Rényi complex Bragg-Williams free energy indicate the formation of zeros synonymous with the Hawking-Page transition at the temperature $t = t_{HP}$. As the temperature increases, this zero splits into distinct points, corresponding to thermal radiation-large black hole and small-large black hole phase transitions. Moreover, the depiction of the energy flow vectors reveals the behavior of their associated lines, where distortions and reconnections occur, indicating the aforementioned phase transitions and paints a more dynamic picture of the R\'enyi Hawking-Page phase transitions. Besides, We systematically associated a winding number with the second-order Hawking-Page (HP) phase transition, defining it as the number of sheets in the inverse of a multi-valued analytic function. This analysis revealed that $\mathcal{W}_{HP}=1$ in four-dimensional spacetime, establishing that $\mathcal{W}=1$ is the minimum winding number required for a second phase transition to occur in a R\'enyi charged-flat black hole system. Additionally, we introduced the concept of the maximum number of second-order phase transitions for a R\'enyi charged-flat black hole with a given winding number $\mathcal{W}$, denoted as $N_2=\mathcal{W}$. The disparity between the winding numbers associated with the Hawking-Page and Van der Waals transitions vividly illustrates the difference between these two types of phase transitions. Specifically, in a second-order phase transition, the two phases present are indistinguishable, whereas, in a first-order transition, three distinct phases exist, with one serving as an intermediate phase.

In the canonical ensemble, we employed the same complex analysis framework to explore the intricate structure of the Van der Waals (VdW) phase transition. The emergence and dynamics of the complex conjugate zeros, along with the presence of a non-vanishing real zero in the complex free energy representations, highlight a fundamental distinction between first-order (VdW) and second-order (HP) phase transitions. Notably, this non-trivial real zero, representing the small-intermediate black holes transition point, is absent in all second-order phase transitions. This analysis suggests that the difference between these two types of phase transitions extends beyond the existence of an intermediate state. The complex energy flow associated with the Van der Waals transition resembles a doublet flow, and the structure of streamlines, as well as their distortions and reconnections, differs significantly from the Hawking-Page case.

Furthermore, we demonstrated that the Van der Waals phase transition for the R\'enyi charged-flat black hole is characterized by a winding number $\mathcal{W}_{VdW}=3$ in four-dimensional spacetime, which is also the minimum winding number required for a first-order phase transition to occur. We then defined the maximum number of first-order phase transitions for a R\'enyi charged-flat black hole with a given winding number $\mathcal{W}$ as $N_1=\floor{\frac{\mathcal{W}}{3}}$.

Finally, we extended our investigations to arbitrary dimensions and generalized the winding numbers formulas to $d$-dimensional spacetimes. Furthermore, we revealed a holographic-type connection in the Hawking-Page case such that the behavior in $d$-dimensions at very low temperatures $t\ll t_{min}$ is up to a scale and phase factors identical to the $d+1$-dimensional one at very high temperatures $t\gg t_{hp}$. The structure of the Riemann surfaces demonstrates not only the intrinsic topological differences among the same black holes phase transitions in arbitrary dimensions but equally draws attention to topological inequivalence between Hawking-Page and Van der Waals transitions in different dimensions.
%..... VdW \& high dim.....

%---------------------

This novel approach underscores the importance of complex analysis and nonextensive statistics in studying black hole thermodynamics and sets the stage for future research that could further unravel the complexities of gravitational systems and their thermodynamic properties especially a possible equivalence between the Gibbs-Boltzmann statistics and the Rényi one.

% \begin{frame}{}
% 	\centering
%	\animategraphics[autoplay,loop,width=0.7\linewidth]{0.8}{f_abs-}{0}{8}
%\end{frame}

%\newpage
%\appendix
%\numberwithin{equation}{section}

%
\bibliographystyle{unsrt}
\bibliography{RNRiemann.bib} 
\end{document}